\documentclass[12pt]{article}
\usepackage[doublespacing]{setspace}
\usepackage{amsfonts,amssymb,amsmath}
\usepackage{color}
\usepackage[usenames,dvipsnames,svgnames,table]{xcolor}
\usepackage{float}
\usepackage{libertine} 
\usepackage[T1]{fontenc}
\usepackage{libertinust1math}
\usepackage{pict2e}
\usepackage{enumitem}
\usepackage{pdflscape}
\usepackage{placeins}
\usepackage[normalem]{ulem}
\usepackage{lscape}
\usepackage{mathtools}
\usepackage{booktabs}
\usepackage{marginnote}         
\usepackage{versionPO}          
\usepackage{verbatim}           
\usepackage{caption}

\newenvironment{proof}[1][Proof]{\noindent\textbf{#1.} }{\ \rule{0.5em}{0.5em}}
\usepackage{pifont}

\usepackage{threeparttable}

\usepackage{url}
\usepackage{chapterbib} 
\usepackage[sectionbib]{natbib} 
\defcitealias{BradyReport}{Brady Report (1988)} 

\newtheorem{lmm}{Lemma}

\newtheorem{prop}{Proposition}

\newtheorem{col}{Corollary}

\usepackage[demo]{graphicx}
\usepackage{subcaption}
\usepackage[
bookmarks=false,
pdfpagelabels=false,
hyperfootnotes=false,
hyperindex=false,
pageanchor=false,
colorlinks,
citecolor=bleu,
linkcolor=bleu,
urlcolor=bleu
]{hyperref}

\definecolor{bleu}{rgb}{0.06, 0.3, 0.57}
\definecolor{cardinal}{rgb}{0.6, 0.0, 0.0}
\definecolor{rouge}{cmyk}{0, 1,1,0.45}

\excludeversion{notes}     
\includeversion{links}          
\excludeversion{submit}		

\iflinks{}{\hypersetup{draft=true}}

\ifnotes{%
\usepackage[margin=1in,paperwidth=10in,right=2.5in]{geometry}%
\usepackage[textwidth=1.4in,shadow,colorinlistoftodos]{todonotes}%
}{%
\usepackage[margin=1in]{geometry}%
\usepackage[disable]{todonotes}%
}

\makeatletter\let\chapter\@undefined\makeatother 

\DeclareMathOperator{\sgn}{sgn}

\usepackage{soul} 

\graphicspath{{Figures/}}

\makeatletter
\newcommand*{\centerfloat}{%
  \parindent \z@
  \leftskip \z@ \@plus 1fil \@minus \textwidth
  \rightskip\leftskip
  \parfillskip \z@skip}
\makeatother

\global\long\def\E{\mathbb{E}}
\global\long\def\Var{\mathrm{Var}}

\begin{document}

\newcommand{\tit}{Valuation Duration of the Stock Market}
\newcommand{\ack}{
We thank Nick Barberis, Geert Bekaert, Jules van Binsbergen, Zhi Da, Kent Daniel, Joost Driessen, Stefano Giglio, Will Goetzmann, Benjamin Golez, Hui Guo, Bryan Kelly, Ralph Koijen, Lars Lochstoer, Alan Moreira, Riccardo Sabbatucci, Tano Santos, Jos\'{e} Scheinkman, Jessica Wachter, and Michael Weber for in-depth discussions.
}

\title{\tit\ifsubmit{}{\thanks{\ack}}}

\ifsubmit{}{\author{Ye Li\thanks{University of Washington Foster School of Business. E-mail: \href{mailto:liye@uw.edu}{liye@uw.edu}}\and Chen Wang\thanks{University of Notre Dame Mendoza College of Business. E-mail: \href{mailto:chen.wang@nd.edu}{chen.wang@nd.edu}}}}

\date{October 10, 2023}
\maketitle
\thispagestyle{empty}

\begin{abstract}


At the peak of the tech bubble, only 0.57\% of market valuation comes from dividends in the next year. Taking the ratio of total market value to the value of one-year dividends, we obtain a valuation-based duration of 175 years. In contrast, at the height of the global financial crisis, more than 2.2\% of market value is from dividends in the next year, implying a duration of 46 years.
What drives valuation duration? We find that market participants have limited information about cash flow beyond one year. Therefore, an increase in valuation duration is due to a decrease in the discount rate rather than good news about long-term growth. Accordingly, valuation duration negatively predicts annual market return with out-of-sample $R^2$ of 15\%, robustly outperforming other predictors in the literature.
While the price-dividend ratio reflects the overall valuation level, our valuation-based measure of duration captures the slope of the valuation term structure. We show that valuation duration, as a discount rate proxy, is a critical state variable that augments the price-dividend ratio in spanning the (latent) state space for stock-market dynamics.



\end{abstract}

\clearpage
\setcounter{page}{1}

\listoftodos[TO-DO]
\doublespacing


\section{Introduction}


We define valuation duration of the stock market as the ratio of total market capitalization to the price of the dividend in the very next year (the one-year dividend strip price in \citealp*{BBK2012}).
In March 2000, the U.S. stock market has a valuation duration of 175 years, which suggests that at the peak of the dot-com bubble, the U.S. stock market derives 99.4\% of value from dividends beyond the next year. Duration varies significantly over time, with a monthly standard deviation of 1.6 years in the last three decades. At the market bottom during the global financial crisis, the stock market duration is only 46 years  in March 2009 with more than 2.2\% of market value coming from dividends paid within the next year.

What drives the relative valuation of long- vs. short-term dividends?
An increase in market duration represents a steepening of the valuation term structure, with the long-term dividends commanding a higher valuation relative to dividends in the near term. While duration represents the slope, the standard price-dividend ratio captures the overall level of market valuation. What information can we extract from the dynamics of slope and level of the valuation term structure?

We find that market duration and the price-dividend ratio span the state space of equity market.
Stock prices are determined by the conditional expectations of future returns and dividends, so a model of the aggregate stock market requires at least two state variables for return and cash-flow dynamics \citep*[e.g., ][]{LettauWachter2007, Binsbergen2010, KragtDejongDriessen2020}. In other words, the lower bound of the state-space dimension is two. Our analysis shows that two is also the upper bound. Market duration and the price-dividend ratio as the two state variables contain sufficient information for forecasting returns and future dividends. Moreover, market duration is the state variable that corresponds closely to the conditional expected return. We show that this strong connection between market duration and the discount rate is due to the striking difference in cash-flow predictability at short and long horizons.

Identifying state variables is an important task in asset pricing and macro finance as it lays the foundation for various research topics.\footnote{Previous studies on macro dynamics and asset pricing based on consumption, production, behavioral biases, and intermediation frictions have motivated a variety of state variables \citep*[e.g.,][]{Sims1980, Cochrane1991, CampbellAmmer1993, BSV1998, Campbell1999, BHS2001, LettauLudvigson2001, AngPiazzesi2003, BansalYaron2004, Campbell2004c, KaltenbrunnerLochstoer2010, Lettau2011, Gabaix2012, He2013, Wachter2013, Kelly2013b, BrunnermeierSannikov2014, Barberis2015, Muir2017, Campbell2018, Campbell2020, Cieslak2021}.}
Our findings show that beyond the standard price-dividend ratio, valuation duration is another key state variable for the equity market, and it is particularly powerful in revealing the discount-rate dynamics.
Next, we summarize the key steps of our analysis and our contributions to the asset pricing literature in three areas, state-space modeling, return predictability, and cash-flow expectations.

Our analysis starts with a characterization of the equity-market state space.
We explore information directly from the prices of dividend strips rather than select macro or market variables. The logarithm of dividend strip prices at different maturities scaled by the realized dividend are linear functions of (latent) state variables in an exponential-affine setup \citep*[e.g., ][]{LettauWachter2007} 
or through log-linearization \citep[e.g., ][]{Campbell1988, Binsbergen2010}. Therefore, this set of valuation ratios of dividend strips fully reveals the state variables as long as the number of dividend strips is at least as large as the number of state variables and the valuation ratios have linearly independent state-variable coefficients.

We find that two principal components drive the dynamics of valuation ratios of dividends at different maturities. Importantly, when forecasting market return and dividend growth, two valuation ratios are sufficient to maximize the predictive power. While different pairs of valuation ratios possess different predictive power, the pairs with the strongest predictive power perform just as well as combinations of three or more valuation ratios.
Our findings suggest that the state space is two-dimensional. Next, we show that the log market duration ($dr$), i.e., the logarithm of total market value scaled by one-year dividend price, and the log price-dividend ratio ($pd$) possess a predictive power for both return and dividend growth that is at least as strong as any combination of dividend valuation ratios and thus can serve as the pair of state variables.

A striking finding from our forecasting exercises is that $dr$ alone is sufficient for forecasting returns. Augmenting $dr$ with $pd$ or other valuation ratios of dividend strips does not improve the predictive power. In fact, the return predictive power of $dr$ not only subsumes that of $pd$ and other valuation ratios but also surpasses other predictors in the literature, such as those summarized in \cite{WelchGoyal2007} and more recent studies.
Specifically, $dr$ demonstrates remarkable in-sample and out-of-sample (OOS) return predictability of annual market returns, reporting substantial $R^2$ of 25\% and 15\%, respectively. The OOS $R^2$ is in sharp contrast to the almost zero OOS $R^2$ from $pd$. Moreover, the predictive power of $dr$ survives the \citet{Hodrick1992} adjustment for standard errors, the \citet{Stambaugh1999} adjustment for small-sample bias, and tests for out-of-sample predictive power such as encompassing (ENC) and Clark-West (CW) tests.


These findings are critical for understanding both the discount-rate dynamics and the fluctuation of market duration. If the conditional expected return is a univariate function of $dr$, then we can rearrange the equation and solve $dr$ as a univariate function of the conditional expected return, which implies that the variation in duration is fully discount-rate driven. To analyze the connection between $dr$ and the discount rate, we set up a two-dimensional state space model. Without loss of generality, the (latent) state variables are the conditional expected return and expected dividend growth rate that follow AR(1) processes \citep{LettauWachter2007, Binsbergen2010}. $dr$ and $pd$ are bivariate linear functions of state variables, and vice versa.

In our model, a necessary and sufficient condition for the conditional expected return to be a univariate function of $dr$ (and vice versa) is that the market does not contain information on cash-flow growth beyond the very next year. Intuitively, under this condition, the price of next year's dividends exhausts all the information about future cash flows, so $dr$, which is the log market value minus the log price of next year's dividend, teases out cash-flow information and only contains information about the discount rate, thus becoming a univariate function of the conditional expected return. Given that the expected dividend growth rate follows an AR(1) process, this condition translates into a zero autoregressive coefficient of the expected dividend growth rate: All the relevant information about future cash flows---the current expected growth rate and the history of realized shocks---is not propagated into the future beyond the next year.\footnote{Given a two-dimensional state space, adding more lags to the autoregressive processes of expected return and expected dividend growth is not necessary. If AP(p) or ARMA(p,q) models were required or if other (macroeconomic or market) variables feed into the dynamics of expected return and expected cash-flow growth rate, we should have found the state-space dimension to be higher than two because to span the state space, we must augment the current conditional expectations of next-period return and cash flow growth rate with lagged conditional expectations.}

To estimate the autoregressive coefficient (persistence) of the expected dividend growth rate, we use analyst forecasts to proxy for the expected cash-flow growth and estimate two econometric models of belief dynamics that take advantage of, respectively, analysts' short-term and long-term forecasts. As an alternative method, we also fit a state-space model to dividend data to estimate the persistence of expected dividend growth. The message from our findings is consistent: Cash-flow growth expectation lacks persistence. This provides an explanation of the strong connection between market duration and expected return.

We also show analytically that return prediction errors should comove with the value of the autoregressive coefficient of cash-flow growth expectations. To test this prediction, we conduct a rolling-window estimation. In each window, we estimate the autoregressive coefficient of the expected cash-flow growth and the forecast error from predicting returns with $dr$. Both the in-sample and out-of-sample forecasting errors comove with the value of the autoregressive coefficient, suggesting that when the market contains limited information about cash-flow growth beyond one year, the mapping between $dr$ and the expected return strengthens.

In our model, a zero autoregressive coefficient of the expected cash-flow growth rate fully captures the fact that the market contains limited information about cash-flow growth beyond the very next year. The emphasis on this autoregressive coefficient is warranted because the expected dividend growth rate follows an AR(1) process. While our analysis of the state-space dimension supports this model specification, we still go beyond our model to characterize market participants' cash-flow expectations.
First, we show that dividend and earnings growth within one year are highly predictable, especially by analysts' forecasts.\footnote{This can be explained by firms offering forward guidance on near-term performance.}
The $R^2$ is 73\% from forecasting the near year's earnings growth. In contrast, cash-flow predictability beyond one year is weak.
For growth from the next one to two years, the in-sample $R^2$ of our best prediction model is only 8\%. It declines to $7\%$ for growth between the second and third years.

Our findings on the striking difference between short- and long-term cash-flow predictability lend support to $dr$ as a discount-rate proxy.
\cite{Cochrane2011} points out that the price-dividend ratio, $pd$, may predict future returns if cash flows are not predictable. We show that short-term cash flows are in fact highly predictable but long-term cash flows are not. This finding then suggests that the discount-rate proxy should be market duration, $dr$, which captures the relative valuation of long- vs. short-term cash flows and reflects the slope of the valuation term structure, rather than $pd$, which reflects the overall level of valuation. Consider a market timing strategy based on the return predictive power of $dr$. Because $dr$ has a negative predictive coefficient, the strategy reduces market exposure when duration increases
and increases market exposure when duration decreases. The strategy delivers a Sharpe ratio of $0.58$, achieving an improvement of 55\% over the buy-and-hold strategy \citep{CampbellThompson2008}. It bets against the valuation of long-term cash flows rather than the overall level of valuation that is reflected in $pd$.

Beyond time-varying discount rate, return predictability typically entertains an alternative interpretation based on mispricing. The profitability of $dr$-based market timing strategy
can be explained by market participants' lack of information on long-term growth and the associated over- or under-valuation of cash flows at long horizons. Moreover, the market timing strategy also avoids betting against the valuation of short-term cash flows that market participants are informed about. When the valuation of long-term cash flows rises, the resultant increase in market duration signals exuberance, while when the valuation of long-term cash flows declines, market participants may be overly pessimistic about long-term growth. This interpretation is consistent with the emphasis in \cite*{Bordalo2023} on market participants' errors in forecasting long-term growth as a key driver of asset prices.

\paragraph{Literature.} Asset prices are typically modeled as functions of state variables.
One approach to selecting state variables is through economic theories (e.g., \citealp{Campbell1999}, \citealp{BansalYaron2004}). Our paper contributes to the alternative approach of state space modeling \citep[e.g., ][]{Duffee2002jf, DaiSingleton2003}. Using the valuation ratios of dividend strips, we empirically analyze the dimension of the state space and provide support to the widely adopted assumption of a two-dimensional state space \citep*[e.g.,][]{LettauWachter2007, Binsbergen2010, KragtDejongDriessen2020}. Moreover, we show that our valuation-based measure of duration and the commonly used price-dividend ratio map out the state space of stock market. Our work builds upon the previous research on measuring the prices of dividend strips \citep*{BBK2012, BHKV2013, Binsbergen2015,CejnekRandl2016, CejnekRandl2020, CejnekRandlZechner2021, GolezJackwerth2023, GiglioKellyKozak2023}. While this literature emphasizes characterizing the term structure of equity risk premium, our paper instead focuses on the term structure of equity valuation ratios and how to use valuation ratios to map out state variables.\footnote{There is an extensive literature on the term structure of equity risk premium \citep*[e.g., ][]{LettauWachter2007, Lettau2011, Hansen2008, Croce2014, Belo2015,Ai2018,BackusBoyarchenkoChernov2018, Bansal2021, Goncalves2021jf,Gormsen2021,Boguth2022}.}

Our approach of extracting state variables from disaggregated valuation ratios is similar in spirit to that in \citet{Kelly2013b} but differs in several crucial ways. \citet{Kelly2013b} decompose the stock market into individual stocks and filter out state variables from the valuation ratios of individual stocks. We decompose the market by the horizon of aggregate cash flows.\footnote{There is a large literature on measuring the cash-flow duration for the aggregate market based on accounting information \citep*{Binsbergen2021, Golez2023} and for individual stocks based on accounting information \citep*{Dechow2004, Da2009, Chen2011, Weber2018, Goncalves2021, Walter2022} or derivative prices \citep{Gormsen2023}. Our paper differs by constructing a valuation-based measure of duration rather than a measure of duration based on the term structure of corporate cash flows.}
By avoiding idiosyncratic noise in firm-level valuation ratios, we do not rely on statistical filtering to extract state variables. The valuation ratios of dividend strips, our measure of market duration, and the price-dividend ratio are already combinations of state variables. Since our two state variables can be measured in real time and directly from market prices, they are less prone to estimation error and thus offer a real-time 
characterization of the state of equity market.

An application of state variables in the finance literature is to predict asset returns. 
We contribute to the voluminous literature on return predictability by offering a novel predictor based on a relative valuation of short- and long-term cash flows (e.g., \citealp*{Fama1988,Campbell1988,LettauLudvigson2001,BakeWurgler2000,Lewellen2004jfe,WelchGoyal2007,Cochrane2007,AngBekaert2007,Lettau2007,CampbellThompson2008,Kelly2013b,RapachRinggenbergZhou2016,Martin2017,Johnson2019,Chen2022,KellyMalamudZhou2022,Bordalo2023}). The construction of our predictor is simple, and it outperforms the other stock-market return predictors across various metrics.
Moreover, we provide an intuitive explanation on the connection between valuation duration and the conditional expected return based on the term structure of cash-flow predictability.

We show that short-term cash flows are highly predictable but long-term cash flows are not, and this is key to identifying a discount-rate proxy. \cite{Cochrane2011} suggests that the price-dividend ratio, $pd$, predicts future returns and serves as a discount-rate proxy if cash flows are not predictable. But our findings on the striking difference in cash-flow predictability at short- vs. long horizons suggest that the proper discount-rate proxy and the state variable driving the conditional expected return should be market duration, $dr$. Our findings on cash flow predictability at different horizons contribute to the literature on cash-flow predictability \citep*{LarrainYogo2008,Binsbergen2010,Koijen2011,BHKV2013,Chen2013,JagannathanLiu2018,Pettenuzzo2020,Golez2023,Sabbatucci2022,Pruitt2023}

Characterizing the dynamics of expected cash-flow growth rate has always been an important topic in the asset pricing literature \citep*{BansalYaron2004, BeelerCampbell2012,Belo2015,CDJohannesLochstoer2016}. Recently a growing body of work focuses on analyzing market participants' expectation formation. \citet*{AdamNagel2022} summarize the latest developments.\footnote{There is a fast-growing body of literature on firm-level cash flow expectations \citep*{LaPorta1996,Dechow1997,Copeland2004,Da2011,PiotroskiSo2012,Bordalo2019,Bouchaud2019,Binsbergen2022} and expectations of aggregate cash-flow growth \citep*{Chen2013,Delao2020,GaoMartin2021, McCarthyHillenbrand2021,Nagel2022, CharlesFrydmanKilic2023,Bordalo2023b,Vasudevan2023}.} Our findings are most related to \cite*{Bordalo2023}, who document that market participants' errors in forecasting long-term growth are key to understanding asset-price fluctuations. Also related, \cite*{Da2011} find that the disparity between long- and short-term earnings expectations drives the cross-sectional difference in average stock returns.
Our contribution is to show that, first, there exists a striking difference in short- vs. long-term cash flow predictability, and second, such difference guides us towards finding a discount-rate proxy. Moreover, we show that the term structure of cash-flow predictability is tied to the persistence of agents' cash-flow expectations. Our findings echo the recent studies on the importance of agents' perceived persistence of key state variables in explaining belief formation and asset prices \citep{Gabaix2019, Wang2020}.

\section{Valuation Ratios and the State Space} \label{sec:model}
We consider a dynamic economy where the information filtration is driven by a Markov process. Specifically, the state of an economy at time $t$ is summarized by $X_t$, a $K$-by-$1$ vector of state variables. We assume that $X_t$ evolves as a first-order vector autoregression
\begin{equation}
	X_{t+1} = \Pi X_{t} + \sigma _X^{\top} \epsilon_{t+1},
\end{equation}
where $\epsilon_{t+1}$ is a $N$-by-$1$ vector of shocks that capture all the news at $t+1$ and are independent over time with normal distribution $N\left(\mathbf{0}, \Sigma \right)$. Note that since any higher-order vector autoregression can be written as a first-order vector autoregression by expanding the number of state variables, the AR(1) specification is without loss of generality. The autoregressive coefficients are given by $\Pi $, a constant $K$-by-$K$ matrix, and $\sigma _X$ is a $N$-by-$K$ matrix of shock loadings.

The growth rate of dividend from $t$ to $t+1$ has a $N$-by-$1$ shock-loading vector $\sigma _D$,
\begin{equation} \label{eq:div}
	\ln \left(\frac{D_{t+1}}{D_{t}}\right)=g_t+\sigma _D^{\top} \epsilon_{t+1},
\end{equation}
where the time-varying expected dividend growth rate is given by
\begin{equation} \label{eq:g}
	g_t=\phi ^{\top} X_t+\overline{g}-\frac{1}{2}\sigma _D^{\top} \Sigma \sigma _D.
\end{equation}
We allow the state-variable loadings, $\phi $, to be any $K$-by-$1$ vector. If $g_t$ does not depend on a state variable, the corresponding element in $\phi $ is zero; likewise, if a shock does not affect the growth rate of aggregate dividend, the corresponding element in $\sigma _D$ is zero.

No arbitrage implies the existence of a stochastic discount factor
\begin{equation} \label{eq:sdf}
	M_{t+1}=\exp \left\lbrace -r_f-\frac{1}{2}\lambda ^{\top}_t\Sigma \lambda _t -\lambda _t^{\top}\epsilon_{t+1}\right\rbrace ,
\end{equation}
where $r_f$ is the one-period risk-free rate
and the $N$-by-$1$ vector of risk prices, $\lambda _t$, is given by
\begin{equation} \label{eq:lambda}
	\lambda _t=\overline{\lambda }+\theta ^{\top} X_t.
\end{equation}
We do not impose any restrictions on the state-variable loadings of the risk prices in $\lambda _t$.

Let $P^n_t$ denote the time-$t$ price of the dividend paid at $t+n$. The no-arbitrage pricing functional gives a recursive equation for the prices of dividend strips: for $n\geq 1$,
\begin{equation}
	P^n_t=\mathbb{E}_t\left[M_{t+1}P^{n-1}_{t+1}\right],
\end{equation}
with the boundary condition $P^0_t=D_t.$
The log price-dividend ratio of the dividend strip with maturity $n$ is given by
\begin{equation} \label{eq:ey}
	s^n_t\equiv \ln \left(\frac{P^n_t}{D_t}\right) = A\left(n\right)+B\left(n\right)^{\top} X_t,
\end{equation}
where $A\left(n\right)$ and $B\left(n\right)$ are deterministic functions of $n$ given by a system of recursive equations 
with initial conditions $A\left(0\right)=0\text{, and }B\left(0\right)=0$ (see equations (\ref{eq:B_app})-(\ref{eq:A_app}) and derivation in Appendix I).

Given $K$ log price-dividend ratios of strips, $\left\lbrace s^{n_i}_t: \forall i\in \{1,2,...,K\}, n_i\in \{1,2,...,n,...\}\right\rbrace $, with a full-rank loading matrix, $\mathbf{B}\left(\left\lbrace n_i \right\rbrace _{i=1}^K\right)\equiv \left[ B\left(n_1\right), B\left(n_2\right), ..., B\left(n_K\right) \right]^T$, the state space is recovered by
\begin{equation}\label{eq:recov}
	X_t=
	\mathbf{B}\left(\left\lbrace n_i \right\rbrace _{i=1}^K\right)^{-1}
	\left[s^{n_1}_t-A\left(n_1\right),\, ...,\, s^{n_K}_t-A\left(n_K\right)\right]^{\top}
\end{equation}
When the rank condition fails, these valuation ratios can still recover part of the state space. Let $J$ ($<K$) denote the maximum number of log price-dividend ratios with linearly independent loadings $B\left(n\right)$ and $\left\lbrace n_i\right\rbrace _{i=1}^{J}$ denote the corresponding set of maturities. We can write (\ref{eq:recov}) as follows
\begin{equation}
	\mathbf{B}\left(\left\lbrace n_i \right\rbrace _{i=1}^J\right)
	X_t
	=
	\left[s^{n_1}_t - A\left(n_1\right),\,...,\, s^{n_J}_t - A\left(n_J\right)\right]^{\top}.
\end{equation}
The stock market acts as a linear mapping, i.e., $\mathbf{B}\left(\left\lbrace n_i \right\rbrace _{i=1}^J\right)$, that compresses the $K$-dimensional state space of $X_t$ into a $J$-dimensional space generated by the log price-dividend ratios. In sum, a collection of log price-dividend ratios of dividend strips (partially) span the state space. For information embedded in the state variables, we analyze these valuation ratios.



Let $P_t$ denote the total stock market capitalization (i.e., the market price of dividends across all maturities). Following the same method, we solve the log price-dividend ratio of the market
\begin{equation}
	pd_{t}=\ln \left(P_{t}/D_{t}\right)=A + B^\top  X_t\text{,}
\end{equation}
where $A$ and $B$ are constants defined in Appendix I. Stock market duration is defined as the log ratio of total market capitalization to the price of dividends paid in the next year.
\begin{align}
	dr_{t} & \equiv \ln \left(P_{t}/P^1_{t}\right)=\ln \left(P_{t}/D_{t}\right)-\ln \left(P^1_{t}/D_{t}\right)=pd_t - s^1_t \nonumber \\
	       & =A-A(1) + \left(B - B(1)\right)^\top X_t.
\end{align}

Therefore, just like the valuation ratios of dividend strips, $pd_t$ and $dr_t$ are also driven by the state variables $X_t$. By having different coefficients of $X_t$, these two variables reveal different information about the state space. Next, we analyze the dimension of the state space using $dr_t$, $pd_t$, and the valuation ratios of dividend strips with different maturities. After showing empirically that the state space is two-dimensional, we argue that $pd_t$ and $dr_t$, span the state space and contain sufficient information for forecasting stock market returns and cash-flow growth.

\section{The Dimension of State Space} \label{sec:dimension}
\subsection{Variable construction and summary statistics}
\paragraph{Constructing dividend strip prices.}
Let $P^{n}_t$ denote the price of the dividend paid in year $n$. First, we calculate $P^{n+}_t$, the price of dividends that are paid after the first $n$ years. Under the risk-neutral measure,
\begin{equation}
	P^{n+}_t
	=e^{-n r_f}\mathbb{E}^{RN}_t\left[\sum_{\tau =1}^{+\infty }e^{-\tau r_f}D_{t+n+\tau }\right]
	=e^{-n r_f}\mathbb{E}^{RN}_t\left[\mathbb{E}^{RN}_{t+n}\left[\sum_{\tau =1}^{+\infty }e^{-\tau r_f}D_{t+n+\tau }\right]\right],
\end{equation}
where the expectation operator, $\mathbb{E}^{RN}_{t+n}\left[\cdot \right]$, was inserted under the law of iterated expectations. Note that the (ex-dividend) stock price at $t+n$ is
\begin{equation}
	S_{t+n}=\mathbb{E}^{RN}_{t+n}\left[\sum_{\tau =1}^{+\infty }e^{-\tau r_f}D_{t+n+\tau }\right],
\end{equation}
so we have
\begin{equation}
	P^{n+}_t=e^{-n r_f}\mathbb{E}^{RN}_t\left[S_{t+n}\right].
\end{equation}
The first component, $e^{-n r_f}$, is ${ZCB}^n_t$, the price of a zero-coupon bond with maturity $n$. The second component is the risk-neutral expectation of stock price, i.e., the futures price, $F^n_t$ \citep{DuffieTextbook}.

We construct the price of dividend strips
using zero-coupon bond prices and stock index futures prices. First, we calculate $P^{1}_t$, the price of the dividend paid in the next year,
\begin{equation}
	P^{1}_t=P_t - P^{1+}_t,
\end{equation}
as the difference between the price of all dividends, i.e., the current stock price $P_t$, and the price of dividends paid after the next year. Following the same method, we calculate the price of dividends paid in the next six months, $P^{0.5}$ from $P_t - P^{0.5+}_t$. In our empirical analysis, we use the valuation ratios of dividend strips with maturity $1$ and $0.5$, i.e., $s^1=\ln(P^1/D_t)$ and $s^{0.5}=\ln(P^{0.5}/D_t)$, and the valuation ratio of dividends paid beyond one year, $s^{1+}=\ln(P^{1+}/D_t)$. Our analysis in the previous section shows that these valuation ratios are different linear combinations of state variables.

\paragraph{Data and summary statistics} For futures prices, we use S\&P 500 index futures, which are the most actively traded stock futures. We obtain S\&P 500 futures prices from Datastream.\footnote{We obtain the daily settlement prices for the S\&P 500 futures. For return and cash-flow prediction at the monthly frequency, we use the settlement price of the last trading day of each month. The maturities of the traded futures contracts vary over time, so to obtain futures prices with constant maturity, we apply the shape-preserving piecewise cubic interpolation to complete the futures curve. The results using linear interpolation are similar. } 
We obtain the zero-coupon bond prices from the Fama-Bliss database. The return and level of the S\&P 500 index are obtained from CRSP. The dividend data is from S\&P Global and obtained via the updated dataset of \citet{WelchGoyal2007}. Our final dataset is from January 1988 to December 2019. The sample starts in 1988 to have high-quality dividend data from S\&P Global and, importantly, a sufficiently liquid futures market without structural changes.\footnote{\citet*{WangMichalskiJordanMoriarty1994JFM} identify structural changes of liquidity in the S\&P 500 futures market in the pre-1987 period, during the market crash, and in the post-1987 period.} After the market crash of October 1987, regulators overhauled several trade-clearing protocols.\footnote{The stock market crash in October 1987 reveals anomalous trading in the futures market that was primarily driven by portfolio insurance (\citetalias{BradyReport}). According to the New York Stock Exchange: ``In response to the market breaks in October 1987 and October 1989, the New York Stock Exchange instituted circuit breakers to reduce volatility and promote investor confidence. By implementing a pause in trading, investors are given time to assimilate incoming information and the ability to make informed choices during periods of high market volatility.''}
Lastly, Fama-French factors at the monthly frequency are obtained from Ken French's website.

\begin{table}[!t]
	\caption[Summary Statistics]{Summary Statistics

		\footnotesize This table reports the number of observations, mean, standard deviation, minimum, maximum, quartiles, and monthly autocorrelation ($\rho $) of the main variables in this paper, including our main return predictor, $dr$ (``duration''),
		the price-dividend ratio $pd$ of the S\&P 500 index,
		the filtered series for demeaned expected returns and dividend growth following \citet{Binsbergen2010} $\mu^{F}$ and $g^{F}$,
		the single predictive factor extracted from 100 book-to-market and size portfolios from \citet{Kelly2013b} $KP$,
		short-term dividend strip price to dividend ratio (0.5 year and 1 year) $\log(P^{0.5}/D)$ and $\log(P^{1}/D)$,
		long-term dividend strip price to dividend ratio (beyond 1 year) $\log(P^{1+}/D)$,
		one-month and one-year log returns of the S\&P 500 index $r^{S\&P}_{t+1/12} $ and $r^{S\&P}_{t+1} $,
		one-month and one-year log market returns from Fama-French market portfolio $r^{MKT}_{t+1/12}$ and $r^{MKT}_{t+1}$,
		and the 1-year dividend growth rate of S\&P 500 index and the Fama-French market portfolio $\log(D_{t+1}/D_t)$ and $\log(D^{MKT}_{t+1}/D^{MKT}_t)$.
		Our sample is monthly observations 1988:01--2019:12.
	}
	\centering
	\footnotesize
	\renewcommand{\arraystretch}{1.3}{
		\begin{tabular}{lrrrrrrrrrr}
	\toprule
	                                                          & obs & mean   & std   & min    & 25\%   & 50\%   & 75\%   & max    & $\rho$ \\ \midrule
	$dr_t$                                                    & 384 & 4.027  & 0.494 & 2.952  & 3.727  & 4.044  & 4.208  & 6.632  & 0.919  \\
	$pd_t$                                                    & 384 & 3.883  & 0.289 & 3.239  & 3.656  & 3.930  & 4.047  & 4.524  & 0.985  \\
	$\mu^{F}_t$                                               & 384 & -0.039 & 0.024 & -0.091 & -0.051 & -0.041 & -0.024 & 0.010  & 0.991  \\
	$KP_t$                                                    & 384 & -0.504 & 0.073 & -0.725 & -0.562 & -0.482 & -0.450 & -0.378 & 0.955  \\
	$s_t^{0.5}\equiv\log(P^{0.5}/D)$                          & 384 & 0.095  & 0.157 & -0.568 & 0.046  & 0.126  & 0.187  & 0.429  & 0.929  \\
	$s_t^{1}\equiv\log(P^{1}/D)$                              & 384 & 0.009  & 0.041 & -0.184 & -0.015 & 0.013  & 0.034  & 0.108  & 0.022  \\
	$s_t^{1+}\equiv\log(P^{1+}/D)$                            & 384 & 3.863  & 0.297 & 3.204  & 3.629  & 3.913  & 4.030  & 4.521  & 0.985  \\
	$r^{S\&P}_{t+1/12} $                                      & 384 & -0.142 & 0.280 & -2.241 & -0.210 & -0.098 & 0.016  & 0.393  & 0.766  \\
	$r^{S\&P}_{t+1}$                                          & 384 & -0.819 & 0.281 & -2.629 & -0.883 & -0.768 & -0.666 & -0.280 & 0.604  \\
	$r^{MKT}_{t+1/12}$                                        & 384 & 0.009  & 0.042 & -0.187 & -0.016 & 0.014  & 0.036  & 0.108  & 0.051  \\
	$r^{MKT}_{t+1}$                                           & 384 & 0.096  & 0.159 & -0.554 & 0.036  & 0.128  & 0.194  & 0.440  & 0.924  \\
	$\Delta d_{t+1}\equiv\log(D_{t+1}/D_t)$                   & 384 & 0.059  & 0.070 & -0.237 & 0.025  & 0.068  & 0.112  & 0.168  & 0.994  \\
	$\Delta d^{MKT}_{t+1}\equiv\log(D^{MKT}_{t+1}/D^{MKT}_t)$ & 384 & 0.058  & 0.081 & -0.207 & 0.018  & 0.051  & 0.107  & 0.262  & 0.962  \\
	$g_t^{F}$                                                   & 384 & 0.019  & 0.059 & -0.233 & -0.002 & 0.031  & 0.056  & 0.132  & 0.939  \\
	\bottomrule
\end{tabular}
	}
	\label{tb:sum}
\end{table}

We construct market duration, $dr_t$, as the logarithm of the ratio of S\&P 500 market capitalization to the price of dividends paid in the next year. The traditional price-dividend ratio is the logarithm of the ratio of S\&P 500 market capitalization to the realized dividend in the last year. Table \ref{tb:sum} reports the summary statistics of $dr_t$, the traditional price-dividend ratio $pd_t$, the valuation ratios ($s^{0.5}$, $s^1$, and $s^{1+}$), the monthly return of S\&P 500 ($r^{S\&P}_{t+1/12}$ where in the subscript $1/12$ denotes one month or $1/12$ of a year), the annual return of S\&P 500 ($r^{S\&P}_{t+1}$), and for comparison, the monthly and annual returns of the Fama-French market portfolio (MKT) ($r^{MKT}_{t+1/12}$ and $r^{MKT}_{t+1}$). Our sample includes monthly observations until 2019, i.e., before the extreme market fluctuations during the Covid-19 pandemic. Our baseline analysis focuses on the returns and dividends of the S\&P 500 index because we construct the strip prices using the S\&P 500 futures data.\footnote{Previous studies of return predictability \citep[e.g.,][]{AngBekaert2007} also use S\&P 500 Index as a market proxy.} For robustness, we also report results using Fama-French market portfolio returns and dividends. Note that we include $\mu^F$ and $KP$, the return predictors from \citet{Binsbergen2010} and \citet{Kelly2013b}, respectively, because these variables are also constructed to extract information on state variables. To highlight our contribution, we will compare our analysis with that in \citet{Binsbergen2010} and \citet{Kelly2013b}.


\begin{figure}[!t]
	\begin{center}
		\includegraphics[scale=0.47]{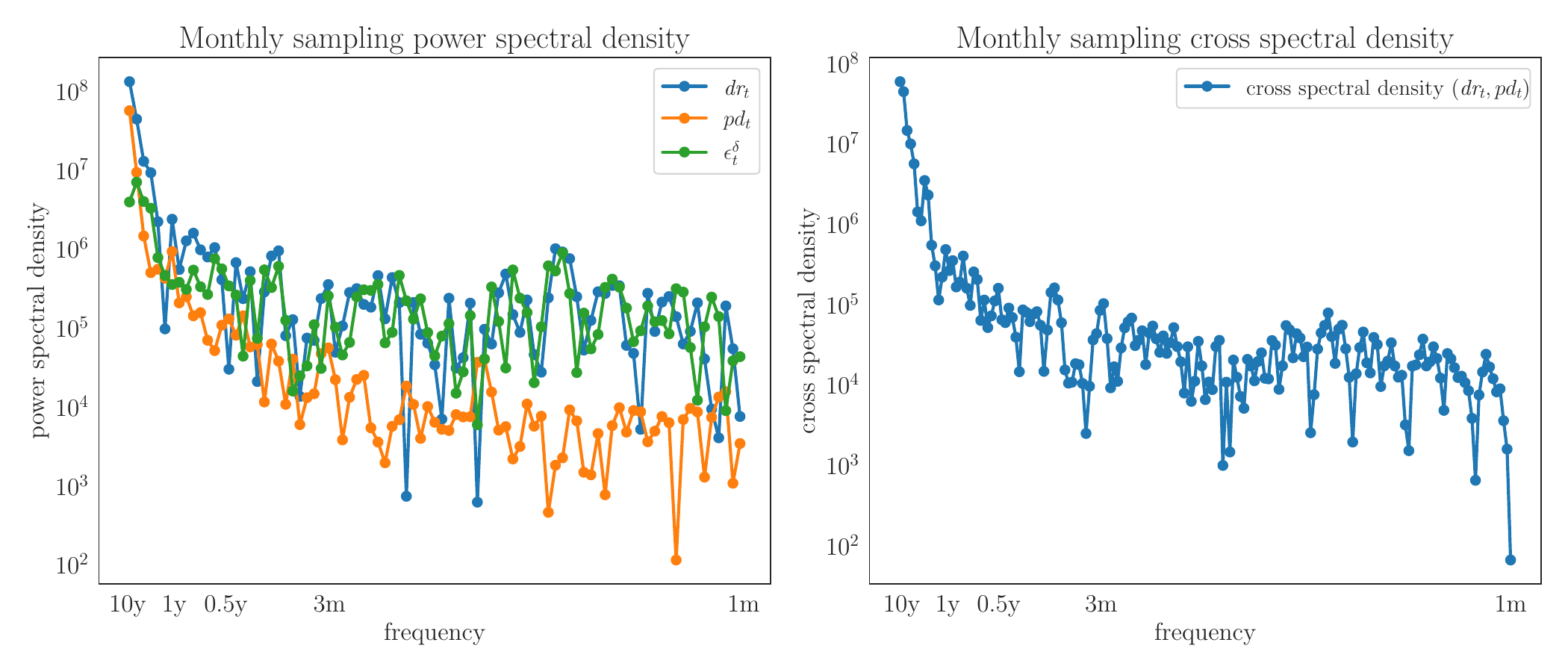}
	\end{center}
	\vspace{-0.25in}
	\caption[fig: spectral analysis]{{Spectrum and Cross-spectrum of Duration $dr$ and Price-Dividend Ratio $pd$.}

		\footnotesize
		The left panel shows the estimated spectral densities of $dr_t$, $pd_t$, and the residuals of $dr_t$ after projecting on $pd_t$ ($\epsilon ^{pr}_t$).
		The integral of spectral density is equal to the variance.
		The horizontal line starts from zero and ends at $\pi $, but is labeled with the corresponding length of a cycle.
		The right panel shows the cross-spectral density between $dr_t$ and $pd_t$.
		The integral of cross-spectral density is equal to the covariance.}
	\label{fig:spectrum}
\end{figure}

The mean of stock market duration $dr_t$ is $4.027$, which translates into $56=\exp(4.027)$ years, meaning that the total market value is $56$ times the valuation of dividends in the next year. $dr_t$ has a wide range of variation, with a minimum of $2.952$ (i.e., 19 years) in November 1988 right before the 1990-1991 recession and a maximum of 6.632 (i.e., 759 years) near the end of the dot-com boom (November 2000). $dr_t$ has a lower monthly autocorrelation (``$\rho $'') than $pd_t$.

$dr_t$ and $pd_t$ are correlated but contain distinct information. As shown in the cross-spectrum in Figure \ref{fig:spectrum}, the correlation of 0.87 is mainly from low-frequency movements.
Panel A of Figure \ref{fig:spectrum} shows the spectrum of $dr_t$, $pd_t$, and $\epsilon^{dr}_t$ (the residual from linearly projecting $dr_t$ on $pd_t$). The area under the spectrum curve is the variance, so the figure provides a variance decomposition in the frequency domain. On the horizontal axis, instead of showing the frequencies from zero to $\pi $, we mark the corresponding length of the cycle for easier interpretation. Once orthogonalized to $pd_t$, $dr_t$'s residual varies mainly at annual or higher frequencies. Panel B plots the cross-spectrum of $dr_t$ and $pd_t$. The integral is the covariance between $dr_t$ and $pd_t$. The correlation between $dr_t$ and $pd_t$ is mainly from low frequencies. This indicates that it is the high-frequency variation in $dr_t$ that brings information distinct from that revealed by the traditional price-dividend ratio.
This is consistent with the findings in \cite*{KragtDejongDriessen2020} about different state variables fluctuating at different frequencies.
Figure \ref{fig:spectrum_daily} in the Appendix shows the spectrum analysis based on daily data with similar results.

As shown in Section \ref{sec:model}, $dr_t$ and $pd_t$ are essentially different combinations of state variables. Our state-space approach is closely related to \cite{Binsbergen2010}. \cite{Binsbergen2010} use the realized returns and dividends to estimate a latent-state model and filter out the conditional expected return, $\mu^F_t$, and the conditional expected dividend growth rate, $g^F_t$. These filtered variables are also combinations of state variables. We replicate the analysis of \cite{Binsbergen2010}, and throughout our analysis in this paper, we compare our state-space representation via valuation ratios with the information about state space in $\mu^F_t$ and $g^F_t$. \cite{Kelly2013b} also take a state-space approach and use the cross-section of market-to-book ratios of individual stocks to extract the conditional expected return of the aggregate stock market. We have also replicated \cite{Kelly2013b} and include $KP_t$ for comparison.

\subsection{Analyzing the dimension of the state space}

\begin{figure}[!t]
	\centering
	\begin{subfigure}{0.45\textwidth}
		\centering
		\includegraphics[width=\textwidth]{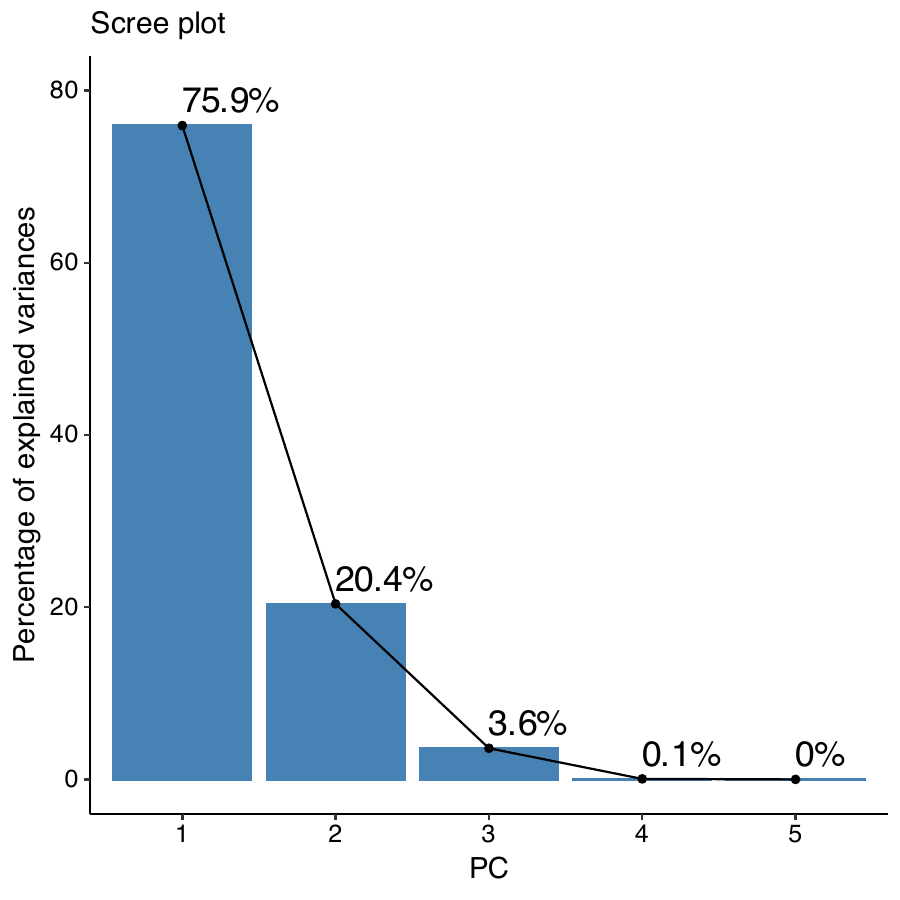}
		\caption{Variance Explained}
	\end{subfigure}
	\hfill
	\begin{subfigure}{0.45\textwidth}
		\centering
		\includegraphics[width=\textwidth]{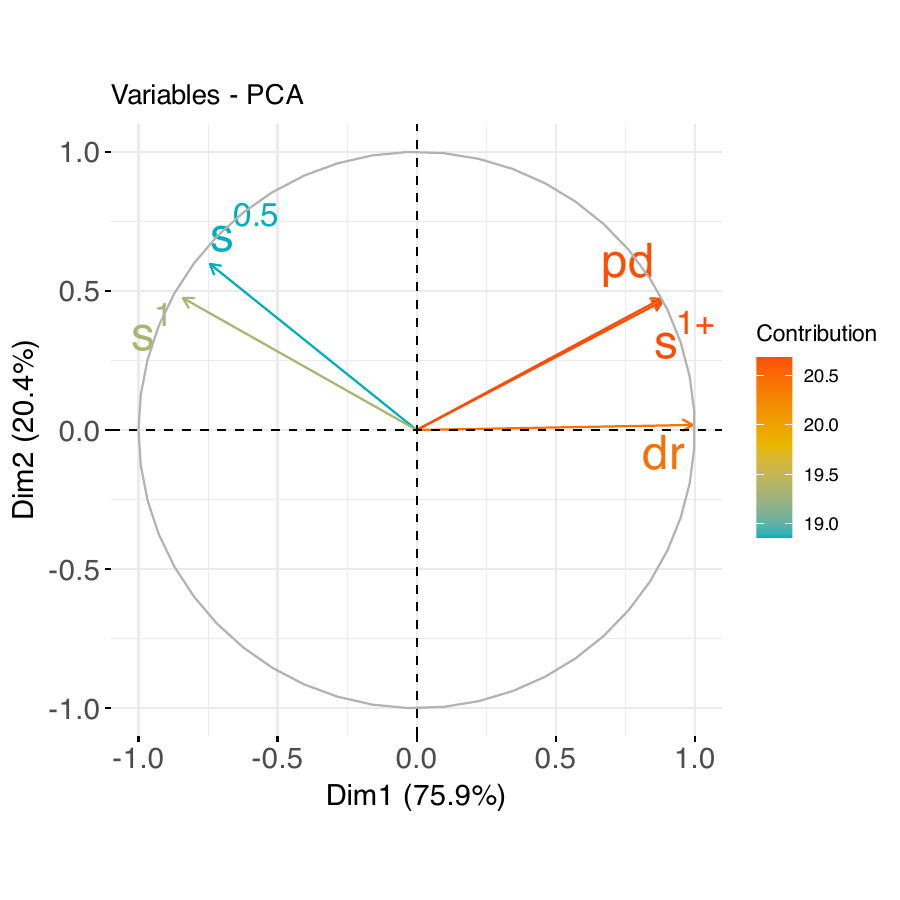}
		\caption{Variable Loading}
	\end{subfigure}
	\caption[pca var5]{Principal Component Analysis of Valuation Ratios

		\footnotesize
		This figure reports the PCA results for $dr$, $pd$, $s^{0.5}_t$, $s^1_t$, and $s^{1+}_t$.
		Panel A plots the variance explained by each principal component.
		Panel B plots the loading of each variable on the first two principal components.
	}
	\label{fig:pca var5}
\end{figure}

We are interested in determining the dimension of the state space. Traditionally, a variety of macroeconomic and financial-market variables have been incorporated into different analytical frameworks as the state variables that generate the relevant information for asset pricing (e.g., \citealp{Cieslak2021}). For example, the market price-dividend ratio, $pd_t$, has been commonly selected as a state variable (\citealp{Muir2017}). However, there has been limited evidence on the dimension of the state space, i.e., how many state variables are needed in an asset pricing framework. Among the theoretical studies, such as \cite{LettauWachter2007} and \cite{Binsbergen2010}, a common assumption is that two state variables are sufficient for characterizing the stock market dynamics and, in particular, for predicting returns and cash-flow growth. In these models, these two state variables are often directly specified as the conditional expectation of return and the conditional expectation of cash-flow growth.

According to our analysis in Section \ref{sec:model}, one way to determine the dimension of the state space is to analyze the collection of valuation ratios, such as those of dividend strips, $pd$, and our measure of market duration $dr$. In Panel A of Figure \ref{fig:pca var5}, we report the results from principal component analysis (PCA) of $dr$, $pd$, and valuation ratios of dividend strips with maturities of six months, one year, and of dividends paid beyond one year. The first two components account for 96.3\% of total variance. In correspondence with our theoretical analysis in Section \ref{sec:model}, we show in Panel B of Figure \ref{fig:pca var5} that these valuation ratios have different loadings on the two principal components, labeled as Dim1 (dimension 1) and Dim2 (dimension 2).

The results in Figure \ref{fig:pca var5} indicate that the state space, mapped out by the valuation ratios, is likely to be two-dimensional, and that we may pick any two valuation ratios, for example, the pair of $pd$ and $dr$, to span the state space. However, as pointed out by \cite{Kelly2015}, a shortcoming of PCA analysis is that the information embedded in the principal components may not be the information most relevant for objects of interest, which, in asset pricing studies, are cash-flow and return dynamics. Therefore, we also take a predictive regression approach to analyze the state space. The expected return and expected dividend growth rate are driven by the state variables. By projecting future returns and dividend growth rates on the valuation ratios, we are able to evaluate how many valuation ratios are needed to achieve the highest predictive power and thereby analyze the dimension of the underlying state space based on information relevant to cash-flow and return dynamics.

\begin{figure}[!t]
	\centering
	\includegraphics[width=\textwidth]{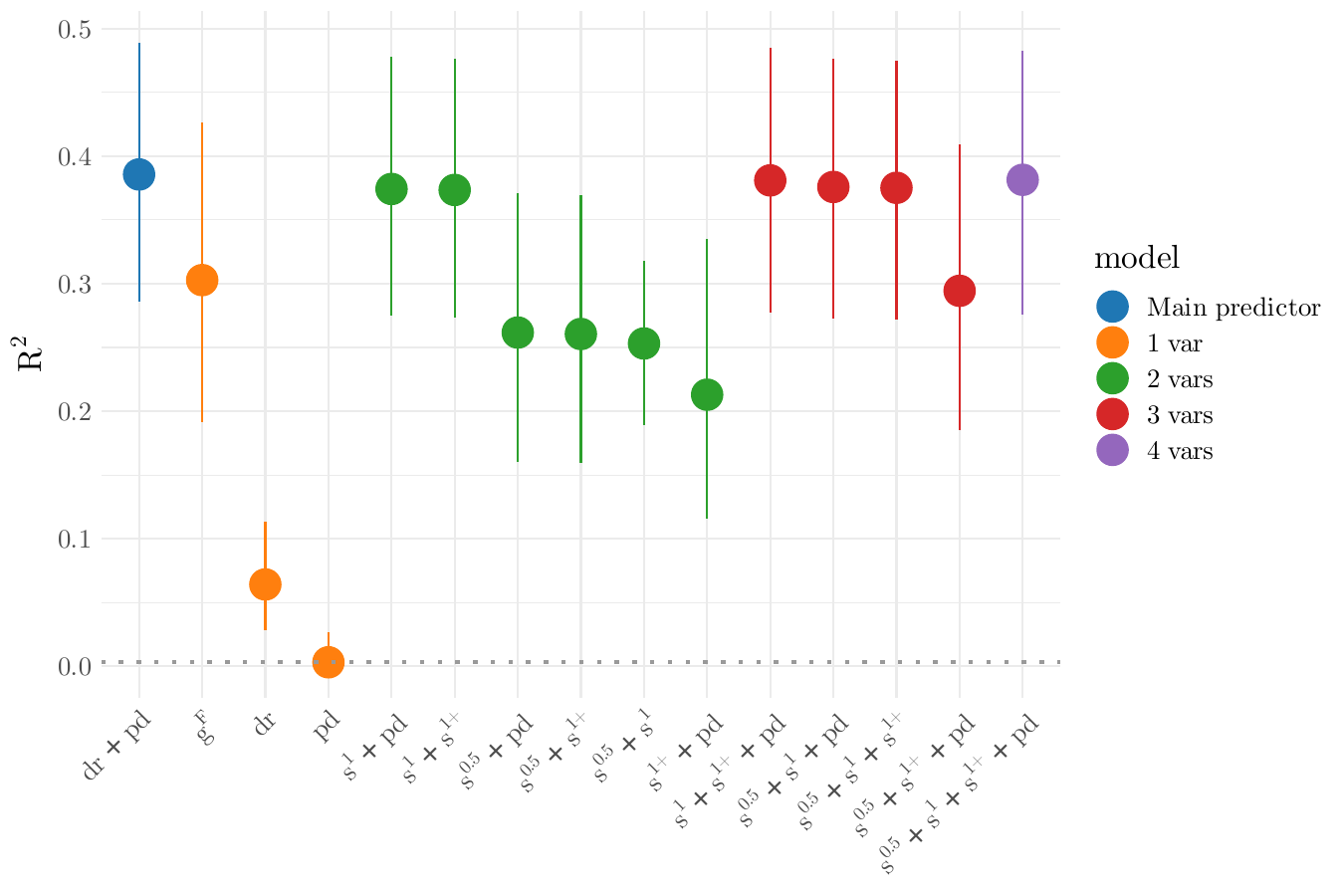}
	\caption[fig: bootstrap r2 dividend growth]{$R^2$ from Dividend Growth Predictive Regressions with Bootstrapped Confidence Interval.

	\footnotesize
	This figure reports $R^2$ from one-year S\&P 500 Index dividend growth predictive regressions with bootstrapped confidence interval.
	The predictors include our main predictor -- a linear combination of `duration' and 1-year dividend strip price to dividend ratio $dr+s^{1}$,
	the price-dividend ratio $pd$,
	short-term (0.5- and 1-year) dividend strip price to dividend ratio ($pd^{0.5}$ and $pd^1$),
	long-term (beyond 1-year) dividend strip price to dividend ratio ($pd^{1+}$),
	the filtered series for demeaned dividend growth following \citet{Binsbergen2010} $g^{F}$,
	the single predictive factor extracted from 100 book-to-market and size portfolios from \citet{Kelly2013b} $KP$,
	and all combinations of $pd$, $pd^{0.5}$, $pd^1$ and $pd^{1+}$.
	Each bar corresponds to one separate predictive regression.
	The range of the bar represents 95\% confidence intervals of the predictive $R^2$ obtained using the bootstrap method.
	}
	\label{fig:bootstrap dg}%
\end{figure}

In Figure \ref{fig:bootstrap dg}, we report the $R^2$ of predicting the annual dividend growth of the S\&P 500 index in the next year with different combinations of valuation ratios. We report the detailed regression results in Table \ref{tab:c42 dg} in the Appendix. Our predictive regression is run on monthly observations. In the first specification, the predictors include $dr$ and $pd$, which achieve the highest adjusted $R^2$. In the next specification, for comparison, we show the forecasting performance of $g^F$, the predictor from \cite{Binsbergen2010} who develop a latent state model and filter out the conditional expectation of growth rate $g^F$.

In the third specification in Figure \ref{fig:bootstrap dg}, we include our market duration measure $dr$ alone. Interestingly, $dr$ shows limited return predictive power in comparison with the combination of $dr$ and $pd$. This is surprising as one would expect that when $dr$ increases, i.e., the stock market assigns a higher valuation to long-run dividends than the short-run dividends, the expected growth rate of dividends should rise accordingly. In the fourth specification, we show that the traditional price-dividend ratio also has limited cash-flow predictive power.

The weak standalone predictive power of $dr$ and $pd$ stands in contrast with the strong predictive power of $dr$ and $pd$ combined in the first specification. This again indicates that to fully capture the information embedded in the state variables, two valuation ratios are needed. In the subsequent specifications, we show that different pairs of valuation ratios exhibit different cash-flow predictive power. This shows the importance of taking a predictive regression approach in determining the dimension of the state space rather than simply relying on PCA of valuation ratios. Any given pair of valuation ratios fully spans the two principal components, as indicated in the linearly independent principal-component loadings of different valuation ratios in Panel B of Figure \ref{fig:pca var5}. However, different pairs of valuation ratios may still contain different information about return and cash-flow dynamics.

In the last five specifications in Figure \ref{fig:bootstrap dg}, we show that three or four valuation ratios do not outperform two valuation ratios in forecasting dividend growth. The adjusted $R^2$ we report is in-sample $R^2$, which does not reflect poor out-of-sample performance due to potential overfitting from adding more predictors; in other words, when competing with $dr$ and $pd$, the three or four valuation ratios do not have a mechanical disadvantage due to our choice of performance metric. Overall, our results indicate that two valuation ratios (in particular, the combination of $dr$ and $pd$) are the necessary minimum and sufficient for forecasting dividend growth.

\begin{figure}[!t]
	\centering
	\includegraphics[width=\textwidth]{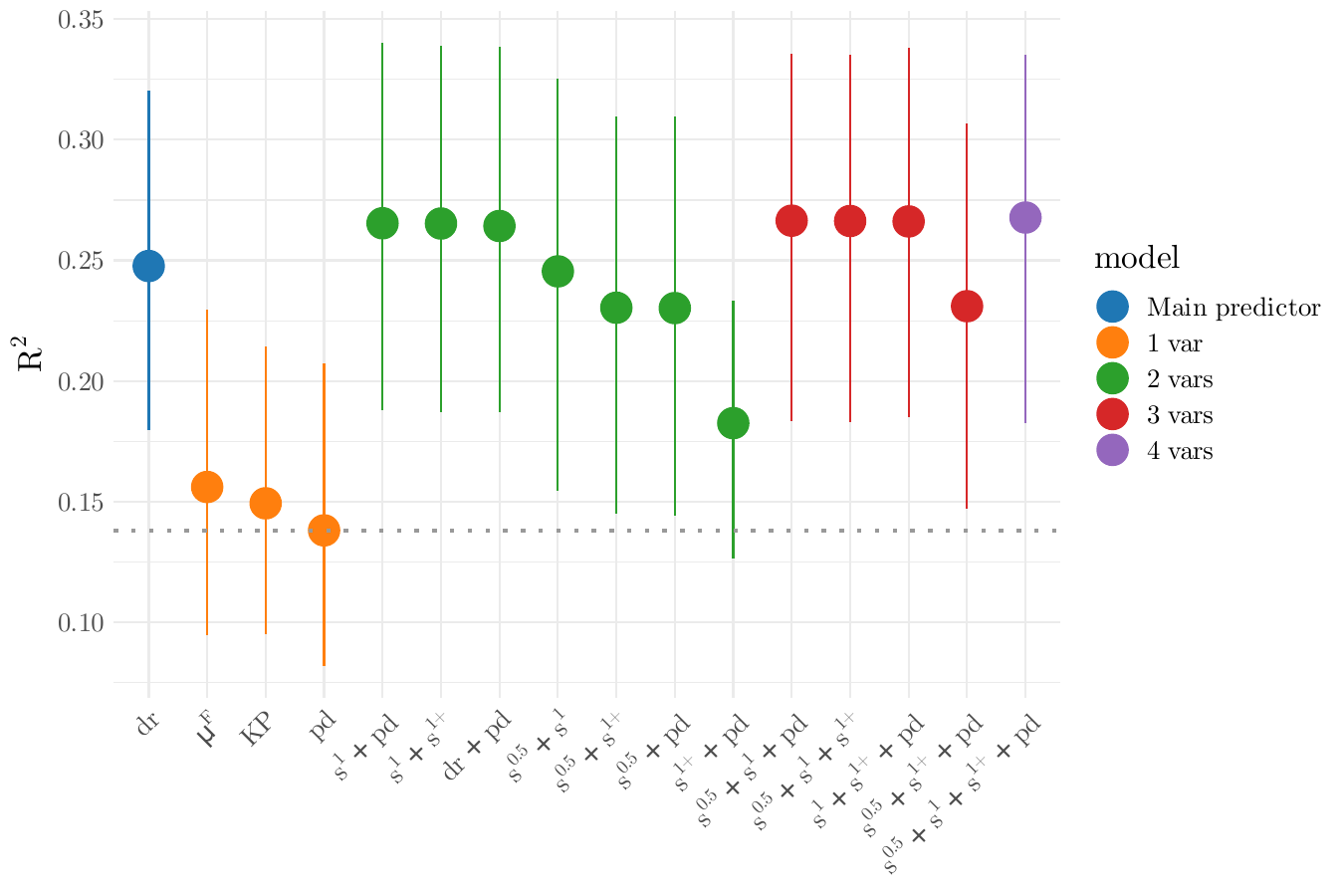}
	\caption[fig: bootstrap r2 return]{$R^2$ from Return Predictive Regressions with Bootstrapped Confidence Interval.

		\footnotesize
		This figure reports $R^2$ from one-year S\&P 500 Index return predictive regressions with bootstrapped confidence interval.
		The predictors include our main predictor `duration' $dr$,
		the price-dividend ratio $pd$,
		short-term (0.5- and 1-year) dividend strip price to dividend ratio ($pd^{0.5}$ and $pd^1$),
		long-term (beyond 1-year) dividend strip price to dividend ratio ($pd^{1+}$),
		the filtered series for demeaned expected returns following \citet{Binsbergen2010} $\mu^{F}$,
		the single predictive factor extracted from 100 book-to-market and size portfolios from \citet{Kelly2013b} $KP$,
		and all combinations of $pd$, $pd^{0.5}$, $pd^1$ and $pd^{1+}$.
		Each bar corresponds to one separate predictive regression.
		The range of the bar represents 95\% confidence intervals of the predictive $R^2$ obtained using the bootstrap method.
	}
	\label{fig:bootstrap ret}%
\end{figure}

Next, we examine the dimension of the state space by forecasting returns. In Figure \ref{fig:bootstrap ret}, we report the $R^2$ of predicting the annual return of S\&P 500 with different combinations of price-dividend ratios. Our predictive regression is run monthly.  We report the detailed regression results in Table \ref{tab:c42 ret} in the Appendix. The conclusion is similar to that in cash-flow prediction. Having three or more price-dividend ratios does not improve predictability relative to the best performance of combinations of two price-dividend ratios. In particular, combining $dr$ and $pd$ achieves the best forecasting performance. These results again suggest that the state space is two-dimensional, and to capture the information embedded in state variables, we need $dr$ and $pd$ for the purposes of analyzing both return and cash-flow dynamics.

Interestingly, the predictive power of $dr$ alone is comparable to that of two or more valuation ratios. Moreover, in the second, third, and fourth specifications in Figure \ref{fig:bootstrap ret}, we find that $\mu^F$, the return predictor in \cite{Binsbergen2010}, $KP$, the return predictor in \cite{Kelly2013b}, and the traditional price-dividend ratio all underperform $dr$ in our sample period.
Overall, our results suggest a close connection between $dr_t$ and the conditional expected return. This finding is critical for understanding what drives the variation of market duration and, vice versa, what drives the expected return. We will explore this topic in the next section.

Before we move on to Section \ref{sec:prediction}, we briefly discuss the rich information content of the pair $dr$ and $pd$ beyond forecasting return and cash-flow growth. In Table \ref{tb:predict macro} in the Appendix, we report the $R^2$ from forecasting a variety of macroeconomic and financial-market variables with $dr$ and $pd$, only $dr$, only $pd$, and for illustration purposes, $dr$ in combination with $s^{1}$ (the log price-dividend ratio of one-year dividend strip), and only $s^{1}$. The pair $dr$ and $pd$ demonstrate the strongest and most consistent predictive power when forecasting variables related to financial intermediaries' balance-sheet capacity (with the $R^2$ ranging from 30\% to 40\%). When forecasting macroeconomic variables related to business-cycle dynamics, $dr$ and $pd$ have an $R^2$ consistently above 20\%. Moreover, the $R^2$ from forecasting sentiment proxies is consistently above 10\%. Finally, combining $dr$ and $pd$ outperforms other specifications.

Overall, our results suggest that $dr$ and $pd$ together provide not only relevant information for understanding asset-pricing dynamics but also other key objects in macro-finance. Through our analysis, we hope to demonstrate the importance of incorporating market duration, $dr$, as a key state variable in macro-finance studies.

\section{Duration and Expected Return} \label{sec:prediction}
A striking finding from our forecasting exercises is that market duration alone is quite sufficient for predicting returns. Augmenting $dr$ with $pd$ or other valuation ratios does not significantly improve the performance. Therefore, $dr$ captures the combination of state variables that drive the conditional expected return.
Next, we provide further evidence on the return predictive power of $dr$. As shown in Section \ref{sec:model}, the loadings of $dr$ on state variables depend on various parameters. We find that $dr$ and the conditional expected return coincide in their state-variable loadings.


\subsection{Univariate return prediction}
We run the standard predictive regression to forecast market return in the next twelve months:
\begin{equation}
	r_{t+1}=\alpha +\beta dr _t+\epsilon_{t+1} \text{,}
	\label{eq:unireg}
\end{equation}
Because we use overlapping monthly data, we adopt \cite{NeweyWest1987} standard errors with 18 lags to account for the moving-average structure induced by overlap \citep{CP2005}. We also calculate \cite{Hodrick1992} standard errors. \cite{Hodrick1992} shows that GMM-based autocovariance correction \citep[e.g.,][]{NeweyWest1987} may have poor small-sample properties. Under the serial correlation in the error term, another concern is the bias induced by the persistence of the predictor.\footnote{The persistence of a return predictor can cause small-sample bias \citep{NelsonKim1993, Stambaugh1999} and spurious regression \citep*{Ferson2003}.} While $dr_t$ has an autocorrelation below that of the traditional price-dividend ratio, $pd_t$, we still report the adjusted estimate of $\beta $ following \cite{Stambaugh1999}. In the appendix (Table \ref{tb:ivx}), we also report the IVX-Wald test of predictive power \citep*{Kostakis2015} that explicitly accounts for the predictor persistence.

The adjusted $R^2$ measures in-sample fitness. Several studies raised concerns over the out-of-sample performance of return predictors \citep{Bossaerts1999, WelchGoyal2007}. To address these issues, we report the out-of-sample $R^2$ and two formal tests of out-of-sample performance. We calculate out-of-sample forecasts as a real-time investor, using data up to time $t$ in the predictive regression to estimate $\beta $, which is then multiplied by the time-$t$ value of the predictor to form the forecast. Out-of-sample forecasting starts from December 1997, when we have at least ten years of data.
Out-of-sample $R^2$ is
\[
	R^2_{OOS}=1-\frac{\sum _t \left(r_{t+1}-\hat{r}_{t+1}\right)^2}{\sum _t \left(r_{t+1}-\overline{r}_t\right)^2} \text{,}
\]
where $\hat{r}_{t+1}$ is the forecast value and $\overline{r}$ is the average of twelve-month returns (the first is January-December 1998).
The out-of-sample $R^2$ lies in the range $\left(-\infty ,1\right]$, where a negative number means that a predictor provides a less accurate forecast than the historical mean.

We report the p-value of two out-of-sample performance tests, ``$ENC$'' and ``$CW$''. $ENC$ is the encompassing forecast test derived by \cite{ClarkMcCracken2001}, which is widely used in the literature. We test whether the predictor has the same out-of-sample forecasting performance as the historical mean and compare the value of the statistic with critical values calculated by \cite{ClarkMcCracken2001} to obtain a p-value range. \cite{ClarkWest2007} adjust the standard MSE t-test statistic to produce a modified statistic ($CW$) that has an asymptotic distribution well approximated by the standard normal distribution, so for $CW$, we report the precise p-value. 

\begin{table}[!t]
	\caption[tab2: annual return prediction - S\&P500 return]{Annual Return Prediction

		\footnotesize This table reports the results of predictive regression (equation (\ref{eq:unireg})). The left-hand side variable is the return of the S\&P 500 index in the next twelve months.
		We consider four right-hand side variables (i.e., predictors), $dr_t$, $pd_t$, filtered series for expected returns following \citet{Binsbergen2010} $\mu^{F}$,
		and the single predictive factor extracted from 100 book-to-market and size portfolios from \citet{Kelly2013b} $KP$.
		The $\beta $ estimate is reported, followed by \cite{NeweyWest1987} t-statistic (with 18 lags), \cite{Hodrick1992} t-statistic, the coefficient adjusted for \cite{Stambaugh1999} bias, and the in-sample adjusted $R^2$. We run the regression monthly.
		Starting from December 1997, we form out-of-sample forecasts of return in the next twelve months by estimating the regression with data up to the current month and use the forecasts to calculate out-of-sample $R^2$, ENC test (\citealp{ClarkMcCracken2001}), and the p-value of CW test (\citealp{ClarkWest2007}).
		Our monthly sample is 1988:01--2019:12.
	}
	\centering
	\footnotesize
	\begin{tabular}{p{6cm}p{1.5cm}p{1.5cm}p{1.5cm}p{1.5cm}p{1.5cm}}
		\toprule
		                                                              & \multicolumn{5}{c}{$r_{t+1}$}                                           \\\cmidrule{2-6}
		{}                                                            & (1)                           & (2)      & (3)     & (4)     & (5)      \\
		\midrule
		$ dr_t$                                                       & -0.156                        &          &         &         & -0.228   \\
		\quad \quad \textit{\small Hodrick t}                         & [-3.354]                      &          &         &         & [-2.924] \\
		\quad \quad \textit{\small  Newey-West t}                     & (-4.499)                      &          &         &         & (-3.517) \\
		\quad \quad \textit{\small  Stambaugh bias adjusted $\beta $} & -0.146                        &          &         &         &          \\
		$pd_t$                                                        &                               & -0.199   &         &         & 0.141    \\
		                                                              &                               & [-2.367] &         &         & [1.721]  \\
		                                                              &                               & (-2.747) &         &         & (1.209)  \\
		                                                              &                               & -0.189   &         &         &          \\
		$\mu^{F}_{t}$                                                 &                               &          & 2.584   &         &          \\
		                                                              &                               &          & [2.313] &         &          \\
		                                                              &                               &          & (2.804) &         &          \\
		                                                              &                               &          & 2.594   &         &          \\
		$KP_t$                                                        &                               &          &         & 0.895   &          \\
		                                                              &                               &          &         & [2.960] &          \\
		                                                              &                               &          &         & (2.857) &          \\
		                                                              &                               &          &         & 0.905   &          \\\addlinespace[1ex]
		$N$                                                           & 384                           & 384      & 384     & 384     & 384      \\
		$R^2$                                                         & 0.248                         & 0.138    & 0.156   & 0.149   & 0.264    \\
		OOS $R^2$                                                     & 0.146                         & 0.004    & -0.032  & 0.041   & 0.180    \\
		ENC                                                           & 2.968                         & 0.833    & 0.651   & 2.978   & 5.985    \\
		$p(ENC)$                                                      & $<$0.05                       & $>$0.10  & $>$0.10 & $<$0.05 & $<$0.01  \\
		$p(CW)$                                                       & 0.022                         & 0.200    & 0.303   & 0.031   & 0.021    \\
		\bottomrule
	\end{tabular}
	\label{tb:uncondUS}
\end{table}

Table \ref{tb:uncondUS} presents the results. Column (1) shows that the market duration, $dr$, demonstrates a striking degree of return predictive power. The in-sample estimation generates a predictive $R^2$ reaching $24.8\%$.\footnote{\cite*{FSW1997JF} discuss the potential data mining issues that arise from researchers searching among potential regressors. They derive a distribution of the maximal $R^2$ when $k$ out of $m$ potential regressors are used as predictors and calculate the critical value for $R^2$, below which the prediction is not statistically significant. For instance, when $m = 50$, $k = 5$, and the number of observations is 250, the $95\%$ critical value for $R^2$ is 0.164.}
Out-of-sample forecasts deliver an $R^2$ of $14.6\%$, significantly outperforming the historical mean as shown by the p-values of $ENC$ and $CW$.\footnote{In our calculation of out-of-sample $R^2$ starts from Dec. 1997 (after the first ten years of data). Figure \ref{fig:R2split} in the Appendix reports the out-of-sample $R^2$ for different start dates and compares the OOS $R^2$ of $dr$ with that of $pd$.} The predictive coefficient is also large in magnitude, indicating high volatility of the conditional expected return. A decrease of $dr_t$ by one standard deviation adds 7.7\% to the expected return. Both Newey-West and Hodrick t-statistics are significant at least at the $1\%$ level.

Column (2) of Table \ref{tb:uncondUS} reports the results for $pd_t$. The predictive power of $pd_t$ is much weaker than $dr_t$ in all aspects. Its in-sample $R^2$ is almost half of that of $dr_t$ and $pd_t$ barely exhibits any out-of-sample predictive power with $R^2$ equal to $0.4\%$. In both $ENC$ and $CW$ tests, $pd_t$ fails to beat the historical mean with any statistical significance. Its coefficient is smaller in magnitude than that of $dr_t$. A decrease in $pd_t$ by one standard deviation leads to an increase of expected return by 5.8\%, implying a less volatile expected return than the one from $dr_t$. The IVX-Wald test of \citet*{Kostakis2015} in Table \ref{tb:ivx} in the Appendix also supports the significant predictive power of $dr_t$ while rejecting the predictive power of $pd_t$.


Next, we compare $dr_t$ with two return predictors that are conceptually related. \cite{Binsbergen2010} extract information about state variables that drive the conditional expected return and expected cash-flow growth by estimating a latent-state model.
Our approach differs as we do not estimate or filter the state variables but instead rely on $dr_t$, $pd_t$, and valuation ratios of dividend strips. In particular, the construction of our preferred return predictor, $dr_t$, only requires the total market capitalization and the price of dividends paid in the next year, which are directly observed from the market rather than estimated. In Column (3) of Table \ref{tb:uncondUS}, we follow the procedure in \cite{Binsbergen2010} to construct their return predictor, $\mu^F_t$. While $\mu^F_t$ slightly outperforms $pd_t$, its predictive power is weaker than that of $dr_t$ across different metrics. Note that \cite{Binsbergen2010} conduct analysis with a time period different from ours. Our sample period is constrained by the availability and reliability of index futures data and ends before the Covid-19 era of extreme downward and upward market movements.

While \cite{Binsbergen2010} use the aggregate data on realized returns and dividend growth, \cite{Kelly2013b} demonstrate another filtering method that utilizes the cross-section of market-to-book ratios of individual stocks.
Individual stocks' market-to-book ratios map out the state variables that drive the aggregate market but, as we show in Appendix I.2, these valuation ratios contain noise that is orthogonal to the expected market return. \cite{Kelly2013b} use the method of partial least squares to reduce the noise. Our approach differs as we avoid dealing with such noise. $dr_t$, $pd_t$, and the valuation ratios of dividend strips of the aggregate market differ in their loadings on the aggregate state variables but do not contain idiosyncratic noise. In our approach, the challenge is to find the valuation ratio or a combination of valuation ratios whose state-variable loadings coincide with those of the conditional expected return and thereby achieve the highest predictive power. In the next subsection, we characterize the necessary and sufficient condition for $dr_t$ to be such a return predictor. Following the procedure in \cite{Kelly2013b}, we construct their return predictor, denoted by $KP_t$. In column (4) of Table \ref{tb:uncondUS}, we report the prediction results. $KP_t$ significantly outperforms $pd_t$ but still underperforms $dr_t$ across different metrics, such as Newey-West t-statistic, Hodrick t-statistic, in-sample $R^2$, out-of-sample $R^2$, $ENC$, $CW$, and, IVX-Wald test reported in Table \ref{tb:ivx} in the Appendix.

\begin{table}[!t]
	\caption[corr predictors]{Correlation between Return Predictors

		\footnotesize This table reports the correlation matrix of four main return predictors.
		$dr$ is our main return predictor, ``duration.''
		$pd$ is the price-dividend ratio of the S\&P 500 index.
		$\mu^{F}$ is the filtered series for demeaned expected returns, following \citet{Binsbergen2010},
		$KP$ is the single predictive factor extracted from 100 book-to-market and size portfolios from \citet{Kelly2013b}.
		Data sample: 1988:01--2019:12.
	}
	\centering
	\footnotesize
	
\begin{tabular}{lrrrr}
    \toprule
              & $dr$   & $pd$   & $\mu_{F}$ & $KP$ \\\midrule
    $dr$      & 1      &        &           &      \\
    $pd$      & 0.873  & 1      &           &      \\
    $\mu_{F}$ & -0.892 & -0.967 & 1         &      \\
    $KP$      & -0.565 & -0.496 & 0.468     & 1    \\\bottomrule
\end{tabular}

	\label{tb:corr predictors}
\end{table}

While these return predictors are correlated as shown in Table \ref{tb:corr predictors}, their predictive power differs significantly with market duration outperforming the rest. In the appendix, we demonstrate the robustness of our results by repeating the analysis with different forecasting targets. In Table \ref{tb:uncondUS excess}, we replace the S\&P 500 annual return with the excess S\&P 500 annual return, and in Table \ref{tb:uncondUS mkt} and \ref{tb:uncondUS mktrf}, we consider the Fama-French market portfolio return and excess return, respectively. Finally, we also show that market duration demonstrates superior return predictive power at a monthly horizon. Our baseline results are reported in \ref{tb:uncondUS monthly sp}, and see Table \ref{tb:uncondUS monthly sp-rf} for results on predicting monthly S\&P 500 excess return. Table \ref{tb:uncondUS monthly mkt} and Table \ref{tb:uncondUS monthly mkt-rf} report the results on predicting the monthly Fama-French market portfolio return and excess return, respectively.

\begin{figure}[!t]
	\centering
	\includegraphics[scale=0.47]{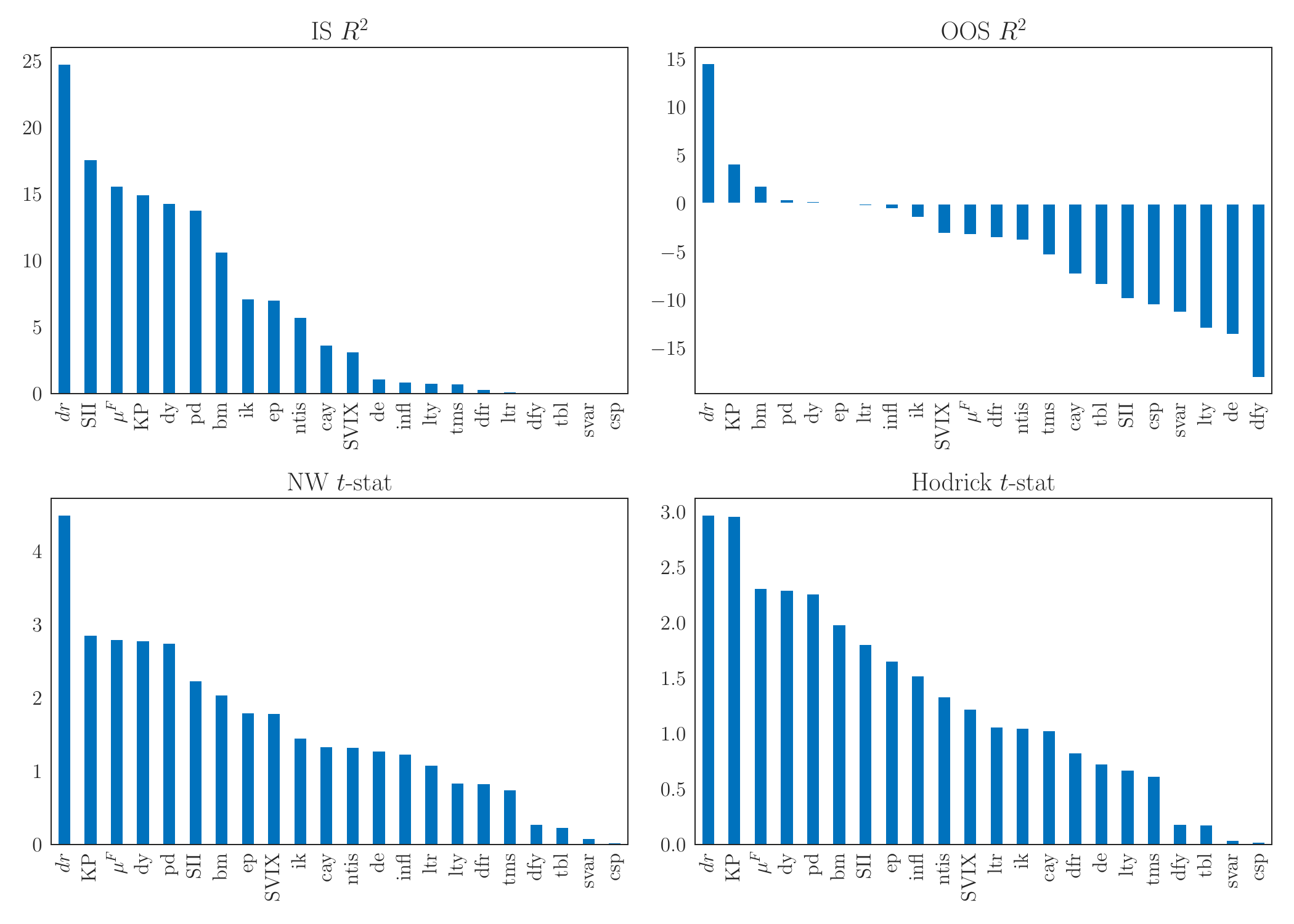}
	\caption[figure comparison: s\&p return]{{Comparison with Alternative Return Predictors.}

		\footnotesize
		This graph compares the 1-year return predictive power between $dr_t$ and other commonly studied predictors in our sample period.
		Panel A reports the in-sample adjusted $R^2$. Panel B reports the out-of-sample $R^2$. Negative out-of-sample $R^2$ indicates that the predictive power is below the historical mean. Panel C reports the absolute values of \cite{NeweyWest1987} t-statistic (with an 18-month lag). Panel D reports the absolute values of \cite{Hodrick1992} t-statistic. Most predictors are from \citet{WelchGoyal2007} and include the price-dividend ratio (pd), the default yield spread (dfy), the inflation rate (infl), stock variance (svar), the cross-section premium (csp), the dividend payout ratio (de), the long-term yield (lty), the term spread (tms), the T-bill rate (tbl), the default return spread (dfr), the dividend yield (dy), the long-term rate of return (ltr), the earnings-to-price ratio (ep), the book to market ratio (bm), the investment-to-capital ratio (ik), the net equity expansion ratio (ntis), the percent equity issuing ratio (eqis), and the consumption-wealth-income ratio (cay). SII is the short interests index from \citet*{RapachRinggenbergZhou2016} (1988-2014). SVIX is an option-implied lower bound of the 1-year equity premium from \cite{Martin2017} (1996-2012). KP is the single predictive factor extracted from 100 book-to-market and size portfolios from \citet{Kelly2013b}. BK is the filtered series for expected returns following \citet{Binsbergen2010}.}
	\label{tb:comp}
\end{figure}

So far, we have compared the return predictive power of $dr_t$ with that of $pd_t$, which together with $dr_t$ span the state space. We also considered $\mu^F_t$ and $KP_t$ that are conceptually related. Figure \ref{tb:comp} compares $dr$ with other predictors proposed in the literature, including the price-dividend ratio (pd), the default yield spread (dfy), the inflation rate (infl), stock variance (svar), the cross-section premium (csp), the dividend payout ratio (de), the long-term yield (lty), the term spread (tms), the T-bill rate (tbl), the default return spread (dfr), the dividend yield (dy), the long-term rate of return (ltr), the earnings-to-price ratio (ep), the book to market ratio (bm), the investment-to-capital ratio (ik), the net equity expansion ratio (ntis), the percent equity issuing ratio (eqis), the consumption-wealth-income ratio (cay), the short interests index (SII), and the option-implied lower bound of 1-year equity premium (SVIX).\footnote{Note that the dividend yield (dy) is not the inverse of price-dividend ratio ($pd$) because in the denominator of dy is the lagged market value (not the current value).} Most predictors are studied in a return predictability survey by \cite{WelchGoyal2007}, and others are proposed more recently, such as the short interest index, ``SII'' in \citet*{RapachRinggenbergZhou2016}, and SVIX \citep{Martin2017}. In the comparison, we also include $pd_t$, $\mu^F_t$, and $KP_t$ from Table \ref{tb:uncondUS}.\footnote{When constructing $KP_t$, we estimate the partial least squares model following the procedure in \cite{Kelly2013b}. In Figure \ref{tb:comp_KP_original}, We report the comparison of return predictive power using the authors' original parameter estimates rather than our own estimates.}

In Figure \ref{tb:comp}, we report in-sample (``IS'') $R^2$, out-of-sample (``OOS'') $R^2$, the absolute values of Newey-West, and Hodrick t-statistics. $dr$ outperforms other predictors in all aspects. Among the alternatives, $KP$, $pd$, and the book-to-market ratio (``$bm$'') deliver the most consistent performance across the four metrics while $\mu^F$ exhibits strong performance except for OOS $R^2$. The performance of other return predictors is not quite consistent across different metrics.

In Table \ref{tb:corr_predictors}, we report the correlation between $dr $ and these predictors. Besides $pd$, $\mu^F$, and $KP$, all the other predictors with correlation above 50\% or below -50\%, are all valuation ratios, such as the market-to-book ratio and price-earnings ratio. This is consistent with the intuition in \cite{Kelly2013b} and with our emphasis on using valuation ratios to capture information embedded in the state variables. In the appendix, we compare $dr$ with other predictors for alternative forecasting targets, such as S\&P 500 excess annual return (Figure \ref{tb:comp_exc}), Fama-French market portfolio annual return (Figure \ref{tb:comp_mkt}), and Fama-French market portfolio excess annual return (Figure \ref{tb:comp_mkt_exc}). As shown in Figure \ref{tb:comp_KP_original}, market duration exhibits the strongest predictive power across the four metrics and such performance is consistent across all forecasting targets.

\subsection{A two-dimensional state space model} \label{subsec:model1}
The findings on return predictability are critical for understanding the structure of both the conditional expected return and market duration. Next, we show that the conditional expected return is a univariate function of $dr$ (and vice versa) when the market does not contain information on future cash-flow growth beyond the next year. Intuitively, under this condition, the price of next year's dividends exhausts all information about future cash-flow growth, so $dr$, which is the logarithm of market value minus the logarithm of the price of next year's dividend, teases out cash-flow information and only contains information about the discount rate.

Stock prices are determined by the conditional expectations of future returns and dividends over different horizons \citep{Campbell1988}. Therefore, a model of the aggregate stock market requires at least two state variables that capture the conditional expected return and dividend growth; in other words, the lower bound of the state-space dimension is two. Our analysis so far shows that two is also the upper bound. Guided by our findings in Section \ref{sec:dimension}, we simplify the general model in Section \ref{sec:model} by specifying a two-dimensional state space.
As in \cite{LettauWachter2007} and \cite{Binsbergen2010}, one state variable drives the conditional expectation of the dividend growth rate, and the second state variable drives the conditional expected return through the price of risk. Specifically, $g_t$
is given by
\begin{equation} \label{eq:g1}
	g_t=z_t + \overline{g} - \frac{1}{2}\sigma _D^{\top} \Sigma \sigma _D,
\end{equation}
where $z_t$ has the following law of motion
\begin{equation} \label{eq:g1_z}
	z_{t+1}=\rho_z z_{t} + \sigma_z^{\top }\epsilon_{t+1}\, .
\end{equation}
The second state variable, $y_t$, with a law of motion
\begin{equation}
	y_{t+1}=\rho_y y_{t} + \sigma_y^{\top }\epsilon_{t+1}\, ,
\end{equation}
drives the price of risk $\lambda_t$, so equation (\ref{eq:lambda}) becomes
\begin{equation} \label{eq:lambda1}
	\lambda _t=\overline{\lambda }+y_{t},
\end{equation}
and the stochastic discount factor (SDF) is given by
\begin{equation} \label{eq:sdf1}
	M_{t+1}=\exp \left\lbrace -r_f-\frac{1}{2}\lambda_t^2(\sigma_{\lambda}^{\top}\Sigma\sigma_{\lambda})^2-\lambda_t\sigma_{\lambda}^{\top}\epsilon_{t+1}\right\rbrace .
\end{equation}
In a two-dimensional state space where one state variable, $z_t$, is assigned to drive the expected dividend growth rate, the price of risk, $\lambda_t$, must be unidimensional, driven by $y_t$, and accordingly, shocks enter into the SDF as a scalar, i.e., the linear combination given by $\sigma_{\lambda}$. Any shock can be priced. The price of risk for the $n$-th shock is $\lambda_t\sigma_{\lambda}(n)$, where $\sigma_{\lambda}(n)$ is the $n$-th element of $\sigma_{\lambda}$.

As in Section \ref{sec:model}, the $N$-by-$1$ shock vector $\epsilon_{t+1}$ contains all information at $t+1$. The variables' shock loadings may differ, for example, $\sigma_z\neq \sigma_y$. $z_t$ and $y_t$ can be correlated through their overlapping exposure to shocks.
Also note that, in comparison with the general model in Section \ref{sec:model}, the state-variable loadings of $g_t$ and $\lambda_t$, i.e., $\phi $ and $\lambda_t$ are set to one. This simplification can be done as $g_t$ and $\lambda_t$ load on one state variable, $z_t$ and $y_t$, respectively.

In Appendix I, we solve the log price-dividend ratio of the aggregate market
\begin{equation}
	pd_t=A_{pd} + B_{pd}y_t + C_{pd}z_t,
\end{equation}
where $A_{pd}$, $B_{pd}$, and $C_{pd}$ are constant. And the log price-dividend ratio of the dividend strip that matures in one year,
\begin{equation}
	s^1_t=A_{1} + B_{1}y_t + C_{1}z_t.
\end{equation}
Therefore, market duration is
\begin{equation}
	dr_t=pd_t - s^1_t = A_{pd}-A_1 + (B_{pd}-B_1)y_t + (C_{pd}-C_1)z_t,
\end{equation}
where $B_{pd}\neq B_{pd}-B_1$ and $C_{pd}\neq C_{pd}-C_1$. Since $dr_t$ and $pd_t$ have different loadings on $z_t$ and $y_t$, we can solve $z_t$ and $y_t$ from $dr_t$ and $pd_t$. Therefore, market duration and the traditional price-dividend ratio span the state space. In particular, $dr_t$ and $pd_t$ contain all the necessary information for forecasting return and dividend growth as we show in Section \ref{sec:dimension}.

\begin{prop}[Spanning]
	The state variables are functions of $dr_t$ and $pd_t$, and vice versa.
\end{prop}

In Appendix I, we show $C_{pd}=1/(1-\kappa_1\rho_z)$ and $C_{1}=1$ and obtain the following result.

\begin{prop}[Discount rate and duration]
	The conditional expected return is a function of $y_t$
	\begin{equation} \label{eq:er}
		\mathbb{E}_t[ r_{t+1} ] = A_{er} + B_{er}y_t,\,
	\end{equation}
	where $A_{er}$ and $B_{er}$ are constant.
	Under $\rho_z=0$, $\mathbb{E}_t[ r_{t+1} ]$ is a univariate function of $dr_t$ and vice versa.
	\label{prop:dr}
\end{prop}

When $\rho_z$, the autoregressive coefficient of expected dividend growth rate $z_t$, is zero, we have $dr_t=A_{pd}-A_1 + (B_{pd}-B_1)y_t$ as the exposure of market duration to $z_t$ is muted. As $y_t$ drives the conditional expected return, $dr_t$ and $\mathbb{E}_t[ r_{t+1} ]$ become univariate functions of one another. Market participants' information at time $t$ about future dividends is summarized by $z_t$ which determines the expected dividend growth rate from $t$ to $t+1$. Therefore, if $z_t$ lacks persistence, the market participants do not have information about dividends beyond $t+1$. By construction, $dr_t$ teases out information about dividend growth up to $t+1$ by deducting $s^1_t$ from $pd_t$, so, intuitively, it contains only the information about the discount rate or expected return under $\rho_z=0$.

\subsection{The persistence of cash-flow growth expectations}
\label{sec:rhoz}
We estimate $\rho_z$, the persistence of expected dividend growth, using two approaches. Our main analysis is based on analyst forecasts as a proxy for market participants' growth expectations. We estimate $\rho_z$ with two econometric models. For robustness, we also estimate $\rho_z$ by fitting the latent state model given by (\ref{eq:g1}) and (\ref{eq:g1_z}) to dividend data. 

Our model
does not require investors to have rational expectations. As a matter of fact, even if the data-generating processes differ from equations (\ref{eq:g1}) and (\ref{eq:g1_z}) in our model, as long as agents in this economy take these equations as the stochastic processes that govern the aggregate dividends, our results on stock and dividend strip valuation hold. Therefore, to estimate $\rho_z$, the persistence of the expected dividend growth rate, we examine market participants' subjective expectations---the analyst forecasts obtained from IBES.

A large literature
has demonstrated that analyst forecasts reflect analysts' beliefs as their compensation
are tightly linked to forecast precision, and that analyst forecasts are likely to represent the market participants' beliefs more broadly \citep*[e.g.,][]{Mikhail1999, Cooper2001, Bradshaw2004}. Admittedly, analyst forecasts may be distorted due to various incentive and institutional frictions (e.g., \citealp{Gu2003}; \citealp{Malmendier2007, Malmendier2014}).
However, the bias is contained as long as such frictions do not vary over time systematically in a way that correlates with the analysts' true beliefs.
Moreover, we will supplement our estimate of $\rho_z$ based on analyst forecasts with an alternative estimate from a standard filtering approach applied to the state space model given by equation (\ref{eq:g1}) and (\ref{eq:g1_z}).

Recent studies using analysts' cash-flow expectations have made substantial progress in explaining a variety of phenomena in asset pricing.\footnote{For example, \citet*{Bordalo2019} find that analyst forecasts of firms' long-term earnings growth overreact to news about fundamentals, leading to predictable cross-sectional return variations. \citet{Delao2020} show that analyst short-term earnings forecasts, aggregated to the S\&P 500 index level, explain a substantial share of the price-dividend ratio's variation. \citet{Nagel2022} propose a model where agents form beliefs of cash-flow growth based on their experienced growth. Such beliefs give rise to a sizable and counter-cyclical equity premium. \citet*{Bordalo2023} incorporate analyst forecasts of short- and long-term earnings of the S\&P 500 index into a dividend discount model with a constant discount rate, which generates a ``synthetic" index price that closely resembles the actual price and exhibits ``excess'' volatility.}
Moreover, our focus on the persistence of agents' growth expectations is related to the theoretical and empirical works of \citet{Gabaix2019} and \citet{Wang2020}, who highlight the importance of agents' perceived persistence of key state variables in generating patterns of under and over-reaction in belief formation and asset prices.
\begin{table}[!t]
	\caption[summibes]{Summary Statistics: Cash Flow Growth Forecasts

	\footnotesize This table reports the number of observations, mean, standard deviation, minimum, maximum, quartiles, and monthly autocorrelation ($\rho $) of various measures of cash flow growth expectations.
	$\E^A_t{\Delta e_{t, t+1}}$, $\E^A_t{\Delta e_{t+1, t+2}}$, and $\E^A_t{\Delta e_{t+2, t+3}}$ are forecasts of 1-year earnings growth for fiscal year 1, 2, and 3 provided by IBES Global Aggregate (IGA).
	IGA $\Delta e_{t}$ and Compustat $\Delta e_{t}$ are the actual 1-year earnings growth from IGA and Compustat, respectively.
	$\E^A_t{\Delta e_{t, t+1}}$ and $LTG_t$ are forecasts of 1-year and long-term earnings growth that we self-aggregate from the IBES Unadjusted US Summary Statistics File.
	Data sample: 1988:01--2019:12.
	}
	\centering
	\footnotesize
	\begin{tabular}{lrrrrrrrrr}
	\toprule
	                                            & obs & mean  & std   & min    & 25\%   & 50\%  & 75\%  & max   & $\rho$ \\ \midrule
	$\E^A_t{\Delta e_{t, t+1}}$             & 384 & 0.103 & 0.096 & -0.167 & 0.056  & 0.103 & 0.154 & 0.425 & 0.897  \\
	$\E^A_t{\Delta e_{t+1, t+2}}$           & 384 & 0.134 & 0.043 & -0.069 & 0.104  & 0.127 & 0.157 & 0.269 & 0.830  \\
	$\E^A_t{\Delta e_{t+2, t+3}}$           & 384 & 0.130 & 0.036 & 0.052  & 0.100  & 0.122 & 0.159 & 0.217 & 0.953  \\
	IGA $\Delta e_{t, t+1}$                     & 384 & 0.072 & 0.135 & -0.380 & -0.008 & 0.092 & 0.148 & 0.425 & 0.929  \\
	Compustat $\Delta e_{t, t+1}$               & 384 & 0.068 & 0.481 & -2.175 & -0.042 & 0.122 & 0.187 & 2.190 & 0.976  \\
	$LTG_t$                     & 384 & 0.125 & 0.018 & 0.093  & 0.115  & 0.120 & 0.129 & 0.187 & 0.986  \\
	\bottomrule
\end{tabular}
	\label{tb:sum ibes}
\end{table}

As the coverage of dividend forecasts started in 2003 in IBES and is too short for our analysis, we follow the literature and proxy for analysts' expectation of dividend growth with their earnings forecasts (available since 1976).
The following accounting identity connects the earnings and dividends:
$	D_{t} = \text{Earnings}_{t} \times (1-\text{plowback rate}_{t}). $
As documented by \citet*{Pastor2008} and \citet*{Chen2013}, the plowback rate is quite stable. 
Therefore, the growth rates of dividends are empirically close to those of earnings: For $k=1,\, 2,\, $ and $3$ years,
\begin{align}
	\Delta d_{t+k}\equiv \ln\left(\frac{D_{t+k}}{D_{t+k-1}}\right)  \approx \Delta e_{t+k}\equiv \ln\left(\frac{\text{Earnings}_{t+k}}{\text{Earnings}_{t+k-1}}\right).
\end{align}
Note that $\Delta d_{t+k}$ is the one-year growth rate from $t+k-1$ to $t+k$. For example, for $k=2$, $\Delta d_{t+2}$ is the dividend growth rate from $t+1$ to $t+2$. The same notation applies to earnings and earnings growth expectations. Taking the analyst's expectations, $\mathbb{E}^A_t\left(\cdot \right)$, on both sides, we can proxy for expected dividend growth using expected earnings growth:
\begin{align} \label{eq:eg2ge}
	\E_t^A\left(\Delta d_{t+k}\right)\approx\E_t^A\left(\Delta e_{t+k}\right)\approx \ln\left(\frac{\E_t^A\left(\text{Earnings}_{t+k}\right)}{\E_t^A\left(\text{Earnings}_{t+k-1}\right)}\right),
\end{align}
where, for $k=1$, $\E_t^A\left(\text{Earnings}_{t+k-1}\right)=\text{Earnings}_{t}$ is the realized dividend in the current year.\footnote{We switch the order of $\mathbb{E}^A_t\left(\cdot \right)$ and $\ln\left(\cdot \right)$, assuming Jensen's inequality terms are negligible for the purpose of estimating $\rho_z$. This is in line with the model in Section \ref{sec:model}: The growth rates are log-normally distributed, so the ignored terms contain constant variances, which do not affect the estimates of $\rho_z$.} 

IBES Global Aggregates (IGA) provides a forecast of aggregate earnings growth for the S\&P 500 index based on firm-level earnings forecasts. The aggregation procedure weighs individual companies by their market capitalization.\footnote{To deal with the fact that companies have different fiscal year-end, IGA calendarizes all company-level data to a December calendar year before aggregation. This approach follows the Compustat rule. Please refer to ``Thomson Reuters Datastream IBES Global Aggregates Reference Guide'' for more detail.} To transform earnings forecasts to forecasts of growth rates, IGA takes the procedure given by equation (\ref{eq:eg2ge}).
The IGA data is available at a weekly frequency. We consider both weekly and monthly frequencies. For monthly frequency, we take the last weekly observation of each month.
At time $t$, analysts' forecasts of earnings are available at three horizons: one, two, and three years, i.e., $\mathbb{E}_t^A\left(Earnings_{t+1}\right)$, $\mathbb{E}_t^A\left(Earnings_{t+2}\right)$, and $\mathbb{E}_t^A\left(Earnings_{t+3}\right)$.
Moreover, since IGA does not aggregate firm-level long-term growth (LTG) forecasts, we follow the same aggregation procedure to create an index-level LTG forecast that will allow us to estimate $\rho_z$ with alternative methods. The firm-level
LTG forecasts are obtained from the IBES Unadjusted US Summary Statistics File. Table \ref{tb:sum ibes} provides summary statistics. 


Next, we map the expectations of cash-flow growth to the model counterparts and derive a system of equations that can be used to estimate $\rho_z$. First, we acknowledge that the analyst forecasts may not perfectly capture market participants' expectations by adding a noise term between analysts' expectation and the market participants' expectation: For $k=1,2,3$,
\begin{equation}
	\E^A_t\left(\Delta e_{t+k}\right)=\E_t\left(\Delta e_{t+k}\right)+\varepsilon ^A_{t,k},
\end{equation}
where $\E_t\left(\cdot \right)$, is the market participants' expectation. Equations (\ref{eq:g1}) imply
\begin{align*}
	\E^A_t\left(\Delta e_{t+1}\right) & = g+z_t+\varepsilon ^A_{t,1}                                                            \\
	\E^A_t\left(\Delta e_{t+2}\right) & = g+\E_t\left(z_{t+1}\right)+\varepsilon ^A_{t,2}=g+\rho_z z_t +\varepsilon ^A_{t,2}    \\
	\E^A_t\left(\Delta e_{t+3}\right) & = g+\E_t\left(z_{t+2}\right)+\varepsilon ^A_{t,3}=g+\rho_z^2 z_t +\varepsilon ^A_{t,3}.
\end{align*}
Using the first equation to substitute out $z_t$ in the second and third equations, we obtain a system:
\begin{equation} \label{eq:analyst}
	\underset{\equiv \, \mathbf{y}^A_{t}}{\underbrace{\begin{bmatrix}
			\E^A_t\left(\Delta e_{t+2}\right) \\
			\E^A_t\left(\Delta e_{t+3}\right)
		\end{bmatrix}}}
	=
	\left(1-\rho_z\right)g
	+ \rho_z
	\underset{\equiv \, \mathbf{x}^A_{t}}{\underbrace{\begin{bmatrix}
			\E^A_{t}\left(\Delta e_{t+1}\right) \\
			\E^A_{t}\left(\Delta e_{t+2}\right)
		\end{bmatrix}}}
	+
	\underset{\equiv \, \epsilon _t}{\underbrace{\begin{bmatrix}
				\varepsilon ^A_{t,1}-\rho_z \varepsilon^A_{t,0} \\
				\varepsilon ^A_{t,2}-\rho_z \varepsilon^A_{t,1}
			\end{bmatrix}}}.
\end{equation}
Therefore, we estimate $\rho_z$ by regressing $\mathbf{y}^A_{t}$ on $\mathbf{x}^A_{t}$. The identification assumption is that under the econometricians' belief, the expectation of the disturbance $\epsilon_t$ is zero conditional on $\mathbf{x}^A_{t}$.
The deviations of analysts' expectations from market participants' expectations are allowed to be correlated across the starting dates of annual dividend growth, i.e., $t$, $t+1$, and $t+2$.\footnote{The identification of $\rho_z$ is robust to the correlation between these errors and $\epsilon _{t}$ in the model in Section \ref{sec:model}, i.e., the structural shocks to realized dividend, market participants' beliefs on cash-flow dynamics, and their price of risk.}

\begin{table}[!t]
	\centering
	\caption[tab0: rhoz estimated from analysts forecasts]{The Persistence of Expected Cash-Flow Growth from Analyst Forecasts

		\footnotesize This table reports the estimates of $\rho_z$, the autoregressive coefficient of expected cash-flow growth, based on equation (\ref{eq:analyst}). The estimation uses aggregate earnings growth forecasts of the S\&P 500 Index obtained from IGA.
		Columns (1) and (3) report the estimates of $\rho_z$ using monthly data, while columns (2) and (4) report the estimates of $\rho_z$ using weekly data. columns (1) and (2) use earnings growth forecasts for 1,2 and 3 years ahead to estimate the two-equation system (\ref{eq:analyst}), while columns (3) and (4) only use earnings growth forecasts for 1 and 2 years ahead to estimate the first equation in (\ref{eq:analyst}).
		$t$-statistics based on Driscoll-Kraay standard errors with autocorrelation of up to 18 lags are reported in parentheses.
		Data sample: 1988:01--2019:12.
	}
	\footnotesize
	\label{tb:rhoz ibes}

\begin{tabular*}{0.95\textwidth}{l@{\extracolsep{\fill}}rrrr}
    \toprule
    {}            & (1)                                                     & (2)      & (3)      & (4)      \\
    \midrule
    $(1-\rho_z)g$ & 0.122                                                   & 0.129    & 0.141    & 0.133    \\
                  & (16.906)                                                & (13.995) & (15.536) & (16.745) \\
    $\rho_z$      & 0.015                                                   & 0.028    & -0.071   & -0.073   \\
                  & (0.381)                                                 & (0.690)  & (-1.379) & (-1.295) \\\\
    $N$           & 1887                                                    & 768      & 384      & 943      \\
    $R^2$         & 0.001                                                   & 0.003    & 0.025    & 0.028    \\
    Sample        & Monthly                                                 & Weekly   & Monthly  & Weekly   \\
    Periods       & Y1:Y3                                                   & Y1:Y3    & Y1:Y2    & Y1:Y2    \\
    \bottomrule
\end{tabular*}
\end{table}

We estimate equation (\ref{eq:analyst}) with both monthly (columns 1 and 3) and weekly observations (columns 2 and 4) of analyst forecasts. The results are reported in Table \ref{tb:rhoz ibes}, Panel A. In columns (1) and (2), our estimation includes both equations in (\ref{eq:analyst}), while in Column (3) and (4), we only include the first equation, i.e., only using forecasts at one- and two-year horizons for better data quality. Across the specifications, the estimate $\hat{\rho}_z$ is statistically indistinguishable from zero.

\begin{table}[!t]
	\centering
	\caption[tab0: rhoz estimated from analysts forecasts]{The Persistence of Expected Cash-Flow Growth from LTG Forecasts

		\footnotesize This table reports the estimates of $\rho_z^{LT}$, the regression coefficient in
		\[
			\log(1+LTG_t) = \text{const} + \rho^{LT}_z\E_t^A\left[\Delta e_{t+1}\right] + \varepsilon_t,
		\]
		where $LTG_t$ is the long-term growth forecasts (LTG) of the S\&P 500 Index, self-aggregated from stock-level LTG forecasts from the IBES Unadjusted Summary File.
		The short-term forecast, $\E_t^A\left[\Delta e_{t+1}\right]$, is the IGA 1-year earnings growth forecast (IGA $\E_t^A\left[\Delta e_{t+1}\right]$).
		$t$-statistics based on Newey-West standard errors with autocorrelation of up to 18 lags are reported in parentheses.
		Our monthly observations are from 1988:01 to 2019:12.
	}
	\footnotesize
	\label{tb:rhoz ibes ltg}
	\begin{tabular*}{0.4\textwidth}{l@{\extracolsep{\fill}}c}
    \toprule
    {}            & (1)                                                     \\
    \midrule
    {}            & $\log(1+LTG_t)$ \\\midrule
    Intercept     & 0.116                               \\
    & (28.615)                            \\
    $\E_t^A\left[\Delta e_{t+1}\right]$    & 0.017                               \\
    & (0.711)                             \\
    &                                     \\
    $N$           & 384                                 \\
    $R^2$         & 0.011                               \\
    \bottomrule
\end{tabular*}
\end{table}

Next, we consider an alternative way to estimate $\rho_z$ by exploring the relationship between forecasts of short-term and long-term earnings growth (LTG). IBES provides firm-level forecasts of the annualized average growth rate of earnings over the next three to five years and has been adopted in the recent literature on expectation formation (e.g., \citealp{LaPorta1996} and \citealp*{Bordalo2019}).
Given the autoregressive structure in equation (\ref{eq:g1_z}), the expected growth rate from period $n$ to $n+1$ depends on the expected growth rate over the very next period via a coefficient $\rho_z^n$. If $\rho_z$ is zero, then such a coefficient is zero, which implies that the average growth rate over three years and beyond does not depend on the expected growth rate over the next year. Therefore, we regress monthly observations of LTG forecast on $\E_t^A\left[\Delta e_{t+1}\right]$, and denote the regression coefficient by $\rho_z^{LT}$. Our estimate of $\rho_z^{LT}$ is statistically indistinguishable from zero. Consistent with our findings in Table \ref{tb:rhoz ibes}, this implies $\rho_z$.

\begin{table}[!t]
	\centering
	\caption[tab0: rhoz estimated from state-space models]{The Persistence of Expected Cash-Flow Growth from the State-Space Model

		\footnotesize This table reports the estimation results of 1) the unrestricted state-space model given by equations (\ref{eq:g1}) and (\ref{eq:g1_z}) in Section \ref{sec:model}, 2) the restricted state-space model (i.e., $ \rho_z=0 $), and for comparison, 3) the MA(1) model ($ \Delta d_{t+1}=g+\sigma_{D}\varepsilon_{t+1}+\chi \sigma_{D}\varepsilon_{t} $), and 4) the AR(1) model ($ \Delta d_{t+1}=g+\gamma\Delta d_{t}+\sigma_{D}\varepsilon_{t+1} $) of the dividend growth ratres. Panel A uses the annual (non-overlapping) dividend growth of the S\&P 500 index, and Panel B uses the annual (non-overlapping) dividend growth of the Fama-French market portfolio. The log likelihood (``LogL''), AIC, and BIC are reported. $t$-stats are in the squared bracket.}
	\footnotesize
	\begin{tabular}{lccccccccc}
		\toprule
		             & $\hat{\rho }_z$ & $\hat{g}$ & $\hat{\sigma}_{d}$ & $\hat{\sigma}_{z}$ & $\hat{\chi }$ & $\hat{\gamma}$ & LogL  & AIC     & BIC     \\ \midrule
		\multicolumn{10}{@{} l}{Panel A: S\&P 500}                                                                                                        \\ \midrule
		Unrestricted & 0.26            & 0.06      & 0.00               & 0.11               &               &                & 74.44 & -140.88 & -128.97 \\
		             & [0.94]          & [3.01]    & [0.00]             & [1.70]             &               &                &       &         &         \\
		Restricted   &                 & 0.06      & 0.08               & 0.08               &               &                & 71.36 & -136.72 & -127.79 \\
		             &                 & [4.68]    & [0.00]             & [0.00]             &               &                &       &         &         \\
		MA(1)        &                 & 0.06      & 0.10               &                    & 0.41          &                & 76.41 & -146.82 & -137.89 \\
		             &                 & [3.38]    & [13.45]            &                    & [6.11]        &                &       &         &         \\
		AR(1)        &                 & 0.04      & 0.11               &                    &               & 0.26           & 74.50 & -142.99 & -134.06 \\
		             &                 & [3.64]    & [14.90]            &                    &               & [3.51]         &       &         &         \\
		\midrule
		\multicolumn{10}{@{} l}{Panel B: MKT}                                                                                                             \\ \midrule
		Unrestricted & -0.08           & 0.06      & 0.00               & 0.15               &               &                & 43.96 & -79.92  & -69.8   \\
		             & [-0.06]         & [3.86]    & [0.00]             & [0.12]             &               &                &       &         &         \\
		Restricted   &                 & 0.06      & 0.11               & 0.11               &               &                & 43.67 & -81.34  & -73.8   \\
		             &                 & [3.62]    & [0.10]             & [0.10]             &               &                &       &         &         \\
		MA(1)        &                 & 0.06      & 0.15               &                    & -0.09         &                & 44.00 & -82.00  & -74.4   \\
		             &                 & [3.94]    & [6.99]             &                    & [-1.02]       &                &       &         &         \\
		AR(1)        &                 & 0.06      & 0.15               &                    &               & -0.08          & 43.96 & -81.93  & -74.39  \\
		             &                 & [3.89]    & [6.98]             &                    &               & [-0.87]        &       &         &         \\ \bottomrule
	\end{tabular}
	\label{tb: ss_est}
\end{table}





Our final method to estimate $\rho_z$ is to directly estimate the state-space model given by equations (\ref{eq:g1}) and (\ref{eq:g1_z}) with the realized dividend data. Using the standard Kalman filter, we obtain estimates of $\rho_z$. For comparison, we report results for both the S\&P 500 index and the Fama-French market portfolio (``MKT'').\footnote{We obtain dividend data for the Fama-French market portfolio from the CRSP NYSE/NYSEMKT/Nasdaq Value-Weighted Market Index.}
Since the model is set up at annual frequency, we use annual (non-overlapping) dividend growth data. The sample spans 1926 to 2019.\footnote{We also used the longest available S\&P 500 dividend series starting from 1872 and obtained similar results. The results are available upon request.} The results are reported in Table \ref{tb: ss_est}, where Panel A and B are for S\&P 500 and MKT, respectively. In the row ``Unrestricted'' of Panel A and B of Table \ref{tb: ss_est}, the estimates of $\hat{\rho}_z$ are statistically indistinguishable from zero.\footnote{The Kalman filter assumes that the shocks to realized and expected dividend growth are uncorrelated. In the appendix, we demonstrate the robustness of our estimate of $\rho_z$ by considering different values of the correlation, from -0.9 to 0.9, while fixing the volatility of realized-dividend shock at the estimate in Panel A. The estimated $\rho_z$ barely moves with the value of shock correlations in $\left[-0.9, 0.9\right]$ as shown in Figure \ref{fig:ssm_est}.}
The restricted model with $\rho_z=0$ generates similar likelihood and information criteria, indicating that allowing $\rho_z$ to be a free parameter does not significantly improve the model fitness.
We also estimate MA(1) and AR(1) models for comparison and find that the estimates of the autoregressive coefficient, i.e., $\chi $ and $\gamma $ for MA(1) and AR(1), respectively, are statistically indistinguishable from zero. In sum, the state-space approach delivers a similar message as the estimation based on analyst forecasts: The persistence of growth expectation is close to zero.

\begin{table}[!t]
	\centering
	\caption[tab0: rhoz estimated from analysts forecasts]{Time-varying $\rho_z$ Estimates

		\footnotesize This table reports the summary statistics of the rolling-window estimates of $\rho_z$ using weekly observations, where each rolling window spans three years. In each window, we estimate $\rho_z$ following equation (\ref{eq:analyst}), using aggregate earnings growth forecasts of the S\&P 500 Index obtained from IGA. Data sample: 1988:01--2019:12.
	}
	\footnotesize
	\begin{tabular*}{0.93\textwidth}{l@{\extracolsep{\fill}}lllllllll}
    \toprule
    & count   & mean  & std   & min    & 25\%   & 50\%  & 75\%  & max   & $\rho$ \\\midrule
    $\hat{\rho}_{z,t}$ & 384   & 0.025 & 0.157 & -0.260 & -0.066 & 0.001 & 0.074 & 0.791 & 0.975 \\
    \bottomrule
\end{tabular*}
	\label{tb:rhoz_t}
\end{table}

In the last exercise, we conduct a rolling-window estimation of $\rho_z$ following the method in Table \ref{tb:rhoz ibes}. A rolling window contains three years of weekly observations.\footnote{The results are similar if we use alternative window lengths from one to five years (available upon request). Our sample period is 1988--2019. The first estimate of $\rho_z$ uses three years of IGA data starting in 1985.} The summary statistics of the rolling-window estimates are reported in Table \ref{tb:rhoz_t}. Naturally, the model of agents' belief formation may adjust over time, so the estimate of $\hat{\rho}_z$ fluctuates. Overall, the results are similar to those in Table \ref{tb:rhoz ibes}, and in particular, the mean and median of the $\hat{\rho}_z$ series
are close to zero. Our findings suggest that the expected cash-flow growth rate lacks persistence ($\rho_z$ is close to zero), which in our model implies a one-to-one mapping between market duration, $dr$, and expected return. In the next proposition, we characterize more tightly the connection between the value of $\rho_z$ and return predictability. The proof is in the Appendix.

\begin{figure}[h!]
	\centering
	\includegraphics[width=\textwidth]{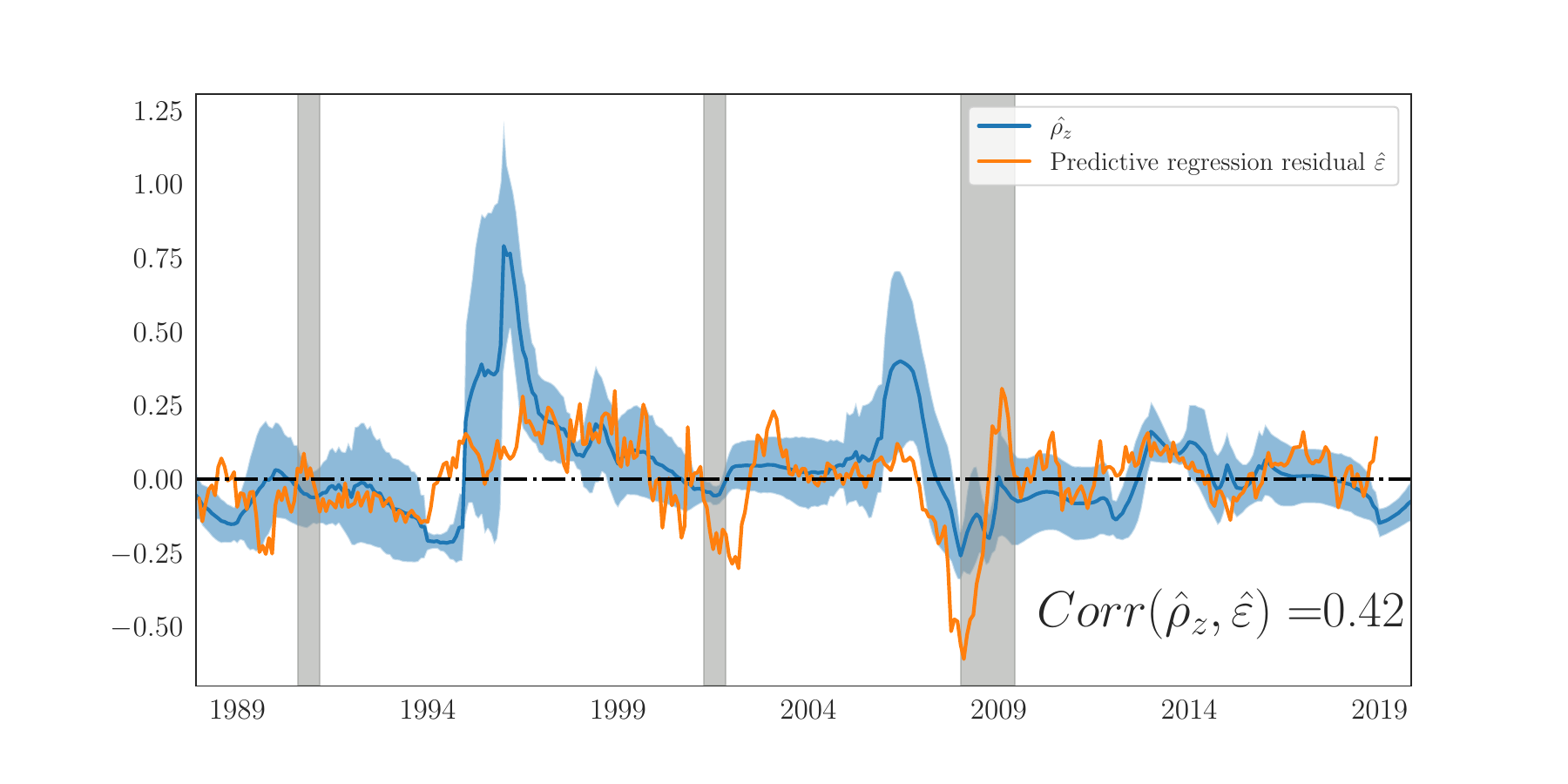}
	\caption[rhoz and prediction errors]{Rolling Estimate of Expected Growth Persistence and Return Prediction Errors

		\footnotesize
		This figure plots the rolling estimate of the autoregressive coefficient of expected cash flow growth, $ \hat{\rho}_{z,t}$, and the return prediction errors using market duration ($dr_t$) as the predictor. $\hat{\rho}_{z,t}$ is estimated using analyst forecasts of S\&P 500 aggregate earnings in rolling regressions with a three-year window. This figure also plots the predictive residuals (denoted by $\varepsilon_t$) from the rolling-sample predictive regressions. The correlation between the two time series is also reported on the graph. Our monthly sample is 1988:01--2019:12.
	}
	\label{fig:rhoz_is_error}%
\end{figure}

\begin{prop}[$\rho_z$ and return forecast errors]
	Let \( \nu_{t+1} \) denote the forecast error when predicting \( r_{t+1} \) with \( dr_t \) and let \( {\rho}_{z} \) denote the autoregressive coefficient of expected cash flow growth $z_t$ in equation (\ref{eq:g1_z}). If \( {\rho}_{z} > 0 \), then \( \nu_{t+1} \) is positive. If \( {\rho}_{z} < 0 \), then \( \nu_{t+1} \) is negative.
\end{prop}

The proposition implies that in a subsample where the estimate of $\rho_z$ is positive (negative), we would expect the return forecasting error to be positive (negative). In Figure \ref{fig:rhoz_is_error}, we plot the rolling-window estimate of $\rho_{z}$ estimates against the return forecasting residuals (denoted by $\varepsilon_t$) from the corresponding rolling window with $dr_t$ as the predictor. The two time series track each other closely, with a correlation of 0.42.\footnote{In Figure \ref{fig:rhoz_oss_error} in the Appendix, we plot $\rho_{z,t}$ against the out-of-sample forecast errors and obtain a similarly positive correlation. We also regress the rolling-window return prediction errors, both in-sample and out-of-sample, on the rolling-window estimate of $\rho_z$ and find a positive regression coefficient (see Table \ref{tb:rhoz_t_regs} in the Appendix).} This provides further evidence of the connection between return predictability and the lack of persistence in expected cash-flow growth.

Market duration $dr$ is the log price-dividend ratio minus the valuation ratio of the one-year dividend strip, which, by construction, teases out market information about near-term growth embedded in the price of the one-year dividend strip. Therefore, $dr$ only contains information about the discount rate, if the market participants possess limited information about growth beyond the next year. In our model, the time-$t$ expectation of growth from $t+2$ onward is a constant under $\rho_z=0$ so the market does not contain information about cash-flow growth beyond the next year under $\rho_z=0$. Next, we step outside of our model given by equations (\ref{eq:g1}) and (\ref{eq:g1_z}) and consider more broadly whether the market contains information about long-term growth.

\subsection{Cash-flow growth predictability: Short horizon vs. long horizon}

\begin{table}[!t]
	\caption[tab2: 1-year dividend and earnings prediction]{One-year Dividend and Earnings Growth Prediction

		\footnotesize This table reports the results of dividend and earnings growth prediction.
		The dependent variables are the 1-year-ahead realized earnings growth from IGA (columns 1-3),  realized dividend growth from Bloomberg (columns 4-6), and realized earnings growth from Compustat (columns 7-9).
		The independent variables are analyst forecasts of one-year earnings growth from IGA ($\E^A_t\left(\Delta e_{t+1}\right)$), duration $dr$, and the price-dividend ratio $pd$.
		$t$-statistics are calculated based on Newey-West standard errors with 18 lags and are reported in parentheses.
		Data sample: 1988:01--2019:12.
	}
	\centering
	\footnotesize
\begin{tabular}{llllllllll}
    \toprule
                                            & \multicolumn{3}{c}{IGA $\Delta e_{t+1}$} & \multicolumn{3}{c}{$\Delta d_{t+1}$} & \multicolumn{3}{c}{Compustat $\Delta e_{t+1}$}                                                                  \\\cmidrule(lr){2-4}\cmidrule(lr){5-7}\cmidrule(lr){8-10}
    {}                                      & {(1)}                                    & {(2)}                                & {(3)}                                          & {(4)}   & {(5)}    & {(6)}    & {(7)}    & {(8)}    & {(9)}    \\\midrule
    Intercept                               & -0.056                                   & -0.426                               & -0.328                                         & 0.028   & -0.312   & -0.295   & 0.122    & 0.478    & 0.394    \\
                                            & (-4.127)                                 & (-1.357)                             & (-2.413)                                       & (1.300) & (-1.526) & (-1.663) & (0.814)  & (0.374)  & (0.352)  \\
    $\E^A_t\left(\Delta e_{t+1}\right)$ & 1.204                                    &                                      & 1.119                                          & 0.326   &          & 0.204    & -0.591   &          & -0.963   \\
                                            & (20.101)                                 &                                      & (15.193)                                       & (2.537) &          & (2.824)  & (-0.629) &          & (-1.067) \\
    $dr_t$                                  &                                          & -0.248                               & -0.110                                         &         & -0.181   & -0.156   &          & -0.333   & -0.452   \\
                                            &                                          & (-2.328)                             & (-2.625)                                       &         & (-3.535) & (-4.188) &          & (-1.129) & (-2.016) \\
    $pd_t$                                  &                                          & 0.385                                & 0.187                                          &         & 0.285    & 0.249    &          & 0.238    & 0.409    \\
                                            &                                          & (2.354)                              & (2.599)                                        &         & (2.782)  & (3.095)  &          & (0.405)  & (0.958)  \\\\
    $N$                                     & 372                                      & 372                                  & 372                                            & 372     & 372      & 372      & 372      & 372      & 372      \\
    $R^2$                                   & 0.731                                    & 0.199                                & 0.769                                          & 0.193   & 0.386    & 0.454    & 0.014    & 0.053    & 0.086    \\
    \bottomrule
\end{tabular}

	\label{tb:1y egdg prediction}
\end{table}

In Section \ref{sec:rhoz}, our focus is on estimating $\rho_z$, the autoregressive coefficient of expected cash-flow growth. In our model, $\rho_z=0$ implies that the market participants do not have information about cash-flow growth beyond the next year. While our analysis of state space in Section \ref{sec:dimension} supports the model specification, we want to step outside of our model and provide more evidence in this subsection on the fact that market information about long-term growth is limited. We do so by characterizing cash-flow predictability at different horizons. Intuitively, if the market participants have information about cash-flow growth at a certain horizon, we should be able to predict such cash-flow growth with information revealed in analysts' forecasts and $dr$ and $pd$ as state variables.

In Table \ref{tb:1y egdg prediction}, we show strong predictability of near-term growth. In column (1), we simply regress the realized one-year growth rate of aggregate earnings from firms covered by IGA on the ex ante analysts' forecast. The $R^2$ is 0.73, so analysts and market participants in general are able to forecast near-term cash-flow growth very well. In column (2), we use our pair of state variables, $dr$ and $pd$, to forecast cash-flow growth and obtain a $R^2$ of 0.20. Combining the information in $dr$ and $pd$ with the analysts' forecast in column (3), the in-sample prediction $R^2$ rises to 0.77. In the other columns of Table \ref{tb:1y egdg prediction}, we replace the forecasting target. Naturally, cash-flow predictability declines in the other cases because the predictor (IGA analyst forecast, in particular) targets the earnings growth of IGA-covered firms rather than that of all firms covered by the Bloomberg database (columns 4-6) or that of all Compustat firms (columns 7-9). Overall, our findings suggest that market participants are informed of near-term cash-flow growth. One explanation is that firms tend to provide forward guidance on their earnings outlook, increasingly so in recent years.

\begin{table}[!t]
	\caption[tab2: beyond 1-year earnings prediction]{Weaker Earnings Growth Predictability beyond One Year

		\footnotesize This table reports results of regressions that predict earnings growth at various horizons.
		The dependent variables are realized earnings growth from IGA of next year (columns 1-3), between the first and second years (columns 4-6), and between the second and third years (columns 7-9).
		The independent variables are analysts' forecasts of one-year earnings growth from IGA (IGA $\E^A_t\left(\Delta e_{t+1}\right)$), the self-aggregated long-term earnings growth forecasts ($LTG_t$) of the S\&P 500 Index, the market duration $dr$ and the price-dividend ratio $pd_t$.
		$t$-statistics calculated based on Newey-West standard errors with 18 lags are reported in parentheses.
		Data sample: 1988:01--2019:12.
	}
	\centering
	\footnotesize

\begin{tabular}{lrrrrrrrrr}
    \toprule
                                            & \multicolumn{3}{c}{IGA $\Delta e_{t, t+1}$} & \multicolumn{3}{c}{IGA $\Delta e_{t+1, t+2}$} & \multicolumn{3}{c}{IGA $\Delta e_{t+2, t+3}$}                                                                   \\\cmidrule(lr){2-4}\cmidrule(lr){5-7}\cmidrule(lr){8-10}
    {}                                      & {(1)}                                       & {(2)}                                         & {(3)}                                         & {(4)}    & {(5)}    & {(6)}    & {(7)}    & {(8)}    & {(9)}    \\
    \midrule
    Intercept                               & 0.053                                       & -0.426                                        & -0.309                                        & 0.325    & -0.063   & -0.037   & 0.185    & 0.513    & 0.496    \\
                                            & (1.360)                                     & (-1.357)                                      & (-2.786)                                      & (2.446)  & (-0.211) & (-0.141) & (1.596)  & (1.577)  & (1.685)  \\
    IGA $\E^A_t\left(\Delta e_{t+1}\right)$ & 1.222                                       &                                               & 1.174                                         & 0.030    &          & -0.047   & -0.319   &          & -0.274   \\
                                            & (20.312)                                    &                                               & (15.749)                                      & (0.192)  &          & (-0.348) & (-2.594) &          & (-1.947) \\
    $LTG_t$                                 & -0.889                                      &                                               & -1.344                                        & -2.155   &          & -2.295   & -0.711   &          & -0.295   \\
                                            & (-3.171)                                    &                                               & (-3.092)                                      & (-2.186) &          & (-1.578) & (-0.768) &          & (-0.234) \\
    $dr_t$                                  &                                             & -0.248                                        & -0.077                                        &          & -0.144   & -0.103   &          & 0.096    & 0.068    \\
                                            &                                             & (-2.328)                                      & (-1.568)                                      &          & (-2.878) & (-2.319) &          & (1.449)  & (1.215)  \\
    $pd_t$                                  &                                             & 0.385                                         & 0.189                                         &          & 0.181    & 0.207    &          & -0.216   & -0.166   \\
                                            &                                             & (2.354)                                       & (2.797)                                       &          & (1.690)  & (2.351)  &          & (-1.666) & (-1.399) \\\\
    $N$                                     & 372                                         & 372                                           & 372                                           & 360      & 360      & 360      & 348      & 348      & 348      \\
    $R^2$                                   & 0.74                                        & 0.20                                          & 0.78                                          & 0.08     & 0.08     & 0.14     & 0.07     & 0.07     & 0.11     \\
    \bottomrule
\end{tabular}
	\label{tb:123y egdg prediction}
\end{table}

Next, we examine the predictability of cash-flow growth at longer horizons. So far, our analysis of growth expectations has been based on the autoregressive model given by equations (\ref{eq:g1}) and (\ref{eq:g1_z}), and the focus has been on $\rho_z$, the persistence parameter, and its connection with the return predictive power of market duration $dr$. In this model, we have the time-$t$ expected growth from $t+1$ to $t+2$, $\E_t[z_{t+2}]=\rho_z\E_t[z_{t+1}]=\rho_z^2 z_t$, so market information about long-term growth is contained in the expected growth rate over the next period. In the following exercises, we expand the set of potential predictors of long-term growth. In particular, we consider the LTG forecast, the aggregated long-term growth forecast over three to five years.

In Table \ref{tb:123y egdg prediction}, we first predict earnings growth over the next year from columns (1) to (3), and consistent with our findings in Table \ref{tb:1y egdg prediction}, predictability is strong, which suggests that market participants are informed about short-term growth. In columns (4) to (6), we use the same variables to predict earnings growth from $t+1$ to $t+2$ and find that predictability declines dramatically. Comparing columns (1) and (4), the predictive $R^2$ declines from 0.74 to 0.08. Our forecasting exercise in columns (7) to (9) delivers the same message. The predictability of growth over an even longer horizon, i.e., from $t+2$ to $t+3$, is even weaker. In Figure \ref{fig:bootstrap iga earnings} in the Appendix, we report the results of alternative predictive models that take advantage of more valuation ratios (and potentially richer information embedded in the state space). Our conclusion remains robust.

Overall, our findings suggest that market participants are not informed about long-term growth, but regarding short-term growth, market information, whether from analysts' forecasts or our state variables, is quite rich. These results on the term structure of cash-flow predictability shed light on the strong connection between market duration $dr$ and expected return. By construction, $dr$ deducts the valuation ratio of the one-year dividend strip from the price-dividend ratio $pd$ and thereby teases out market information about near-term growth.

One may form a market timing strategy guided by our findings on $dr$ as a return predictor. Specifically, when market duration increases, the strategy shorts the market, and when market duration decreases, the strategy adds to the long position. This strategy bets against the market valuation of long-duration cash flows. An out-of-sample $R^2$ of $14.6\%$ in column (1) of Table \ref{tb:uncondUS} implies that the Sharpe ratio of this market timing strategy is $0.58$, which is much higher than the Sharpe ratio from other return predictors, for example, $0.37$ in \citet{CampbellThompson2008}. In the appendix, we show how to calculate the Sharpe ratio based on the out-of-sample $R^2$.

\section{Conclusion}
We construct a valuation-based measure of stock market duration as the ratio of total market capitalization to the price of dividends paid in the very next year. An increase in market duration implies the valuation of long-term cash flows increases relative to that of short-term cash flows. While market duration ($dr$) represents the slope of the valuation term structure, the traditional price-dividend ratio ($pd$) represents the level. After establishing the connection between state variables and valuation ratios of dividend strips, we demonstrate that the state space is two-dimensional. Through return and cash-flow predictions, we find that $dr$ and $pd$ span the state space and may serve as a pair of state variables for asset pricing and in macro-finance studies.

Our findings not only establish market duration as a key state variable of important applications but also shed light on what drives market duration. We find a tight link between market duration and expected return. $dr$ exhibits strong return predictive power, outperforming other predictors.
Such a connection between $dr$ and the discount rate emerges when market information on long-term cash flows is limited, and empirically, we document the supporting evidence of a sharp decline of cash-flow predictability once the forecasting horizon goes beyond one year.

\cite{Cochrane2011} points out how return predictability emerges from the lack of cash-flow predictability and the price-dividend ratio should forecast future returns. We find that cash flows in the near term are in fact highly predictable, but cash flows over the long term are not. Therefore, our measure of market duration is a better return predictor than the price-dividend ratio. Our findings support a market timing strategy of betting against market duration. When the valuation of long-term cash flows rises relative to that of near-term cash flows about which market participants are more informed, it is likely due to a lower discount rate or exuberance over long-term growth \citep*{Bordalo2023}.

\clearpage

\clearpage
\singlespacing
\nocite{LarrainYogo2008}
\bibliography{library}

\clearpage
\renewcommand*{\thepage}{A.\arabic{page}}
\renewcommand{\thefigure}{A.\arabic{figure}}
\renewcommand{\thetable}{A.\arabic{table}}
\setcounter{page}{1}
\setcounter{figure}{0}
\setcounter{table}{0}
\singlespacing

\section*{Appendix I: Derivation} \label{app:derive}
\subsection*{I.1\quad Solving the valuation ratios}
The price-dividend ratio of the dividend strip with maturity $n$, $P_{n,t}/D_t$, satisfies the following recursive equation
\begin{equation}
	\frac{P_{n,t}}{D_t}=\mathbb{E}_t\left[M_{t+1}\frac{D_{t+1}}{D_t}\frac{P_{n-1,t+1}}{D_{t+1}}\right].
\end{equation}
We conjecture that
\begin{equation}
	\ln \left(\frac{P_{n,t}}{D_t}\right) = A\left(n\right)+B\left(n\right)^T X_t.
\end{equation}

Substituting this expression and expressions of stochastic discount factor and dividend growth into the recursive equation, we have
\begin{align}
	  & \exp \left\lbrace A\left(n\right)+B\left(n\right)^{\top }X_{t} \right\rbrace \nonumber \\
	= &
	\mathbb{E}_t\left[\exp \left\lbrace
	-r_f-\frac{1}{2}\lambda_t^{\top }\Sigma \lambda_t - \lambda_t^{\top }\epsilon_{t+1} + g_t + \sigma_D^{\top }\epsilon_{t+1}
	+A\left(n-1\right)+B\left(n-1\right)^{\top }X_{t+1}
	\right\rbrace \right] \nonumber                                                            \\
	= &
	\mathbb{E}_t\left[\exp \left\lbrace
		g_t-r_f-\frac{1}{2}\lambda_t^{\top }\Sigma \lambda_t
		+A\left(n-1\right)+B\left(n-1\right)^{\top }\Pi X_t
		+\left(\sigma_D-\lambda_t+\sigma_X B\left(n-1\right)\right)^{\top }\epsilon_{t+1}
	\right\rbrace \right] \nonumber                                                            \\
	= & \exp \left\lbrace
	g_t-r_f-\frac{1}{2}\lambda_t^{\top }\Sigma \lambda_t
	+A\left(n-1\right)+B\left(n-1\right)^{\top }\Pi X_t
	\color{white}\right\rbrace \nonumber                                                       \\
	  & \color{white}\left\lbrace \color{black}
	+\frac{1}{2}\left(\sigma_D-\lambda_t+\sigma_X B\left(n-1\right)\right)^{\top }\Sigma \left(\sigma_D-\lambda_t+\sigma_X B\left(n-1\right)\right)
	\right\rbrace \nonumber                                                                    \\
	= & \exp \left\lbrace
	g_t-r_f+A\left(n-1\right)+B\left(n-1\right)^{\top }\Pi X_t
	-\left(\sigma_D+\sigma_X B\left(n-1\right)\right)^{\top }\Sigma \lambda _t
	\color{white}\right\rbrace \nonumber                                                       \\
	  & \color{white}\left\lbrace \color{black}
	+\frac{1}{2}\left(\sigma_D+\sigma_X B\left(n-1\right)\right)^{\top }\Sigma \left(\sigma_D+\sigma_X B\left(n-1\right)\right) \right\rbrace
\end{align}
The coefficients on $X_t$ should match $B\left(n\right)$ on the left hand side, so we have
\begin{equation} \label{eq:B_app}
	B\left(n\right)=\left(\Pi ^{\top } - \theta \Sigma \sigma_X \right) B\left(n-1\right)
	+\phi - \theta \Sigma \sigma_D - \gamma.
\end{equation}
The constants must sum up to $A\left(n\right)$ on the left hand side, so we have
\begin{align} \label{eq:A_app}
	A\left(n\right)= & A\left(n-1\right)+\overline{g}-\overline{r}
	-\left(\sigma_D+\sigma_X B\left(n-1\right)\right)^{\top }\Sigma \overline{\lambda }+                                                  \\
	                 & \frac{1}{2}\left(\sigma_D+\sigma_X B\left(n-1\right)\right)^{\top }\Sigma \left(\sigma_D+\sigma_X B\left(n-1\right)\right). \nonumber
\end{align}
The fact that $P^0_t=D_t$ implies the boundary conditions, $A\left(0\right)=B\left(0\right)=0$, which pins down a solution of $A\left(n\right)$ and $B\left(n\right)$.

Finally, we solve the log price-dividend ratio of the aggregate stock market. We conjecture
\begin{equation}
	pd_{t}=\ln \left(P_{t}/D_{t}\right)=A + B^T X_t\text{,}
\end{equation}
and proceed to solve $A$ and $B$. Following \cite{Campbell1988}, we log-linearize the stock market return
\begin{align}
	r^{mkt}_{t+1}
	= & \kappa _{0} + \kappa _{1} pd_{t+1} - pd_{t}+\Delta d_{t+1} \nonumber \\
	= & \kappa _{0} - \left(1-\kappa _{1} \right)A
	- B^{\top } \left(\mathbf{I} -\kappa _{1} \Pi \right) X_t + g_{t} + \left(\kappa _{1} \sigma _{X} B + \sigma _{D}\right)^{\top } \epsilon_{t+1}
\end{align}
Under the no-arbitrage condition, we have
\begin{equation}
	1=\mathbb{E}_t\left[M_{t+1}\exp(r^{mkt}_{t+1})\right].
\end{equation}
We follow the same method of matching undetermined coefficients in the analysis of dividend strip valuation ratios and solve
\begin{align}
	A= & \frac{1}{1-\kappa _1}\left[
		\overline{g}-\overline{r}
		+\kappa _{0}
		-\left(\kappa _1 \sigma _X B + \sigma _{D} \right)^{\top }\Sigma \overline{\lambda }
		+\frac{1}{2}\left(\kappa _1 \sigma _X B \right)^{\top }\Sigma \left(\kappa _1 \sigma _X B \right)
		+\left(\kappa _1 \sigma _X B \right)^{\top }\Sigma \sigma _{D}
	\right ]                                                                                                                                            \\
	B= & \left(\mathbf{I}-\kappa _{1}\Pi ^{\top }-\kappa _1 \theta \Sigma \sigma _X \right)^{-1}\left(\phi -\theta \Sigma \sigma _{D}  -\gamma \right).
\end{align}

\subsection*{I.2\quad Valuation ratios from the cross section}
Consider an individual stock $i$. The dividend dynamics of firm $i$ depend not only on the aggregate state variables, $X_t$, but also on the firm $i$-specific state variables, $Z_{i,t}$, that is $K_i$-dimensional and independent from $X_t$. Without loss of generality, we assume that $Z_{i,t}$ evolves as a first-order vector autoregression
\begin{equation}
	Z_{i,t+1}=\Omega Z_{i,t} + \sigma ^{\top}_{i,Z} \upsilon _{i,t+1},
\end{equation}
where $\upsilon _{i,t+1}$ is a $N_i$-by-1 vector of $i$-specific news that has a normal distribution $N\left(\mathbf{0},\Sigma _i\right)$ and is independent over time and independent from the aggregate shocks $\epsilon _{t+1}$. We use subscript $i$ to differentiate firm $i$ from the aggregate variables (without subscript $i$) and other firms (with subscript $j\neq i$).

The dividend growth rate of firm $i$ loads on the aggregate and idiosyncratic shocks
\begin{equation}
	\ln \left(\frac{D_{i,t+1}}{D_{i,t}}\right) = g_{i,t} + \sigma _{i,D}^{\top} \epsilon_{t+1} + \sigma _{i, \upsilon }^{\top} \upsilon _{i,t+1},
\end{equation}
where the expected dividend growth rate is given by
\begin{equation}
	g_{i,t}=\phi _{i}^{\top} X_t + \delta _{i}^{\top} Z_{i,t} + \overline{g}_{i}-\frac{1}{2}\sigma _{i,D}^{\top}\Sigma \sigma _{i,D} - \frac{1}{2}\sigma _{i,\upsilon }^{\top}\Sigma _{i}\sigma _{i,\upsilon },
\end{equation}
which loads on the aggregate state variables, $X_t$, and firm $i$-specific state variables, $Z_{i,t}$.

The ratio of firm $i$'s dividend strip price, $P^n_{i,t}$, to firm $i$'s current dividend is
\begin{equation}
	\frac{P^n_{i,t}}{D_{i,t}}=\exp \left\lbrace A_i\left(n\right) + B_i\left(n\right)^{\top}X_t + C_i\left(n\right)^{\top} Z_{i,t} \right\rbrace ,
\end{equation}
where $A_i\left(n\right)$, $B_i\left(n\right)$, and $C_i\left(n\right)$ are firm $i$-specific, deterministic functions of $n$ given by the recursive equations
\begin{align}
	B_i\left(n\right)= & \left(\Pi ^{\top } - \theta \Sigma \sigma_X \right) B_i\left(n-1\right) +\phi_i - \gamma - \theta \Sigma \sigma_{i,D}.                                                                                         \\
	C_i\left(n\right)= & \Omega ^{\top} C_i\left(n-1\right)+\delta _i                                                                                                                                                                   \\
	A_i\left(n\right)= & A_i\left(n-1\right)+\overline{g}_i-\overline{r}
	-\left(\sigma_{i,D}+\sigma_X B_i\left(n-1\right)\right)^{\top }\Sigma \overline{\lambda }+\frac{1}{2}\left(\sigma_{i,D}+\sigma_X B_i\left(n-1\right)\right)^{\top }\Sigma  \nonumber                                                \\
	                   & \left(\sigma_{i,D}+\sigma_X B_i\left(n-1\right)\right)+\frac{1}{2}\left(\sigma_{i,\nu }+\sigma_{i,Z} C_i\left(n-1\right)\right)^{\top }\Sigma_i \left(\sigma_{i,\nu }+\sigma_{i,Z} C_i\left(n-1\right)\right).
\end{align}
with the initial conditions
\begin{equation}
	A_i\left(0\right)=0\text{,\, }B_i\left(0\right)=0\text{,\, and\, }C_i\left(0\right)=0.
\end{equation}

The price of firm $i$'s stock, $P_{i,t}$, is the sum of all its dividend strips
\begin{equation}
	\frac{P_{i,t}}{D_{i,t}}=\sum _{n=1}^{+\infty }\frac{P^n_{i,t}}{D_{i,t}}=\sum _{n=1}^{+\infty } \exp \left\lbrace A_i\left(n\right) + B_i\left(n\right)^{\top}X_t + C_i\left(n\right)^{\top} Z_{i,t} \right\rbrace .
\end{equation}
In Appendix I, we use the log-linearization method of \cite{Campbell1988} to solve an approximate exponential-affine form, so the log price-dividend ratio of stock $i$ is
\begin{equation}
	\ln \left(\frac{P_{i,t}}{D_{i,t}}\right)\approx A_i + B_i^{\top}X_t + C_i^{\top} Z_{i,t}.
\end{equation}
Because $Z_{i,t}$ is independent from $X_t$, recovering the state space $X_t$ using individual stocks' price-dividend ratio brings in noise. In a forecasting context, \cite{Kelly2013b} deal with this issue using partial least squares, which is a method to compress the cross-section of valuation ratios into signals (about the state variables) that are most relevant for the forecasting targets.

\subsection*{I.3\quad Solving the two-dimensional state space model}
We conjecture that the market price-dividend ratio is exponential-affine in the state variables, so the log ratio is
\[
	pd_t=\ln \left(S_t/D_t\right)=A+B y_t+C z_t\text{.}
\]
Next, we use the log-linearization of \cite{Campbell1988}, i.e.,
\[
	r_{t+1}=\kappa _{0}+\kappa _{1}pd_{t+1}-pd_{t}+\Delta d_{t+1} \text,
\]
and substitute this log market return into the no-arbitrage condition
\[
	\E_t\left[M_{t+1}\exp \{ r_{t+1} \}\right]=1\text{.}
\]
to obtain
\begin{equation}
	\E_t\left[ \exp \left\lbrace -r_f - \frac{1}{2}\lambda_t^2(\sigma_{\lambda}^{\top}\Sigma\sigma_{\lambda})^2-\lambda_t\sigma_{\lambda}^{\top}\epsilon_{t+1} + \kappa _{0}+\kappa _{1}pd_{t+1}-pd_{t}+\Delta d_{t+1} \right\rbrace   \right] = 1
\end{equation}
Using the conjecture of $pd_t$ and $pd_{t+1}$ and the specification of $g_t$ and $\Delta d_{t+1}$, we obtain
\begin{align}
	\E_t & \left[ \exp \left\lbrace {\color{white}\frac{1}{1}}-r_f - \frac{1}{2}\lambda_t^2(\sigma_{\lambda}^{\top}\Sigma\sigma_{\lambda})^2-\lambda_t\sigma_{\lambda}^{\top}\epsilon_{t+1} + \kappa _{0} - A - B y_t - C z_t + z_t + \overline{g} - \frac{1}{2}\sigma _D^{\top} \Sigma \sigma _D + \sigma _D^{\top} \epsilon_{t+1} \right. \right. \nonumber \\
	     & \left. \left. +\kappa _{1}A +\kappa _{1}B(\rho_y y_{t} + \sigma_y^{\top }\epsilon_{t+1}) +\kappa _{1}C(\rho_z z_{t} + \sigma_z^{\top }\epsilon_{t+1}) {\color{white}\frac{1}{1}}\right\rbrace \right] = 1
\end{align}
For the conjecture of $pd_t$ functional form to hold, the coefficient on $z_t$ is zero, so we obtain
\begin{equation}
	C=\frac{1}{1-\kappa_1\rho_z}
\end{equation}
Collecting all terms with shocks at $t+1$ and using the moment-generating function, we obtain
\begin{align}
	 & \E_t\left[ \exp \left\lbrace
	-\lambda_t\sigma_{\lambda}^{\top}\epsilon_{t+1} + \sigma _D^{\top} \epsilon_{t+1} + \kappa _{1}B\sigma_y^{\top }\epsilon_{t+1} + \kappa _{1}C\sigma_z^{\top }\epsilon_{t+1}
	\right\rbrace \right]  = \exp \left\lbrace \frac{1}{2}\lambda_t^2(\sigma_{\lambda}^{\top}\Sigma\sigma_{\lambda})^2 \right.                                                                                                                                                         \\
	 & \left. - (\sigma _D + \kappa _{1}B\sigma_y + \kappa _{1}C\sigma_z)^{\top }\Sigma\sigma_{\lambda}\lambda_t + \frac{1}{2}(\sigma _D + \kappa _{1}B\sigma_y + \kappa _{1}C\sigma_z)^{\top }\Sigma(\sigma _D + \kappa _{1}B\sigma_y + \kappa _{1}C\sigma_z) \right\rbrace \nonumber
\end{align}
Substituting this expression into the no-arbitrage condition, we obtain
\begin{align}
	 & \exp \left\lbrace -r_f + \kappa _{0} - A - B y_t - C z_t + z_t + \overline{g} - \frac{1}{2}\sigma _D^{\top} \Sigma \sigma _D - (\sigma _D + \kappa _{1}B\sigma_y + \kappa _{1}C\sigma_z)^{\top }\Sigma\sigma_{\lambda}(\overline{\lambda }+y_{t}) \right. \nonumber \\
	 & \left. +\kappa _{1}A +\kappa _{1}B\rho_y y_{t} +\kappa_{1}C\rho_z z_{t} + \frac{1}{2}(\sigma _D + \kappa _{1}B\sigma_y + \kappa _{1}C\sigma_z)^{\top }\Sigma(\sigma _D + \kappa _{1}B\sigma_y + \kappa _{1}C\sigma_z)\right\rbrace = 1
\end{align}
For the conjecture of $pd_t$ functional form to hold, the coefficient on $y_t$ is zero, so we obtain
\begin{equation}
	B = -\frac{(\sigma _D + \kappa _{1}C\sigma_z)^{\top }\Sigma\sigma_{\lambda}}{1 + \kappa _{1}\sigma_y^{\top}\Sigma\sigma_{\lambda} - \kappa_1\rho_y}
\end{equation}
Finally, all the constant terms should add up to zero, so we obtain
\begin{equation}
	A=\frac{\overline{g}-r_f + \kappa_0 - \frac{1}{2}\sigma _D^{\top} \Sigma \sigma _D + \frac{1}{2}(\sigma _D + \kappa _{1}B\sigma_y + \kappa _{1}C\sigma_z)^{\top }\Sigma(\sigma _D + \kappa _{1}B\sigma_y + \kappa _{1}C\sigma_z - 2\sigma_{\lambda}\overline{\lambda })}{1-\kappa_1}
\end{equation}
In the main text, to clarify the notations, we use $A_{pd}$, $B_{pd}$, and $C_{pd}$ to denote $A$, $B$, and $C$ above, respectively.

Next, we solve the time-$t$ log price-dividend ratio of the dividend strip that matures at $t+1$. The no-arbitrage condition dictates
\begin{equation}
	\mathbb{E}_t\left[M_{t+1}\frac{D_{t+1}}{P^1_t}\right]=1,\,
\end{equation}
or equivalently
\begin{equation}
	\mathbb{E}_t\left[M_{t+1}\frac{D_{t+1}}{D_t}\frac{D_t}{P^1_t}\right]=\mathbb{E}_t\left[M_{t+1}\exp\left\lbrace g_t+\sigma _D^{\top}\epsilon_{t+1} - s^1_t\right\rbrace \right]=1,\,
\end{equation}
so we obtain
\begin{equation}
	\E_t\left[ \exp \left\lbrace -r_f - \frac{1}{2}\lambda_t^2(\sigma_{\lambda}^{\top}\Sigma\sigma_{\lambda})^2-\lambda_t\sigma_{\lambda}^{\top}\epsilon_{t+1} + g_t + \sigma _D^{\top}\epsilon_{t+1} - s^1_{t} \right\rbrace \right] = 1.
\end{equation}
We conjecture
\[
	s^1_t=A_1 + B_1 y_t + C_1 z_t.
\]
Substituting this conjecture, the specification of $g_t$, and the specification of $\lambda_t$ into the no-arbitrage condition, we obtain
\begin{equation}
	\small \E_t\left[ \exp \left\lbrace -r_f - \frac{1}{2}(\overline{\lambda}+y_t)^2(\sigma_{\lambda}^{\top}\Sigma\sigma_{\lambda})^2-(\overline{\lambda}+y_t)\sigma_{\lambda}^{\top}\epsilon_{t+1} + z_t + \overline{g} - \frac{1}{1}\sigma_D^{\top}\Sigma\sigma_D + \sigma _D^{\top}\epsilon_{t+1} - A_1 - B_1 y_t - C_1 z_t \right\rbrace \right] = 1. \nonumber
\end{equation}
Using the moment-generating function to simplify the expression, we obtain
\begin{equation}
	\exp \left\lbrace -r_f + z_t + \overline{g} - A_1 - B_1 y_t - C_1 z_t - \sigma_{\lambda}^{\top}\Sigma\sigma_D(\overline{\lambda}+y_t) \right\rbrace = 1.
\end{equation}
For the conjecture of $s^1_t$ functional form to hold, the coefficient of $z_t$ and the coefficient of $y_t$ must be zero, so we obtain
\begin{equation}
	C_1=1,
\end{equation}
and
\begin{equation}
	B_1=-\sigma_{\lambda}^{\top}\Sigma\sigma_D.
\end{equation}
Finally, the constant terms add up to zero, so we obtain
\begin{equation}
	A_1=\overline{g}-r_f-\sigma_{\lambda}^{\top}\Sigma\sigma_D\overline{\lambda}
\end{equation}

Finally, we solve the conditional expected market return. First, we start with $\mathbb{E}_t[ r_{t+1} ] = \kappa_{0}+\kappa _{1}\mathbb{E}_t[ pd_{t+1} ]-pd_{t}+g_t$. Using the expression of $pd_{t+1}$, $pd_t$, and $g_t$, and the specifications of law of motion of $z_t$ and $y_t$, we obtain
\begin{align}
	\mathbb{E}_t[ r_{t+1} ] = & \kappa_{0}-(1-\kappa_1)A + \overline{g} -\frac{1}{2}\sigma_D^{\top}\Sigma\sigma_D - (1-\kappa_1\rho_y)By_t.
\end{align}
We collect the constant terms into $A_{er}$ and define the coefficient of $y_t$ to be $B_{er}$.

\subsection*{I.4\quad Proof of Proposition 3 on $\rho_z$ and return forecasting errors}

%

\begin{proof}
	We know that the expected return is a function of the price of risk $y_t$:
	\[
		\mathbb{E}_t[ r_{t+1} ] = A_{er} + B_{er}y_t,
	\]
	and that
	\[
		dr_t= A_{pd}-A_1 + (B_{pd}-B_1)y_t + (C_{pd}-C_1)z_t.
	\]
	Combining the two equations, we have
	\begin{align}
		\mathbb{E}_t[ r_{t+1} ] & = A_{er} + \frac{B_{er}}{B_1-B_{pd}}\left[dr_t-A_{pd}+A_1-(C_{pd}-C_1)z_t\right] \\
		                        & = \text{const.}+ \frac{B_{er}}{B_1-B_{pd}}\left[dr_t-(C_{pd}-C_1)z_t\right]
		\label{eq:dr pred}
	\end{align}

	If $\rho_z=0$, $\mathbb{E}_t[ r_{t+1}] = \text{const.}+ \frac{B_{er}}{B_1-B_{pd}}dr_t$. The forecast error is a white noise independent of time-$t$ variables:
	\[
		\nu_{t+1} = r_{t+1}- \mathbb{E}_t[ r_{t+1}] = \epsilon_{t+1}.
	\]
	However, if $\rho_z\neq 0$ but the investor still uses equation (\ref{eq:dr pred}) to forecast $t+1$ return, the forecast error is then
	\begin{align*}
		\nu_{t+1} & = r_{t+1}- \left[\text{const.}+ \frac{B_{er}}{B_1-B_{pd}}dr_t\right] =r_{t+1}- \left[\mathbb{E}_t[r_{t+1}]+\frac{B_{er}(C_{pd}-C_1)}{B_1-B_{pd}}z_t\right] \\
		          & = \epsilon_{t+1}-\frac{B_{er}(C_{pd}-C_1)}{B_1-B_{pd}}z_t = \epsilon_{t+1}-\frac{B_{er}}{B_1-B_{pd}} \left(\frac{1}{1-\kappa_1 \rho_z}-1\right) z_t.
	\end{align*}
	The correlation between $\hat{\rho}_{z,t}$ and $\nu_{t+1}$ is therefore
	\[
		Corr({\rho}_{z,t}, \nu_{t+1}) = -\frac{B_{er}}{B_1-B_{pd}}Corr\left({\rho}_{z,t}, \left(\frac{1}{1-\kappa_1 \rho_{z,t}}-1\right) z_t \right)
	\]
	Based on our findings on return predictability, $dr_t$ negatively predicts future returns. Therefore, the coefficient of $dr_t$ in equation (\ref{eq:dr pred}), $\frac{B_{er}}{B_1-B_{pd}}$, is negative. Under this condition, we obtain
	\begin{align*}
		\sgn\left(Corr({\rho}_{z,t}, \nu_{t+1})\right) & = \sgn\left(Cov\left({\rho}_{z,t}, \left(\frac{1}{1-\kappa_1 \rho_{z,t}}-1\right) z_t\right)\right)                                                                                                           \\
		                                               & =\sgn\left(\mathbb{E}\left(\frac{\kappa_1\rho_{z,t}^2 z_t}{1-\kappa_1 \rho_{z,t}}\right)- \mathbb{E}\left(\rho_{z,t}\right)\mathbb{E}\left(\frac{\kappa_1\rho_{z,t} z_t}{1-\kappa_1 \rho_{z,t}}\right)\right)
	\end{align*}
	As demonstrated by the rolling estimation results in Table \ref{tb:rhoz_t} in Section \ref{sec:rhoz}, $\rho_{z,t}$ on average is close to zero (see also Table \ref{tb:rhoz_t}, we have $\mathbb{E}\left(\hat{\rho}_{z,t}\right)\approx 0$. Using 1-year earnings growth forecasts from IBES Global Aggregate (IGA) as a proxy for $z_t$ and $\kappa_1=0.98$, we calculate the estimate of $\mathbb{E}\left(\frac{\kappa_1\hat{\rho}_{z,t}^2 z_t}{1-\kappa_1 \hat{\rho}_{z,t}}\right)$ in our sample to be $0.005626$ with p-value $<0.01$, which implies
	\begin{align*}
		\sgn\left(Corr({\rho}_{z,t}, \nu_{t+1})\right) = \sgn\left(\mathbb{E}\left(\frac{\kappa_1\rho_{z,t}^2 z_t}{1-\kappa_1 \rho_{z,t}}\right)\right) >0.
	\end{align*}
\end{proof}

\subsection*{I.5\quad Deriving the Sharpe ratio of market-timing strategy}
Following \citet{CampbellThompson2008}, we assume that the excess return can be decomposed as follows:
\[
	r_{t+1}=\mu+x_{t}+\varepsilon_{t+1}
\]
where $\mu$ is the unconditional mean. The predictor $x_{t}$ has mean 0 and variance $\sigma_{x}^{2}$, independent from the error term $\varepsilon_{t+1}$. For simplicity, we assume that the mean-variance investor has a relative risk aversion coefficient $\gamma=1$. When using $x_{t}$ to time the market, the investor allocates
\[
	\alpha_{t}=\frac{\mu+x_{t}}{\sigma_{\varepsilon}^{2}}
\]
to the risky asset and on average earns an excess return of
\[
	\E\left(\alpha_{t}r_{t+1}\right)=\E\left(\frac{\left(\mu+x_{t}\right)\left(\mu+x_{t}+\varepsilon_{t+1}\right)}{\sigma_{\varepsilon}^{2}}\right)=\frac{\mu^{2}+\sigma_{x}^{2}}{\sigma_{\varepsilon}^{2}}
\]
The variance of the market-timing strategy is
\begin{align*}
	\Var\left(\alpha_{t}r_{t+1}\right) & =\Var\left[\frac{\left(\mu+x_{t}\right)\left(\mu+x_{t}+\varepsilon_{t+1}\right)}{\sigma_{\varepsilon}^{2}}\right]
\end{align*}
The (squared) market-timing Sharpe ratio $s_{1}^2$ can be written as
\[
	s_{1}^{2}=\frac{\left[\E\left(\alpha_{t}r_{t+1}\right)\right]^{2}}{\Var\left(\alpha_{t}r_{t+1}\right)}=A\cdot\frac{\mu^{2}+\sigma_{x}^{2}}{\sigma_{\varepsilon}^{2}}
\]
where $A$ is a constant that depends on $Var\left[\left(\mu +x_t\right)\left(\mu +x_t+\varepsilon _{t+1}\right)\right]$ and $ (\mu^2+\sigma_{x}^2)/\sigma_{\varepsilon}^2 $.

Given the buy-and-hold Sharpe ratio $s_{0}$,
\[
	s_{0}^{2}=\frac{\mu^{2}}{\sigma_{x}^{2}+\sigma_{\varepsilon}^{2}}
\]
and the predictive regression $R^{2}$,
\[
	R^{2}=\frac{\sigma_{x}^{2}}{\sigma_{x}^{2}+\sigma_{\varepsilon}^{2}}\text{,}
\]
we obtain the relationship between the buy-and-hold and market-timing Sharpe ratios as
\begin{gather*}
	s_{1}^{2}=A\cdot\frac{\mu^{2}+\sigma_{x}^{2}}{\sigma_{\varepsilon}^{2}}=A\cdot\frac{\mu^{2}+\sigma_{x}^{2}}{\left(\sigma_{x}^{2}+\sigma_{\varepsilon}^{2}\right)\left(1-R^{2}\right)}=A\cdot\frac{s_{0}^{2}+R^{2}}{1-R^{2}}
\end{gather*}
When the predictor has no predictive power, we know that $ R^2=0 $ and $ s_0=s_1 $. We therefore pin down the constant $ A=1 $ and obtain
\begin{gather}
	s_{1}=\sqrt{\frac{s_{0}^{2}+R^{2}}{1-R^{2}}}\text{.}
	\label{eq:sharpe}
\end{gather}

Using data back to 1871, \citet{CampbellThompson2008} obtain a long-term estimate of the market buy-and-hold Sharpe ratio (``$s_0$'') of $0.37$ (annualized). If a mean-variance investor uses the information from $dr$ to construct a market-timing strategy, with an out-of-sample $R^2$ of 14.6\%, she would obtain a Sharpe ratio (``$s_1$'') of $0.58$, representing a 54.7\% improvement over the Sharpe ratio achieved by the buy-and-hold approach.

\clearpage
\section*{Appendix II: Additional Tables and Figures}
\begin{figure}[H]
	\begin{center}
		\includegraphics[scale=0.47]{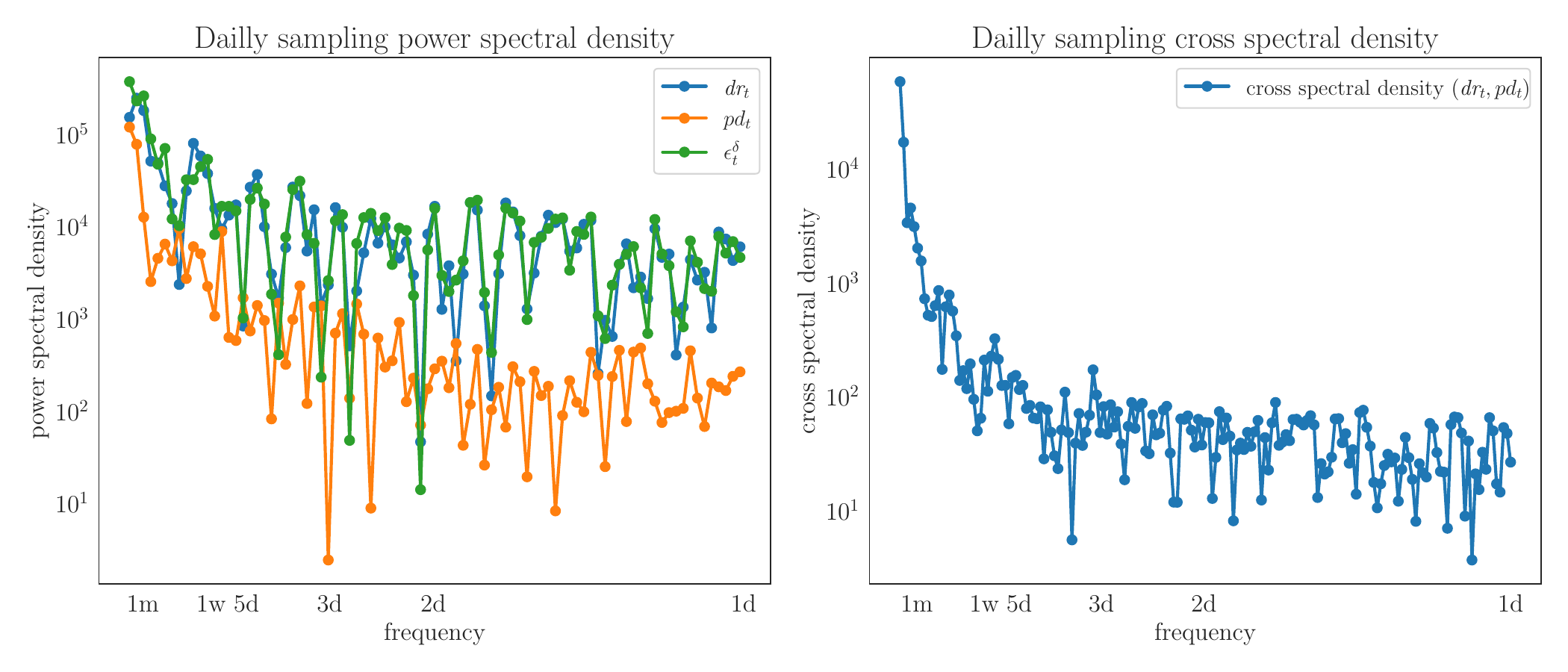}
	\end{center}
	\vspace{-0.25in}
	\caption[fig: spectral analysis daily]{{Spectrum and Cross-spectrum of $dr$ and $pd$ (Daily Frequency).}

		\footnotesize
		The left panel shows the estimated spectral densities of $dr_t$, $pd_t$, and the residuals of $dr_t$ after projecting on $pd_t$ ($\epsilon ^{pr}_t$). The integral of spectral density is equal to the variance. The horizontal line starts from zero and ends at $\pi $, but is labeled with the corresponding length of a cycle. The right panel shows the cross-spectral density between $dr_t$ and $pd_t$. The integral of cross-spectral density is equal to the covariance.}
	\label{fig:spectrum_daily}
\end{figure}

\clearpage
\begin{table}[!hbp] \centering
	\caption[tab: BMA dividend growth]{Predicting One-year Dividend Growth Using Different Combinations of Scaled Dividend Prices

		\footnotesize
		This table reports results from one-year dividend growth predictive regressions of the S\&P 500 Index.
		The predictors are various combinations of scaled dividend prices, which include
		our main predictor $dr$,
		the price-dividend ratio $pd$,
		short-term (0.5- and 1-year) dividend strip price to dividend ratio ($pd^{0.5}$ and $pd^1$),
		long-term (beyond 1-year) dividend strip price to dividend ratio ($pd^{1+}$),
		Each column corresponds to one separate predictive regression.
		Data sample: 1988:01--2019:12.
	}
	\label{tab:c42 dg}
	\resizebox{\textwidth}{!}{
		\begin{tabular}{lccccccccccccc}
			\toprule
			           & {(1)}                                             & {(2)}        & {(3)}         & {(4)}         & {(5)}        & {(6)}         & {(7)}         & {(8)}         & {(9)}        & {(10)}        & {(11)}        & {(12)}        & {(13)}        \\\midrule
			           & \multicolumn{13}{c}{\textit{$\log(D_{t+1}/D_t)$}}                                                                                                                                                                                              \\ \cline{2-14}
			$dr$       & -0.04                                             &              &               &               & 0.02         &               &               & 0.09          &              & -0.18$^{***}$ & -0.18$^{***}$ &               &               \\
			           & (0.03)                                            &              &               &               & (0.04)       &               &               & (0.06)        &              & (0.06)        & (0.06)        &               &               \\
			           &                                                   &              &               &               &              &               &               &               &              &               &               &               &               \\
			$pd$       &                                                   &              &               & 0.07          &              &               & 0.10          &               & 7.70         &               & 0.28$^{**}$   &               & 0.11$^{*}$    \\
			           &                                                   &              &               & (0.07)        &              &               & (0.07)        &               & (5.60)       &               & (0.12)        &               & (0.06)        \\
			           &                                                   &              &               &               &              &               &               &               &              &               &               &               &               \\
			$pd^{0.5}$ &                                                   & -0.02        & 0.14$^{**}$   & 0.14$^{**}$   & 0.14$^{*}$   &               &               &               &              &               &               & -0.04         & -0.04         \\
			           &                                                   & (0.03)       & (0.06)        & (0.06)        & (0.08)       &               &               &               &              &               &               & (0.03)        & (0.03)        \\
			           &                                                   &              &               &               &              &               &               &               &              &               &               &               &               \\
			$s^{1}$    &                                                   & 0.17$^{***}$ &               &               &              & 0.18$^{***}$  & 0.18$^{***}$  & 0.27$^{**}$   &              &               &               & 0.26$^{***}$  & 0.26$^{***}$  \\
			           &                                                   & (0.06)       &               &               &              & (0.06)        & (0.06)        & (0.11)        &              &               &               & (0.07)        & (0.07)        \\
			           &                                                   &              &               &               &              &               &               &               &              &               &               &               &               \\
			$pd^{1+}$  &                                                   &              & 0.07          &               &              & 0.10          &               &               & -7.48        & 0.28$^{**}$   &               & 0.11$^{*}$    &               \\
			           &                                                   &              & (0.06)        &               &              & (0.06)        &               &               & (5.41)       & (0.12)        &               & (0.06)        &               \\
			           &                                                   &              &               &               &              &               &               &               &              &               &               &               &               \\
			Intercept  & 0.21                                              & 0.07$^{***}$ & -0.09         & -0.10         & 0.09         & -0.29         & -0.30         & -0.27         & -0.96        & -0.28         & -0.31         & -0.35         & -0.36         \\
			           & (0.19)                                            & (0.02)       & (0.21)        & (0.22)        & (0.11)       & (0.24)        & (0.25)        & (0.22)        & (0.88)       & (0.23)        & (0.24)        & (0.22)        & (0.22)        \\
			           &                                                   &              &               &               &              &               &               &               &              &               &               &               &               \\
			$N$        & {372}                                             & {372}        & {372}         & {372}         & {372}        & {372}         & {372}         & {372}         & {372}        & {372}         & {372}         & {372}         & {372}         \\
			$R^{2}$    & {0.06}                                            & {0.27}       & {0.26}        & {0.26}        & {0.20}       & {0.37}        & {0.37}        & {0.36}        & {0.21}       & {0.38}        & {0.38}        & {0.41}        & {0.41}        \\\midrule
			           &                                                   &              &               &               &              &               &               &               &              &               &               &               &               \\
			           & \multicolumn{13}{c}{\textit{$\log(D_{t+1}/D_t)$}}                                                                                                                                                                                              \\ \cline{2-14}
			$dr$       & 0.10$^{*}$                                        &              & -0.26$^{***}$ & -0.25$^{***}$ &              & -0.21$^{***}$ & -0.23$^{***}$ & -0.16$^{***}$ &              & -0.23$^{***}$ & -0.24$^{***}$ & -0.25$^{***}$ & -0.24$^{***}$ \\
			           & (0.05)                                            &              & (0.07)        & (0.07)        &              & (0.05)        & (0.06)        & (0.04)        &              & (0.04)        & (0.05)        & (0.06)        & (0.07)        \\
			           &                                                   &              &               &               &              &               &               &               &              &               &               &               &               \\
			$pd$       &                                                   & 3.81         &               & 0.37$^{***}$  & 1.90         &               & 0.33$^{***}$  & 2.05          & 0.36         &               & 0.35$^{***}$  & 0.71          & 2.25          \\
			           &                                                   & (4.26)       &               & (0.11)        & (3.56)       &               & (0.12)        & (3.37)        & (2.96)       &               & (0.10)        & (2.76)        & (3.39)        \\
			           &                                                   &              &               &               &              &               &               &               &              &               &               &               &               \\
			$pd^{0.5}$ & -0.04                                             & 0.10$^{**}$  & -0.03         & -0.04         &              &               &               &               & -0.04        & -0.04         & -0.04         & -0.04         &               \\
			           & (0.03)                                            & (0.04)       & (0.02)        & (0.02)        &              &               &               &               & (0.03)       & (0.03)        & (0.03)        & (0.03)        &               \\
			           &                                                   &              &               &               &              &               &               &               &              &               &               &               &               \\
			$s^{1}$    & 0.35$^{***}$                                      &              &               &               & 0.16$^{***}$ & -0.02         & -0.05$^{**}$  &               & 0.25$^{***}$ & 0.03          & 0.02          &               & -0.08         \\
			           & (0.11)                                            &              &               &               & (0.04)       & (0.05)        & (0.02)        &               & (0.06)       & (0.05)        & (0.05)        &               & (0.05)        \\
			           &                                                   &              &               &               &              &               &               &               &              &               &               &               &               \\
			$pd^{1+}$  &                                                   & -3.65        & 0.36$^{***}$  &               & -1.76        & 0.30$^{***}$  &               & -1.74         & -0.25        & 0.34$^{***}$  &               & -0.34         & -1.87         \\
			           &                                                   & (4.11)       & (0.11)        &               & (3.43)       & (0.10)        &               & (3.25)        & (2.86)       & (0.09)        &               & (2.66)        & (3.22)        \\
			           &                                                   &              &               &               &              &               &               &               &              &               &               &               &               \\
			Intercept  & -0.34$^{*}$                                       & -0.54        & -0.33$^{*}$   & -0.37$^{*}$   & -0.49        & -0.28         & -0.31         & -0.50         & -0.39        & -0.34         & -0.37$^{*}$   & -0.41         & -0.52         \\
			           & (0.20)                                            & (0.68)       & (0.20)        & (0.21)        & (0.58)       & (0.24)        & (0.24)        & (0.55)        & (0.48)       & (0.21)        & (0.21)        & (0.45)        & (0.54)        \\
			           &                                                   &              &               &               &              &               &               &               &              &               &               &               &               \\
			$N$        & {372}                                             & {372}        & {372}         & {372}         & {372}        & {372}         & {372}         & {372}         & {372}        & {372}         & {372}         & {372}         & {372}         \\
			$R^{2}$    & {0.40}                                            & {0.29}       & {0.42}        & {0.43}        & {0.38}       & {0.38}        & {0.38}        & {0.39}        & {0.41}       & {0.42}        & {0.42}        & {0.42}        & {0.39}        \\
			\bottomrule
		\end{tabular}
	}
\end{table}

\clearpage
\begin{table}[!t] \centering
	\caption[tab: BMA returns]{Predicting One-year Returns Using Different Combinations of Scaled Dividend Prices

		\footnotesize
		This table reports results from one-year return predictive regressions of the S\&P 500 Index.
		The predictors are our main predictor $dr$, and various combinations of scaled dividend prices, which include
		the price-dividend ratio $pd$,
		short-term (0.5- and 1-year) dividend strip price to dividend ratio ($pd^{0.5}$ and $pd^1$),
		long-term (beyond 1-year) dividend strip price to dividend ratio ($pd^{1+}$),
		Each column corresponds to one separate predictive regression.
		Data sample: 1988:01--2019:12.
	}
	\label{tab:c42 ret}
	\footnotesize
	\resizebox{\textwidth}{!}{
		\begin{tabular}{lccccccccccccc} \toprule
			           & {(1)}                                   & {(2)}        & {(3)}        & {(4)}        & {(5)}        & {(6)}       & {(7)}       & {(8)}  & {(9)}       & {(10)}      & {(11)}      & {(12)}       & {(13)}       \\ \midrule
			           & \multicolumn{13}{c}{\textit{$r_{t+1}$}}                                                                                                                                                                          \\\cline{2-14}
			$dr$       & -0.16$^{***}$                           &              &              &              &              &             &             &        &             &             &             &              &              \\
			           & (0.04)                                  &              &              &              &              &             &             &        &             &             &             &              &              \\
			           &                                         &              &              &              &              &             &             &        &             &             &             &              &              \\
			$pd$       &                                         & -0.20$^{**}$ &              &              & -0.14$^{*}$  &             & -0.09       & 7.54   &             & -0.07       & 1.44        & -1.41        & -5.40        \\
			           &                                         & (0.10)       &              &              & (0.08)       &             & (0.10)      & (8.81) &             & (0.11)      & (6.81)      & (7.65)       & (7.44)       \\
			           &                                         &              &              &              &              &             &             &        &             &             &             &              &              \\
			$pd^{0.5}$ &                                         &              & -0.14$^{**}$ & 0.18$^{**}$  & 0.18$^{**}$  &             &             &        & -0.13$^{*}$ & -0.13$^{*}$ & 0.16$^{**}$ &              & -0.12        \\
			           &                                         &              & (0.07)       & (0.09)       & (0.09)       &             &             &        & (0.07)      & (0.07)      & (0.08)      &              & (0.08)       \\
			           &                                         &              &              &              &              &             &             &        &             &             &             &              &              \\
			$s^{1}$    &                                         &              & 0.49$^{***}$ &              &              & 0.23$^{**}$ & 0.23$^{**}$ &        & 0.43$^{**}$ & 0.43$^{**}$ &             & 0.24$^{***}$ & 0.50$^{***}$ \\
			           &                                         &              & (0.12)       &              &              & (0.10)      & (0.10)      &        & (0.18)      & (0.18)      &             & (0.06)       & (0.15)       \\
			           &                                         &              &              &              &              &             &             &        &             &             &             &              &              \\
			$pd^{1+}$  &                                         &              &              & -0.13$^{*}$  &              & -0.09       &             & -7.53  & -0.07       &             & -1.54       & 1.29         & 5.22         \\
			           &                                         &              &              & (0.07)       &              & (0.09)      &             & (8.55) & (0.11)      &             & (6.59)      & (7.41)       & (7.19)       \\
			           &                                         &              &              &              &              &             &             &        &             &             &             &              &              \\
			Intercept  & 0.73$^{***}$                            & 0.87$^{**}$  & 0.05         & 0.76$^{***}$ & 0.77$^{***}$ & 0.47        & 0.48        & -0.11  & 0.31        & 0.32        & 0.58        & 0.62         & 0.89         \\
			           & (0.13)                                  & (0.36)       & (0.05)       & (0.26)       & (0.27)       & (0.36)      & (0.38)      & (1.24) & (0.44)      & (0.46)      & (1.01)      & (1.15)       & (1.20)       \\
			           &                                         &              &              &              &              &             &             &        &             &             &             &              &              \\ \midrule
			$N$        & {372}                                   & {372}        & {372}        & {372}        & {372}        & {372}       & {372}       & {372}  & {372}       & {372}       & {372}       & {372}        & {372}        \\
			$R^{2}$    & {0.25}                                  & {0.14}       & {0.31}       & {0.23}       & {0.23}       & {0.26}      & {0.26}      & {0.18} & {0.32}      & {0.32}      & {0.22}      & {0.26}       & {0.33}       \\
			\bottomrule
		\end{tabular}
	}
\end{table}

\clearpage
\begin{table}[!h]
	\caption[table: R2 1-year macro variable prediction]{Forecasting Macroeconomic Variables with Market Duration and Valuation Ratios

		\footnotesize This table reports the $R^2$ of predicting one-year-ahead macroeconomic variables using predictors ($dr$, $pd$, and $s^{1}$) that contain information about the underlying state variables.
		The macroeconomic variables are divided into four categories. 1) Macroeconomic: nominal GDP Growth, Industrial Production Growth (``IP Growth''), Chicago Fed National Activity Index (``CFNAI"), Unemployment Rate, Real Consumption Growth, Total Business Inventories, Nonresidential Fixed Investment (nominal), Residential Fixed Investment (nominal), and GDP Deflator are all from FRED database. 2) Financial: Term Spread and Default Spread (``Baa-Aaa") are from FRED;  Gilchrist-Zakrajšek credit spread (GZ Credit Spread) is from \citet{GilchristZakrajsek2012};
		CAPE is the cyclically adjusted price-earnings ratio from Robert Shiller's website;
		cay is from \citet{LettauLudvigson2001}. 3) Intermediary: Broker/Dealer leverage (``B/D Leverage") is from \citet*{AdrianEtulaMuir2014}; Broker/Dealer 1(5) year average CDS spreads (``B/D 1(5) Year Avg. CDS") is from \citet{GilchristZakrajsek2012}; ROA of banks (``ROA Banks") is from FRED. 4) Uncertainties: CBOE 1-month VIX index (``VIX") and \citet{ChauvetPiger2008}'s smoothed U.S. recession probabilities estimates for given month (``CP Recession") are from FRED; Economics policy uncertainties (``EPU") is from \citet*{BakerBloomDavis2016}; Survey of Professional Forecasters recession probability estimates (``SPF Recession") is from the Philadelphia Fed. 5) Sentiments: Sentiment Index (both raw and orthogonalized against several macro variables), Number of IPOs (``IPO \#") and close-end fund NAV discount (``Close-end Discount") are all from \citet{BakeWurgler2006}.}
	\centering
	\footnotesize
	\resizebox{.72\textwidth}{!}{
		\begin{tabular*}{0.75\textwidth}{l@{\extracolsep{\fill}}lllll}
			\toprule
			{}                                  & $dr + pd$ & $dr$ & $pd$  & $dr + s^{1}$ & $s^{1}$ \\
			\midrule
			\multicolumn{6}{@{} l}{Macroeconomic:}\\
			GDP Growth                          & 0.222         & 0.061    & 0.000 & 0.224             & 0.178    \\
			IP Growth                           & 0.202         & 0.062    & 0.001 & 0.203             & 0.167    \\
			Unemployment Growth                 & 0.335         & 0.062    & 0.001 & 0.325             & 0.229    \\
			Real Consumption Growth             & 0.241         & 0.019    & 0.121 & 0.241             & 0.208    \\
			Business Inventories Growth         & 0.383         & 0.114    & 0.000 & 0.377             & 0.304    \\
			Nonres. Fixed Investment Growth     & 0.366         & 0.055    & 0.005 & 0.361             & 0.240    \\
			CPI Growth                          & 0.311         & 0.227    & 0.311 & 0.302             & 0.071    \\
			&               &          &       &                   &          \\
			\multicolumn{6}{@{} l}{Financial:}    \\
			Baa-Aaa                             & 0.081         & 0.044    & 0.008 & 0.081             & 0.078    \\
			GZ Credit Spread                    & 0.321         & 0.318    & 0.218 & 0.321             & 0.259    \\
			Term Spread                         & 0.122         & 0.001    & 0.039 & 0.120             & 0.023    \\
			CAPE                                & 0.474         & 0.300    & 0.465 & 0.478             & 0.062    \\
			cay                                 & 0.072         & 0.042    & 0.069 & 0.071             & 0.008    \\
			&               &          &       &                   &          \\
			\multicolumn{6}{@{} l}{Intermediary:} \\
			B/D Leverage                        & 0.313         & 0.290    & 0.296 & 0.308             & 0.151    \\
			B/D 1 Year Avg. CDS                 & 0.284         & 0.052    & 0.106 & 0.284             & 0.283    \\
			B/D 5 Year Avg. CDS                 & 0.393         & 0.020    & 0.231 & 0.393             & 0.383    \\
			ROA Banks                           & 0.403         & 0.080    & 0.275 & 0.388             & 0.002    \\
			&               &          &       &                   &          \\
			\multicolumn{6}{@{} l}{Uncertainties:}\\
			VIX                                 & 0.145         & 0.134    & 0.068 & 0.144             & 0.131    \\
			EPU                                 & 0.049         & 0.002    & 0.005 & 0.047             & 0.022    \\
			CP Recession                        & 0.063         & 0.030    & 0.004 & 0.063             & 0.059    \\
			SPF Recession                       & 0.194         & 0.017    & 0.008 & 0.191             & 0.109    \\
			&               &          &       &                   &          \\
			\multicolumn{6}{@{} l}{Sentiments:}   \\
			Sentiment Index                     & 0.131         & 0.131    & 0.099 & 0.131             & 0.094    \\
			Sentiment Index (orth.)             & 0.113         & 0.109    & 0.102 & 0.113             & 0.062    \\
			IPO \#                              & 0.144         & 0.142    & 0.093 & 0.144             & 0.120    \\
			Close-end Discount                  & 0.123         & 0.109    & 0.119 & 0.118             & 0.054    \\
			\bottomrule
		\end{tabular*}
		\label{tb:predict macro}
	}
\end{table}

\clearpage
\begin{table}[!t]
	\caption[table: IVX Wald test]{\citet*{Kostakis2015} IVX-Wald Test

		\footnotesize This table reports test results on the predictive coefficient $ \beta $ in Table (\ref{tb:uncondUS}).
		IVX-Wald is the Wald statistic from \citet*{Kostakis2015} to test $ H_0: \beta=0 $ against $ H_1: \beta\neq0 $.
		$p$-value of the IVX-Wald test is shown in the parentheses.
		The test is designed to be robust to the persistence of the predictor.
		*, **, and *** indicate significance at the 10\%, 5\%, and 1\% levels, respectively.}
	\centering
	\small
	\begin{tabular}{lcccc}
		\toprule
		          & $dr_t$         & $pd_t$  & $\mu^F_t$    & $KP_t$      \\ \midrule
		IVX-Wald  & $ 9.29^{***} $ & 1.56    & $ 2.77^{*} $ & $5.74^{**}$ \\
		$p$-value & (0.002)        & (0.212) & (0.096)      & (0.017)     \\
		\bottomrule
	\end{tabular}
	\label{tb:ivx}
\end{table}

\clearpage
\begin{table}[t]
	\caption[tab2: annual return prediction - S\&P500 excess return]{Annual Excess Return Prediction

		\footnotesize This table reports the results of predictive regression (equation (\ref{eq:unireg})). The left-hand side variable is the excess return of the S\&P 500 index in the next twelve months.
		We consider four right-hand side variables (i.e., predictors), $dr_t$, $pd_t$, filtered series for expected returns following \citet{Binsbergen2010} $\mu^{F}$,
		and the single predictive factor extracted from 100 book-to-market and size portfolios from \citet{Kelly2013b} $KP$.
		The $\beta $ estimate is shown followed by \cite{NeweyWest1987} t-statistic (with 18 lags), \cite{Hodrick1992} t-statistic, the coefficient adjusted for \cite{Stambaugh1999} bias, and the in-sample adjusted $R^2$. We run the regression monthly.
		Starting from December 1997, we form out-of-sample forecasts of return in the next twelve months by estimating the regression with data up to the current month and use the forecasts to calculate out-of-sample $R^2$, ENC test (\citealp{ClarkMcCracken2001}), and the p-value of CW test (\citealp{ClarkWest2007}).
		Data sample: 1988:01--2019:12.
	}
	\centering
	\footnotesize
	\begin{tabular}{p{6cm}p{1.5cm}p{1.5cm}p{1.5cm}p{1.5cm}p{1.5cm}}
		\toprule
		                                                              & \multicolumn{5}{c}{$r^{e}_{t+1}$}                                           \\\cmidrule{2-6}
		{}                                                            & (1)                               & (2)      & (3)     & (4)     & (5)      \\
		\midrule
		$ dr_t$                                                       & -0.146                            &          &         &         & -0.228   \\
		\quad \quad \textit{\small Hodrick t}                         & [-3.178]                          &          &         &         & [-2.945] \\
		\quad \quad \textit{\small  Newey-West t}                     & (-3.867)                          &          &         &         & (-3.571) \\
		\quad \quad \textit{\small  Stambaugh bias adjusted $\beta $} & -0.136                            &          &         &         &          \\
		$pd_t$                                                        &                                   & -0.180   &         &         & 0.161    \\
		                                                              &                                   & [-2.168] &         &         & [1.820]  \\
		                                                              &                                   & (-2.262) &         &         & (1.286)  \\
		                                                              &                                   & -0.170   &         &         &          \\
		$\mu^{F}_{t}$                                                 &                                   &          & 2.293   &         &          \\
		                                                              &                                   &          & [2.033] &         &          \\
		                                                              &                                   &          & (2.205) &         &          \\
		                                                              &                                   &          & 2.303   &         &          \\
		$KP_t$                                                        &                                   &          &         & 0.827   &          \\
		                                                              &                                   &          &         & [2.715] &          \\
		                                                              &                                   &          &         & (2.429) &          \\
		                                                              &                                   &          &         & 0.837   &          \\\addlinespace[2ex]
		$N$                                                           & 384                               & 384      & 384     & 384     & 384      \\
		$R^2$                                                         & 0.219                             & 0.114    & 0.124   & 0.128   & 0.241    \\
		OOS $R^2$                                                     & 0.098                             & -0.040   & -0.096  & 0.005   & 0.138    \\
		ENC                                                           & 1.924                             & 0.296    & 0.021   & 2.175   & 4.539    \\
		$p(ENC)$                                                      & $<$0.10                           & $>$0.10  & $>$0.10 & $<$0.05 & $<$0.05  \\
		$p(CW)$                                                       & 0.058                             & 0.379    & 0.493   & 0.072   & 0.028    \\
		\bottomrule
	\end{tabular}
	\label{tb:uncondUS excess}
\end{table}

\clearpage
\begin{table}[t]
	\caption[tab2: annual return prediction - MKT return]{Annual Return Prediction: Fama-French Market Return

		\footnotesize This table reports the results of predictive regression (equation (\ref{eq:unireg})).
		The left-hand side variable is the market return in the next twelve months from Fama-French.
		We consider four right-hand side variables (i.e., predictors), $dr_t$, $pd_t$, filtered series for expected returns following \citet{Binsbergen2010} $\mu^{F}$,
		and the single predictive factor extracted from 100 book-to-market and size portfolios from \citet{Kelly2013b} $KP$.
		The $\beta $ estimate is shown followed by \cite{NeweyWest1987} t-statistic (with 18 lags), \cite{Hodrick1992} t-statistic, the coefficient adjusted for \cite{Stambaugh1999} bias, and the in-sample adjusted $R^2$. We run the regression monthly.
		Starting from December 1997, we form out-of-sample forecasts of return in the next twelve months by estimating the regression with data up to the current month and use the forecasts to calculate out-of-sample $R^2$, ENC test (\citealp{ClarkMcCracken2001}), and the p-value of CW test (\citealp{ClarkWest2007}).
		Data sample: 1988:01--2019:12.}
	\centering
	\footnotesize
	\begin{tabular}{p{6cm}p{1.5cm}p{1.5cm}p{1.5cm}p{1.5cm}p{1.5cm}}
		\toprule
		                                                              & \multicolumn{5}{c}{$r^{MKT}_{t+1}$}                                           \\\cmidrule{2-6}
		{}                                                            & (1)                                 & (2)      & (3)     & (4)     & (5)      \\
		\midrule
		$ dr_t$                                                       & -0.154                              &          &         &         & -0.222   \\
		\quad \quad \textit{\small Hodrick t}                         & [-3.233]                            &          &         &         & [-2.772] \\
		\quad \quad \textit{\small  Newey-West t}                     & (-4.464)                            &          &         &         & (-3.511) \\
		\quad \quad \textit{\small  Stambaugh bias adjusted $\beta $} & -0.144                              &          &         &         &          \\
		$pd_t$                                                        &                                     & -0.198   &         &         & 0.133    \\
		                                                              &                                     & [-2.302] &         &         & [1.608]  \\
		                                                              &                                     & (-2.706) &         &         & (1.129)  \\
		                                                              &                                     & -0.188   &         &         &          \\
		$\mu^{F}_{t}$                                                 &                                     &          & 2.486   &         &          \\
		                                                              &                                     &          & [2.327] &         &          \\
		                                                              &                                     &          & (2.656) &         &          \\
		                                                              &                                     &          & 2.496   &         &          \\
		$KP_t$                                                        &                                     &          &         & 0.794   &          \\
		                                                              &                                     &          &         & [2.223] &          \\
		                                                              &                                     &          &         & (2.689) &          \\
		                                                              &                                     &          &         & 0.805   &          \\\addlinespace[2ex]
		$N$                                                           & 384                                 & 384      & 384     & 384     & 384      \\
		$R^2$                                                         & 0.236                               & 0.134    & 0.141   & 0.128   & 0.251    \\
		OOS $R^2$                                                     & 0.144                               & 0.022    & -0.023  & -0.001  & 0.181    \\
		ENC                                                           & 3.083                               & 0.963    & 0.598   & 2.483   & 6.163    \\
		$p(ENC)$                                                      & $<$0.05                             & $>$0.10  & $>$0.10 & $<$0.05 & $<$0.01  \\
		$p(CW)$                                                       & 0.017                               & 0.166    & 0.321   & 0.048   & 0.019    \\
		\bottomrule
	\end{tabular}
	\label{tb:uncondUS mkt}
\end{table}

\clearpage
\begin{table}[t]
	\caption[tab2: annual return prediction - MKTRF return]{Annual Return Prediction: Fama-French Market Excess Return

		\footnotesize This table reports the results of predictive regression (equation (\ref{eq:unireg})).
		The left-hand side variable is the market excess return in the next twelve months from Fama-French.
		We consider four right-hand side variables (i.e., predictors), $dr_t$, $pd_t$, filtered series for expected returns following \citet{Binsbergen2010} $\mu^{F}$,
		and the single predictive factor extracted from 100 book-to-market and size portfolios from \citet{Kelly2013b} $KP$.
		The $\beta $ estimate is shown followed by \cite{NeweyWest1987} t-statistic (with 18 lags), \cite{Hodrick1992} t-statistic, the coefficient adjusted for \cite{Stambaugh1999} bias, and the in-sample adjusted $R^2$. We run the regression monthly.
		Starting from December 1997, we form out-of-sample forecasts of return in the next twelve months by estimating the regression with data up to the current month and use the forecasts to calculate out-of-sample $R^2$, ENC test (\citealp{ClarkMcCracken2001}), and the p-value of CW test (\citealp{ClarkWest2007}).
		Data sample: 1988:01--2019:12.}
	\centering
	\footnotesize
	\begin{tabular}{p{6cm}p{1.5cm}p{1.5cm}p{1.5cm}p{1.5cm}p{1.5cm}}
		\toprule
		                                                              & \multicolumn{5}{c}{$r^{MKT, e}_{t+1}$}                                           \\\cmidrule{2-6}
		{}                                                            & (1)                                    & (2)      & (3)     & (4)     & (5)      \\
		\midrule
		$ dr_t$                                                       & -0.144                                 &          &         &         & -0.222   \\
		\quad \quad \textit{\small Hodrick t}                         & [-3.060]                               &          &         &         & [-2.791] \\
		\quad \quad \textit{\small  Newey-West t}                     & (-3.745)                               &          &         &         & (-3.503) \\
		\quad \quad \textit{\small  Stambaugh bias adjusted $\beta $} & -0.134                                 &          &         &         &          \\
		$pd_t$                                                        &                                        & -0.179   &         &         & 0.153    \\
		                                                              &                                        & [-2.108] &         &         & [1.704]  \\
		                                                              &                                        & (-2.199) &         &         & (1.202)  \\
		                                                              &                                        & -0.169   &         &         &          \\
		$\mu^{F}_{t}$                                                 &                                        &          & 2.192   &         &          \\
		                                                              &                                        &          & [2.075] &         &          \\
		                                                              &                                        &          & (2.057) &         &          \\
		                                                              &                                        &          & 2.203   &         &          \\
		$KP_t$                                                        &                                        &          &         & 0.725   &          \\
		                                                              &                                        &          &         & [2.044] &          \\
		                                                              &                                        &          &         & (2.251) &          \\
		                                                              &                                        &          &         & 0.735   &          \\\addlinespace[2ex]
		$N$                                                           & 384                                    & 384      & 384     & 384     & 384      \\
		$R^2$                                                         & 0.206                                  & 0.108    & 0.109   & 0.105   & 0.225    \\
		OOS $R^2$                                                     & 0.099                                  & -0.018   & -0.081  & -0.037  & 0.140    \\
		ENC                                                           & 2.000                                  & 0.376    & -0.047  & 1.700   & 4.656    \\
		$p(ENC)$                                                      & $<$0.10                                & $>$0.10  & >0.10   & <0.10   & $<$0.05  \\
		$p(CW)$                                                       & 0.047                                  & 0.349    & 0.485   & 0.120   & 0.027    \\
		\bottomrule
	\end{tabular}
	\label{tb:uncondUS mktrf}
\end{table}

\clearpage
\begin{figure}[!t]
	\centering
	\includegraphics[scale=0.47]{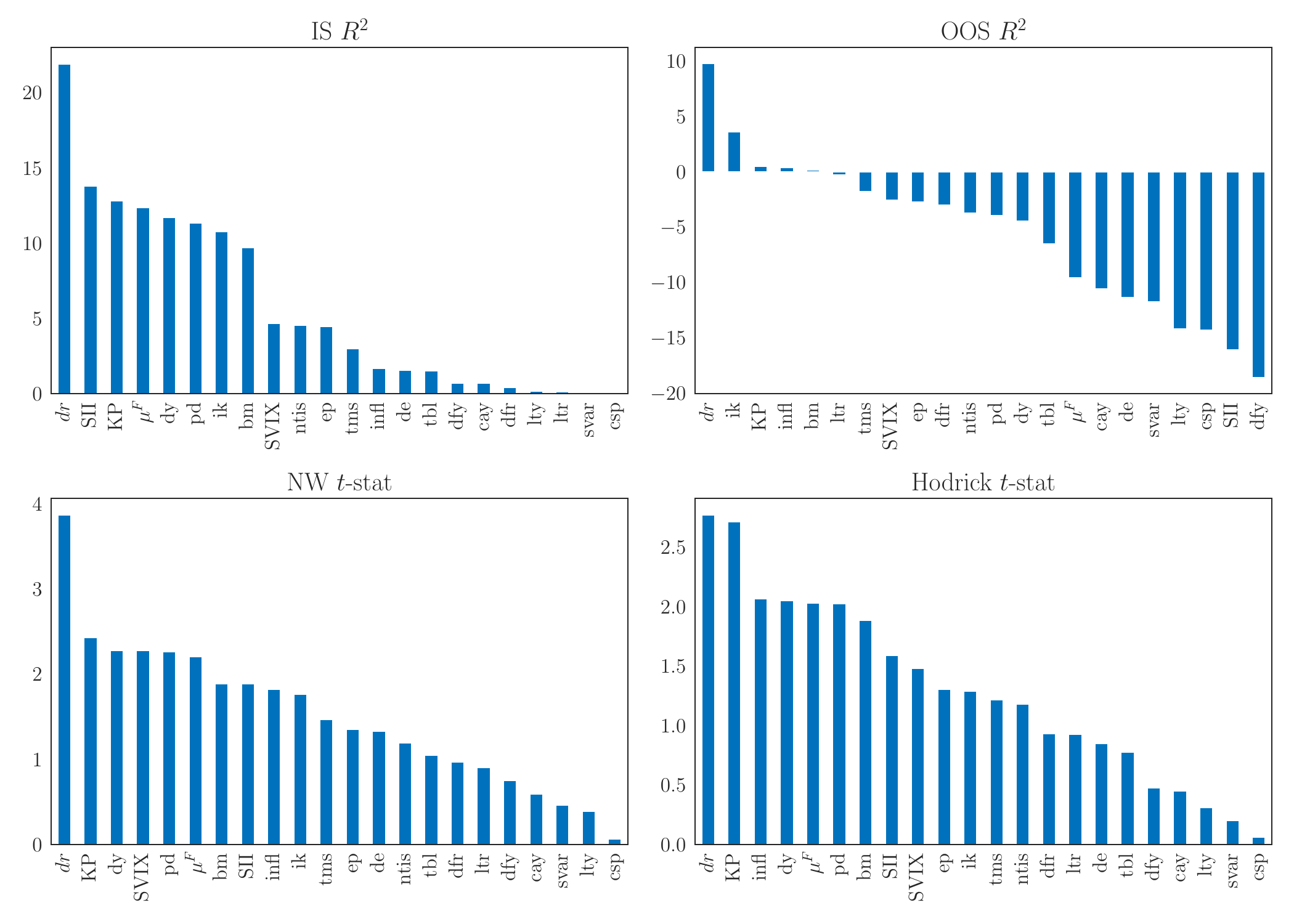}
	\caption[figure comparison: s\&p excess return]{{Comparison with Alternative Return Predictors: Excess Return.}

		\footnotesize
		This graph compares the 1-year return predictive power between $dr_t$ and other commonly studied predictors in our sample period. Panel A reports the in-sample adjusted $R^2$. Panel B reports the out-of-sample $R^2$. Negative out-of-sample $R^2$ indicates that the predictive power is below the historical mean. Panel C reports the absolute values of \cite{NeweyWest1987} t-statistic (with an 18-month lag). Panel D reports the absolute values of \cite{Hodrick1992} t-statistic. Most predictors are from \citet{WelchGoyal2007} and include the price-dividend ratio (pd), the default yield spread (dfy), the inflation rate (infl), stock variance (svar), the cross-section premium (csp), the dividend payout ratio (de), the long-term yield (lty), the term spread (tms), the T-bill rate (tbl), the default return spread (dfr), the dividend yield (dy), the long-term rate of return (ltr), the earnings-to-price ratio (ep), the book to market ratio (bm), the investment-to-capital ratio (ik), the net equity expansion ratio (ntis), the percent equity issuing ratio (eqis), and the consumption-wealth-income ratio (cay). SII is the short interests index from \citet*{RapachRinggenbergZhou2016} (1988-2014). SVIX is an option-implied lower bound of the 1-year equity premium from \cite{Martin2017} (1996-2012). KP is the single predictive factor extracted from 100 book-to-market and size portfolios from \citet{Kelly2013b}. BK is the filtered series for expected returns following \citet{Binsbergen2010}.}
	\label{tb:comp_exc}
\end{figure}

\clearpage
\begin{figure}[!t]
	\centering
	\includegraphics[scale=0.47]{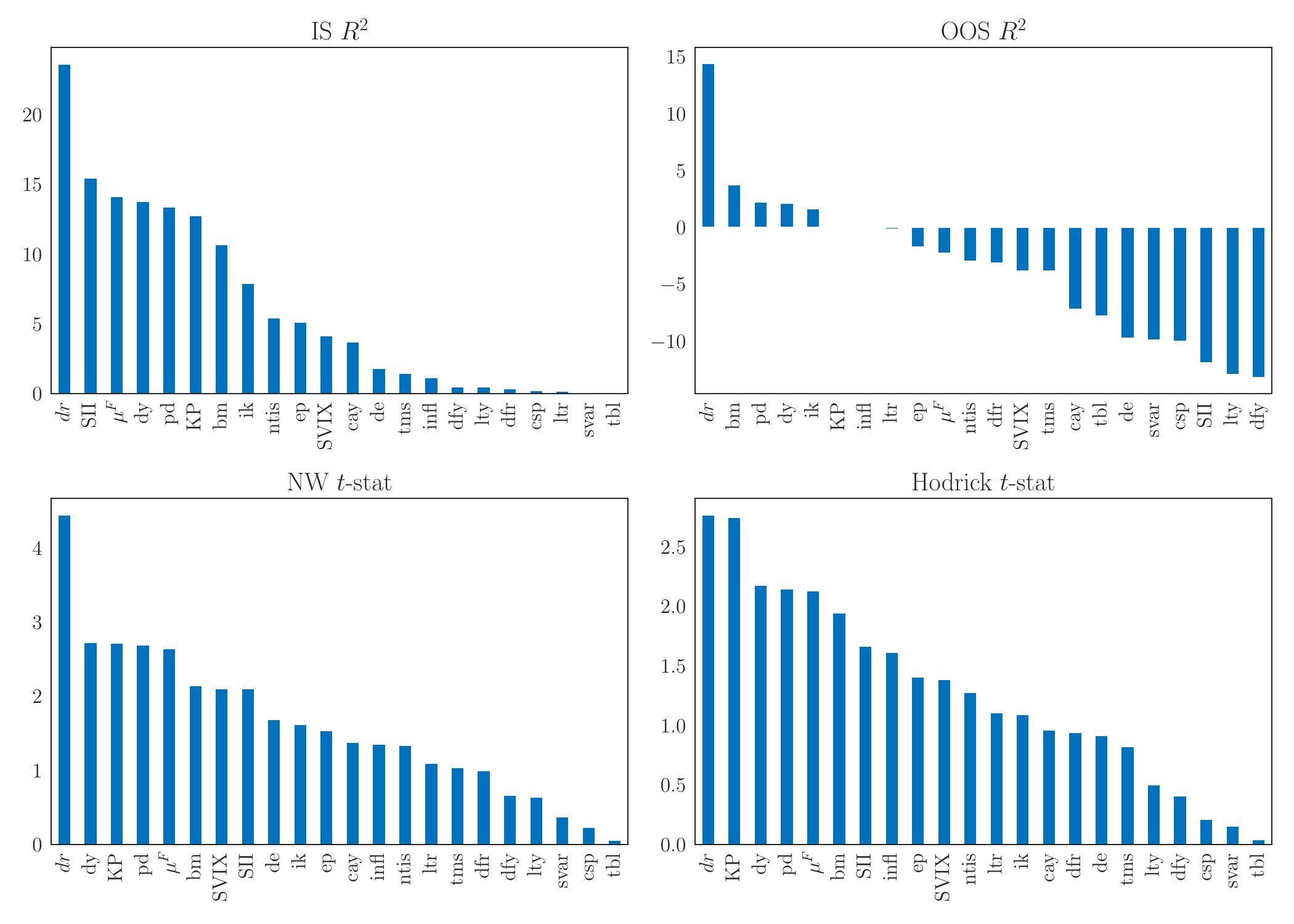}
	\caption[figure comparison: MKT return]{{Comparison with Alternative Return Predictors: Fama-French Market Return.}

		\footnotesize
		This graph compares the 1-year return predictive power between $dr_t$ and other commonly studied predictors in our sample period.
		Panel A reports the in-sample adjusted $R^2$. Panel B reports the out-of-sample $R^2$. Negative out-of-sample $R^2$ indicates that the predictive power is below the historical mean. Panel C reports the absolute values of \cite{NeweyWest1987} t-statistic (with an 18-month lag). Panel D reports the absolute values of \cite{Hodrick1992} t-statistic. Most predictors are from \citet{WelchGoyal2007} and include the price-dividend ratio (pd), the default yield spread (dfy), the inflation rate (infl), stock variance (svar), the cross-section premium (csp), the dividend payout ratio (de), the long-term yield (lty), the term spread (tms), the T-bill rate (tbl), the default return spread (dfr), the dividend yield (dy), the long-term rate of return (ltr), the earnings-to-price ratio (ep), the book to market ratio (bm), the investment-to-capital ratio (ik), the net equity expansion ratio (ntis), the percent equity issuing ratio (eqis), and the consumption-wealth-income ratio (cay). SII is the short interests index from \citet*{RapachRinggenbergZhou2016} (1988-2014). SVIX is an option-implied lower bound of the 1-year equity premium from \cite{Martin2017} (1996-2012). KP is the single predictive factor extracted from 100 book-to-market and size portfolios from \citet{Kelly2013b}. BK is the filtered series for expected returns following \citet{Binsbergen2010}.}
	\label{tb:comp_mkt}
\end{figure}

\clearpage
\begin{figure}[!t]
	\centering
	\includegraphics[scale=0.47]{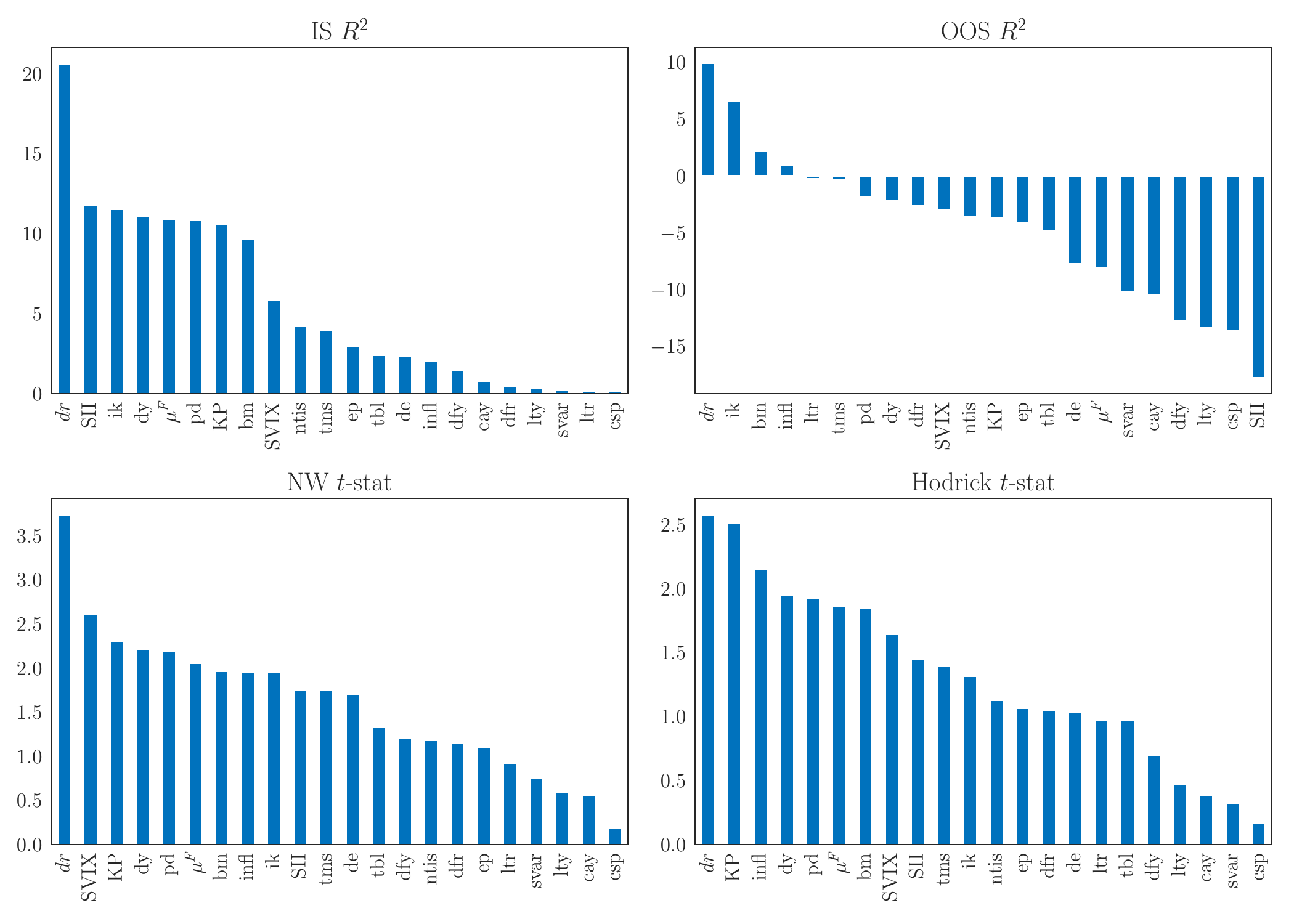}
	\caption[figure comparison: MKT excess return]{{Comparison with Alternative Return Predictors: Fama-French Market Excess Return.}

		\footnotesize
		This graph compares the 1-year return predictive power between $dr_t$ and other commonly studied predictors in our sample period. Panel A reports the in-sample adjusted $R^2$. Panel B reports the out-of-sample $R^2$. Negative out-of-sample $R^2$ indicates that the predictive power is below the historical mean. Panel C reports the absolute values of \cite{NeweyWest1987} t-statistic (with an 18-month lag). Panel D reports the absolute values of \cite{Hodrick1992} t-statistic. Most predictors are from \citet{WelchGoyal2007} and include the price-dividend ratio (pd), the default yield spread (dfy), the inflation rate (infl), stock variance (svar), the cross-section premium (csp), the dividend payout ratio (de), the long-term yield (lty), the term spread (tms), the T-bill rate (tbl), the default return spread (dfr), the dividend yield (dy), the long-term rate of return (ltr), the earnings-to-price ratio (ep), the book to market ratio (bm), the investment-to-capital ratio (ik), the net equity expansion ratio (ntis), the percent equity issuing ratio (eqis), and the consumption-wealth-income ratio (cay). SII is the short interests index from \citet*{RapachRinggenbergZhou2016} (1988-2014). SVIX is an option-implied lower bound of the 1-year equity premium from \cite{Martin2017} (1996-2012). KP is the single predictive factor extracted from 100 book-to-market and size portfolios from \citet{Kelly2013b}. BK is the filtered series for expected returns following \citet{Binsbergen2010}.}
	\label{tb:comp_mkt_exc}
\end{figure}

\clearpage
\begin{figure}[!t]
	\centering
	\includegraphics[scale=0.47]{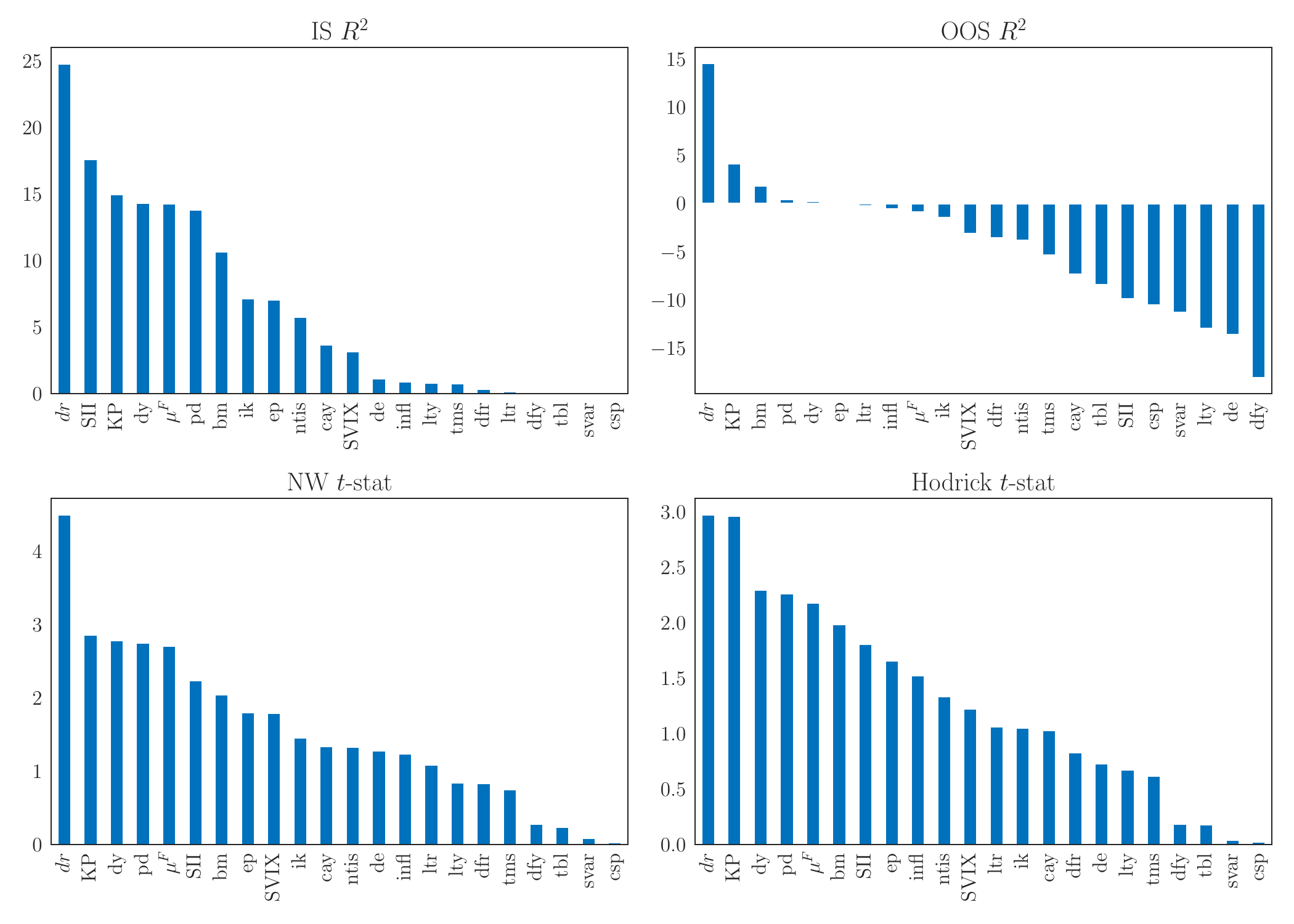}
	\caption[figure comparison: s\&p return - using BK2010 parameters]{{Comparison with Alternative Return Predictors.}

		\footnotesize
		This graph compares the 1-year return predictive power between $dr_t$ and other commonly studied predictors in our sample period.
		Panel A reports the in-sample adjusted $R^2$. Panel B reports the out-of-sample $R^2$. Negative out-of-sample $R^2$ indicates that the predictive power is below the historical mean. Panel C reports the absolute values of \cite{NeweyWest1987} t-statistic (with an 18-month lag). Panel D reports the absolute values of \cite{Hodrick1992} t-statistic. Most predictors are from \citet{WelchGoyal2007} and include the price-dividend ratio (pd), the default yield spread (dfy), the inflation rate (infl), stock variance (svar), the cross-section premium (csp), the dividend payout ratio (de), the long-term yield (lty), the term spread (tms), the T-bill rate (tbl), the default return spread (dfr), the dividend yield (dy), the long-term rate of return (ltr), the earnings-to-price ratio (ep), the book to market ratio (bm), the investment-to-capital ratio (ik), the net equity expansion ratio (ntis), the percent equity issuing ratio (eqis), and the consumption-wealth-income ratio (cay). SII is the short interests index from \citet*{RapachRinggenbergZhou2016} (1988-2014). SVIX is an option-implied lower bound of the 1-year equity premium from \cite{Martin2017} (1996-2012). KP is the single predictive factor extracted from 100 book-to-market and size portfolios from \citet{Kelly2013b}. BK is the filtered series for expected returns following \citet{Binsbergen2010}.
	}
	\label{tb:comp_KP_original}
\end{figure}

\clearpage
\begin{table}[t]
	\caption[tab2b: monthly return prediction - S\&P500 return]{One-month Return Prediction

		\footnotesize This table reports the results of predictive regression (equation (\ref{eq:unireg})). The left-hand side variable is the return of the S\&P 500 index in the next month.
		We consider four right-hand side variables (i.e., predictors), $dr_t$, $pd_t$, filtered series for expected returns following \citet{Binsbergen2010} $\mu^{F}$,
		and the single predictive factor extracted from 100 book-to-market and size portfolios from \citet{Kelly2013b} $KP$.
		The $\beta $ estimate is reported, followed by \cite{Hodrick1992} t-statistic, OLS t-statistic, the coefficient adjusted for \cite{Stambaugh1999} bias, and the in-sample adjusted $R^2$. We run the regression monthly.
		Starting from December 1997, we form out-of-sample forecasts of return in the next twelve months by estimating the regression with data up to the current month and use the forecasts to calculate out-of-sample $R^2$, ENC test (\citealp{ClarkMcCracken2001}), and the p-value of CW test (\citealp{ClarkWest2007}).
		Data sample: 1988:01--2019:12}
	\centering
	\footnotesize
	\begin{tabular}{p{4cm}p{1.5cm}p{1.5cm}p{1.5cm}p{1.5cm}p{1.5cm}}
		\toprule
		                                         & \multicolumn{5}{c}{$r_{t+1/12}$}                                           \\\cmidrule{2-6}
		{}                                       & (1)                              & (2)      & (3)     & (4)     & (5)      \\
		\midrule
		Intercept                                & 0.056                            & 0.067    & 0.017   & 0.018   & 0.036    \\
		                                         & [3.031]                          & [2.163]  & [4.198] & [1.203] & [0.893]  \\
		                                         & (3.302)                          & (2.385)  & (4.158) & (1.251) & (1.130)  \\
		$ dr_t$                                  & -0.012                           &          &         &         & -0.017   \\
		\quad \quad \textit{\small Hodrick t}    & [-2.529]                         &          &         &         & [-1.427] \\
		\quad \quad \textit{\small Newey-West t} & (-2.826)                         &          &         &         & (-2.034) \\
		$pd_t$                                   &                                  & -0.015   &         &         & 0.011    \\
		                                         &                                  & [-1.891] &         &         & [0.530]  \\
		                                         &                                  & (-2.090) &         &         & (0.751)  \\
		$\mu^{F}_{t}$                            &                                  &          & 0.211   &         &          \\
		                                         &                                  &          & [2.224] &         &          \\
		                                         &                                  &          & (2.401) &         &          \\
		$KP_t$                                   &                                  &          &         & 0.019   &          \\
		                                         &                                  &          &         & [0.656] &          \\
		                                         &                                  &          &         & (0.680) &          \\\addlinespace[1ex]
		$R^2$                                    & 0.021                            & 0.011    & 0.015   & 0.001   & 0.022    \\
		OOS $R^2$                                & 0.015                            & 0.004    & 0.007   & -0.012  & 0.005    \\
		ENC                                      & 2.678                            & 1.122    & 1.673   & -0.676  & 2.384    \\
		$p(ENC)$                                 & $<$0.05                          & $>$0.10  & $<$0.10 & $>$0.10 & $<$0.10  \\
		$p(CW)$                                  & 0.018                            & 0.179    & 0.122   & 0.325   & 0.129    \\
		\bottomrule
	\end{tabular}
	\label{tb:uncondUS monthly sp}
\end{table}

\clearpage
\begin{table}[t]
	\caption[tab2b: monthly return prediction - S\&P500 excess return]{One-month Excess Return Prediction

		\footnotesize This table reports the results of predictive regression (equation (\ref{eq:unireg})). The left-hand side variable is the excess return of the S\&P 500 index in the next month.
		We consider four right-hand side variables (i.e., predictors), $dr_t$, $pd_t$, filtered series for expected returns following \citet{Binsbergen2010} $\mu^{F}$,
		and the single predictive factor extracted from 100 book-to-market and size portfolios from \citet{Kelly2013b} $KP$.
		The $\beta $ estimate is reported, followed by \cite{Hodrick1992} t-statistic, OLS t-statistic, the coefficient adjusted for \cite{Stambaugh1999} bias, and the in-sample adjusted $R^2$. We run the regression monthly.
		Starting from December 1997, we form out-of-sample forecasts of return in the next twelve months by estimating the regression with data up to the current month and use the forecasts to calculate out-of-sample $R^2$, ENC test (\citealp{ClarkMcCracken2001}), and the p-value of CW test (\citealp{ClarkWest2007}).
		Data sample: 1988:01--2019:12}
	\centering
	\footnotesize
	\begin{tabular}{p{4cm}p{1.5cm}p{1.5cm}p{1.5cm}p{1.5cm}p{1.5cm}}
		\toprule
		                                         & \multicolumn{5}{c}{$r^{e}_{t+1/12}$}                                           \\\cmidrule{2-6}
		{}                                       & (1)                                  & (2)      & (3)     & (4)     & (5)      \\
		\midrule
		Intercept                                & 0.051                                & 0.059    & 0.013   & 0.014   & 0.026    \\
		                                         & [2.757]                              & [1.878]  & [3.328] & [0.901] & [0.644]  \\
		                                         & (3.016)                              & (2.080)  & (3.316) & (0.938) & (0.814)  \\
		$ dr_t$                                  & -0.011                               &          &         &         & -0.018   \\
		\quad \quad \textit{\small Hodrick t}    & [-2.394]                             &          &         &         & [-1.504] \\
		\quad \quad \textit{\small Newey-West t} & (-2.684)                             &          &         &         & (-2.134) \\
		$pd_t$                                   &                                      & -0.014   &         &         & 0.014    \\
		                                         &                                      & [-1.687] &         &         & [0.670]  \\
		                                         &                                      & (-1.873) &         &         & (0.944)  \\
		$\mu^{F}_{t}$                            &                                      &          & 0.188   &         &          \\
		                                         &                                      &          & [1.967] &         &          \\
		                                         &                                      &          & (2.137) &         &          \\
		$KP_t$                                   &                                      &          &         & 0.015   &          \\
		                                         &                                      &          &         & [0.514] &          \\
		                                         &                                      &          &         & (0.535) &          \\\addlinespace[1ex]
		$R^2$                                    & 0.019                                & 0.009    & 0.012   & 0.001   & 0.021    \\
		OOS $R^2$                                & 0.012                                & 0.001    & 0.003   & -0.013  & 0.003    \\
		ENC                                      & 2.338                                & 0.670    & 1.060   & -0.766  & 2.233    \\
		$p(ENC)$                                 & $<$0.05                              & $>$0.10  & $>$0.10 & $>$0.10 & $<$0.10  \\
		$p(CW)$                                  & 0.038                                & 0.283    & 0.228   & 0.302   & 0.159    \\
		\bottomrule
	\end{tabular}
	\label{tb:uncondUS monthly sp-rf}
\end{table}

\clearpage
\begin{table}[t]
	\caption[tab2b: monthly return prediction - MKT return]{One-month Return Prediction: Fama-French MKT Return

		\footnotesize This table reports the results of predictive regression (equation (\ref{eq:unireg})). The left-hand side variable is the return of the S\&P 500 index in the next month.
		We consider four right-hand side variables (i.e., predictors), $dr_t$, $pd_t$, filtered series for expected returns following \citet{Binsbergen2010} $\mu^{F}$,
		and the single predictive factor extracted from 100 book-to-market and size portfolios from \citet{Kelly2013b} $KP$.
		The $\beta $ estimate is reported, followed by \cite{Hodrick1992} t-statistic, OLS t-statistic, the coefficient adjusted for \cite{Stambaugh1999} bias, and the in-sample adjusted $R^2$. We run the regression monthly.
		Starting from December 1997, we form out-of-sample forecasts of return in the next twelve months by estimating the regression with data up to the current month and use the forecasts to calculate out-of-sample $R^2$, ENC test (\citealp{ClarkMcCracken2001}), and the p-value of CW test (\citealp{ClarkWest2007}).
		Data sample: 1988:01--2019:12}
	\centering
	\footnotesize
	\begin{tabular}{p{4cm}p{1.5cm}p{1.5cm}p{1.5cm}p{1.5cm}p{1.5cm}}
		\toprule
		                                         & \multicolumn{5}{c}{$r^{MKT}_{t+1/12}$}                                           \\\cmidrule{2-6}
		{}                                       & (1)                                    & (2)      & (3)     & (4)     & (5)      \\
		\midrule
		Intercept                                & 0.056                                  & 0.067    & 0.017   & 0.018   & 0.037    \\
		                                         & [2.833]                                & [2.086]  & [4.070] & [1.117] & [0.891]  \\
		                                         & (3.207)                                & (2.332)  & (4.029) & (1.185) & (1.125)  \\
		$ dr_t$                                  & -0.012                                 &          &         &         & -0.017   \\
		\quad \quad \textit{\small Hodrick t}    & [-2.354]                               &          &         &         & [-1.330] \\
		\quad \quad \textit{\small Newey-West t} & (-2.742)                               &          &         &         & (-1.945) \\
		$pd_t$                                   &                                        & -0.015   &         &         & 0.011    \\
		                                         &                                        & [-1.819] &         &         & [0.488]  \\
		                                         &                                        & (-2.044) &         &         & (0.697)  \\
		$\mu^{F}_{t}$                            &                                        &          & 0.208   &         &          \\
		                                         &                                        &          & [2.091] &         &          \\
		                                         &                                        &          & (2.306) &         &          \\
		$KP_t$                                   &                                        &          &         & 0.018   &          \\
		                                         &                                        &          &         & [0.588] &          \\
		                                         &                                        &          &         & (0.626) &          \\\addlinespace[1ex]
		$R^2$                                    & 0.019                                  & 0.011    & 0.014   & 0.001   & 0.021    \\
		OOS $R^2$                                & 0.012                                  & 0.003    & 0.005   & -0.014  & 0.003    \\
		ENC                                      & 2.227                                  & 0.876    & 1.275   & -1.009  & 1.869    \\
		$p(ENC)$                                 & $<$0.05                                & $>$0.10  & $<$0.10 & $>$0.10 & $>$0.10  \\
		$p(CW)$                                  & 0.034                                  & 0.220    & 0.171   & 0.220   & 0.176    \\
		\bottomrule
	\end{tabular}
	\label{tb:uncondUS monthly mkt}
\end{table}

\begin{table}[t]
	\caption[tab2b: monthly return prediction - MKT excess return]{One-month Return Prediction: Fama-French MKT excess Return

		\footnotesize This table reports the results of predictive regression (equation (\ref{eq:unireg})). The left-hand side variable is the excess return of the S\&P 500 index in the next month.
		We consider four right-hand side variables (i.e., predictors), $dr_t$, $pd_t$, filtered series for expected returns following \citet{Binsbergen2010} $\mu^{F}$,
		and the single predictive factor extracted from 100 book-to-market and size portfolios from \citet{Kelly2013b} $KP$.
		The $\beta $ estimate is reported, followed by \cite{Hodrick1992} t-statistic, OLS t-statistic, the coefficient adjusted for \cite{Stambaugh1999} bias, and the in-sample adjusted $R^2$. We run the regression monthly.
		Starting from December 1997, we form out-of-sample forecasts of return in the next twelve months by estimating the regression with data up to the current month and use the forecasts to calculate out-of-sample $R^2$, ENC test (\citealp{ClarkMcCracken2001}), and the p-value of CW test (\citealp{ClarkWest2007}).
		Data sample: 1988:01--2019:12}
	\centering
	\footnotesize
	\begin{tabular}{p{4cm}p{1.5cm}p{1.5cm}p{1.5cm}p{1.5cm}p{1.5cm}}
		\toprule
		                                         & \multicolumn{5}{c}{$r^{MKT, e}_{t+1/12}$}                                           \\\cmidrule{2-6}
		{}                                       & (1)                                       & (2)      & (3)     & (4)     & (5)      \\\midrule
		Intercept                                & 0.051                                     & 0.059    & 0.013   & 0.013   & 0.027    \\
		                                         & [2.576]                                   & [1.810]  & [3.221] & [0.828] & [0.646]  \\
		                                         & (2.926)                                   & (2.034)  & (3.208) & (0.880) & (0.816)  \\
		$ dr_t$                                  & -0.011                                    &          &         &         & -0.018   \\
		\quad \quad \textit{\small Hodrick t}    & [-2.228]                                  &          &         &         & [-1.402] \\
		\quad \quad \textit{\small Newey-West t} & (-2.602)                                  &          &         &         & (-2.042) \\
		$pd_t$                                   &                                           & -0.014   &         &         & 0.013    \\
		                                         &                                           & [-1.621] &         &         & [0.623]  \\
		                                         &                                           & (-1.830) &         &         & (0.885)  \\
		$\mu^{F}_{t}$                            &                                           &          & 0.185   &         &          \\
		                                         &                                           &          & [1.844] &         &          \\
		                                         &                                           &          & (2.049) &         &          \\
		$KP_t$                                   &                                           &          &         & 0.014   &          \\
		                                         &                                           &          &         & [0.453] &          \\
		                                         &                                           &          &         & (0.484) &          \\\addlinespace[1ex]
		$R^2$                                    & 0.017                                     & 0.009    & 0.011   & 0.001   & 0.019    \\
		OOS $R^2$                                & 0.010                                     & 0.000    & 0.001   & -0.014  & 0.001    \\
		ENC                                      & 1.903                                     & 0.443    & 0.699   & -1.082  & 1.718    \\
		$p(ENC)$                                 & $<$0.10                                   & $>$0.10  & $>$0.10 & $>$0.10 & $>$0.10  \\
		$p(CW)$                                  & 0.065                                     & 0.340    & 0.297   & 0.205   & 0.209    \\\bottomrule
	\end{tabular}
	\label{tb:uncondUS monthly mkt-rf}
\end{table}

\clearpage
\begin{table}[!t]
	\caption[tab: correlation among return predictors]{Correlations with Other Return Predictors

		\footnotesize This table shows the correlations of alternative return predictors with both $dr_t$ and $pd_t$ from 1988 to 2019. $\mu^F$ is the filtered demeaned expected return following \citet{Binsbergen2010}. KP is a predictive factor extracted from 100 book-to-market and size portfolios from \citet{Kelly2013b}. Most alternative predictors are from \citet{WelchGoyal2007} that include the default yield spread (dfy), the inflation rate (infl), stock variance (svar), the cross-section premium (csp), the dividend payout ratio (de), the long-term yield (lty), the term spread (tms), the T-bill rate (tbl), the default return spread (dfr), the dividend yield (dy, log difference between current-period dividend and lagged S\&P 500 index price), the long-term rate of return (ltr), the earnings-to-price ratio (ep), the book to market ratio (bm), the investment-to-capital ratio (ik), the net equity expansion ratio (ntis), the percent equity issuing ratio (eqis), and the consumption-wealth-income ratio (cay). SII is the short interests index from \citet*{RapachRinggenbergZhou2016} (1988-2014). SVIX is an option-implied lower bound of 1-year equity premium from \cite{Martin2017} (1996-2012). ZCB1Y is the one-year zero-coupon bond yield from Fama-Bliss.}
	\centering
	\small
	\singlespacing
	\begin{tabular}{lrr}
		\toprule
		{}      & $dr$   & $pd$   \\ \midrule
		$pd$    & 0.873  & 1.000  \\
		$\mu^F$ & -0.892 & -0.967 \\
		KP      & -0.565 & -0.496 \\
		SII     & 0.047  & -0.015 \\
		SVIX    & 0.055  & -0.304 \\
		bm      & -0.778 & -0.826 \\
		cay     & -0.364 & -0.377 \\
		csp     & 0.345  & 0.428  \\
		de      & -0.245 & -0.463 \\
		dfr     & -0.024 & 0.005  \\
		dfy     & -0.078 & -0.273 \\
		$dr$    & 1.000  & 0.873  \\
		dy      & -0.879 & -0.990 \\
		ep      & -0.558 & -0.453 \\
		ik      & 0.664  & 0.657  \\
		infl    & -0.100 & -0.074 \\
		ltr     & -0.000 & -0.056 \\
		lty     & -0.368 & -0.425 \\
		ntis    & -0.074 & 0.076  \\
		svar    & 0.149  & -0.053 \\
		tbl     & -0.175 & -0.243 \\
		tms     & -0.255 & -0.217 \\
		\bottomrule
	\end{tabular}

	\label{tb:corr_predictors}

\end{table}

\clearpage
\begin{figure}[!t]
	\centering
	\includegraphics[width=\textwidth]{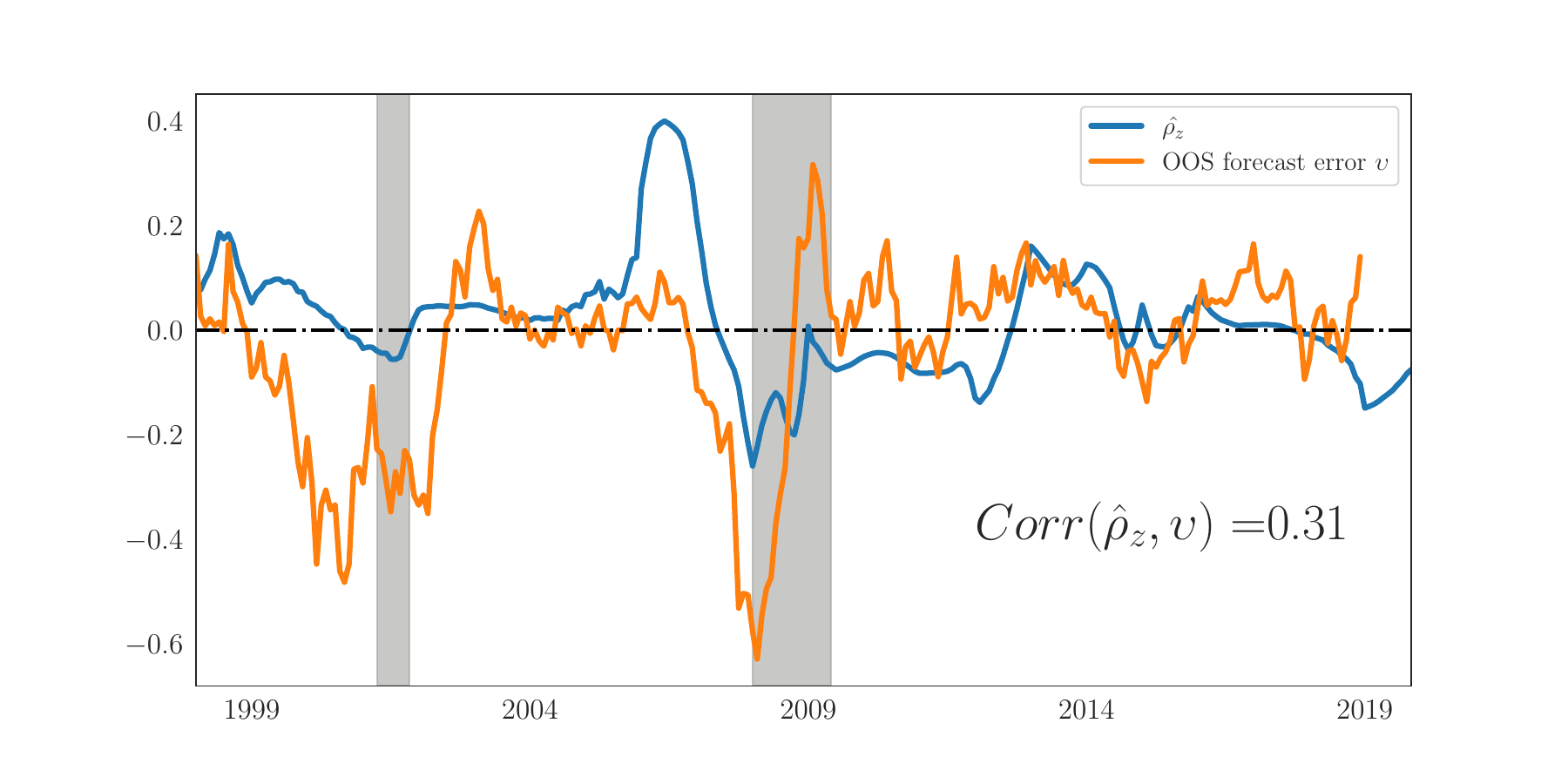}
	\caption[rhoz and prediction errors]{Rolling Estimate of Expected Growth Persistence and Return Prediction Errors: OOS

		\footnotesize
		This figure plots the rolling estimate of the autoregressive coefficient of expected cash flow growth, $ \hat{\rho}_{z,t}$, and the return prediction errors using market duration ($dr_t$) as the predictor. $\hat{\rho}_{z,t}$ is estimated using analyst forecasts of S\&P 500 aggregate earnings in rolling regressions with a three-year window. This figure also plots the out-of-sample forecast errors $\hat{\nu}_t$, which is calculated as the difference between the realized one-year S\&P 500 return and the one-period-ahead return forecast using $dr_t$ as the predictor. The first out-of-sample forecast starts in 1998. The correlation between the two time series is also reported on the graph. Our monthly sample is 1988:01--2019:12.
	}
	\label{fig:rhoz_oss_error}%
\end{figure}

\clearpage
\begin{table}[!t]
	\centering
	\caption[tab0: rhoz estimated from analysts forecasts]{Time-varying $\rho_z$ and Return Predictability

		\footnotesize This table reports results from regressions that link return predictability from Duration ($dr_t$) to the time-varying expected cash-flow growth persistence ($\hat{\rho_z}$).
		The dependent variables are in-sample residuals ($\varepsilon_t$) from return predictive regressions (column 1) and out-of-sample return forecast errors (column 2).
		The independent variable is the time-varying expected cash-flow growth persistence ($\hat{\rho_z}$), estimated in three-year rolling windows.
		$t$-statistics calculated based on Newey-West standard errors with 18 lags are reported in parentheses.
		Data sample: 1988:01--2019:12.
	}
	\footnotesize
	\label{tb:rhoz_t_regs}
	\begin{tabular}{lcc}
    \toprule
    {}                 & $\hat{\varepsilon}_t$ & $\upsilon_t$ \\
    \midrule
    Intercept          & -0.011                & -0.046       \\
                       & (-1.127)              & (-4.076)     \\
    $\hat{\rho}_{z,t}$ & 0.556                 & 0.469        \\
                       & (5.143)               & (4.599)      \\\\
    $N$                & 252                   & 252          \\
    $R^2$              & 0.173                 & 0.094        \\
    \bottomrule
\end{tabular}
\end{table}

\clearpage
\begin{figure}[H]
	\centering
	\begin{subfigure}[b]{.6\textwidth}
		\includegraphics[width=\textwidth]{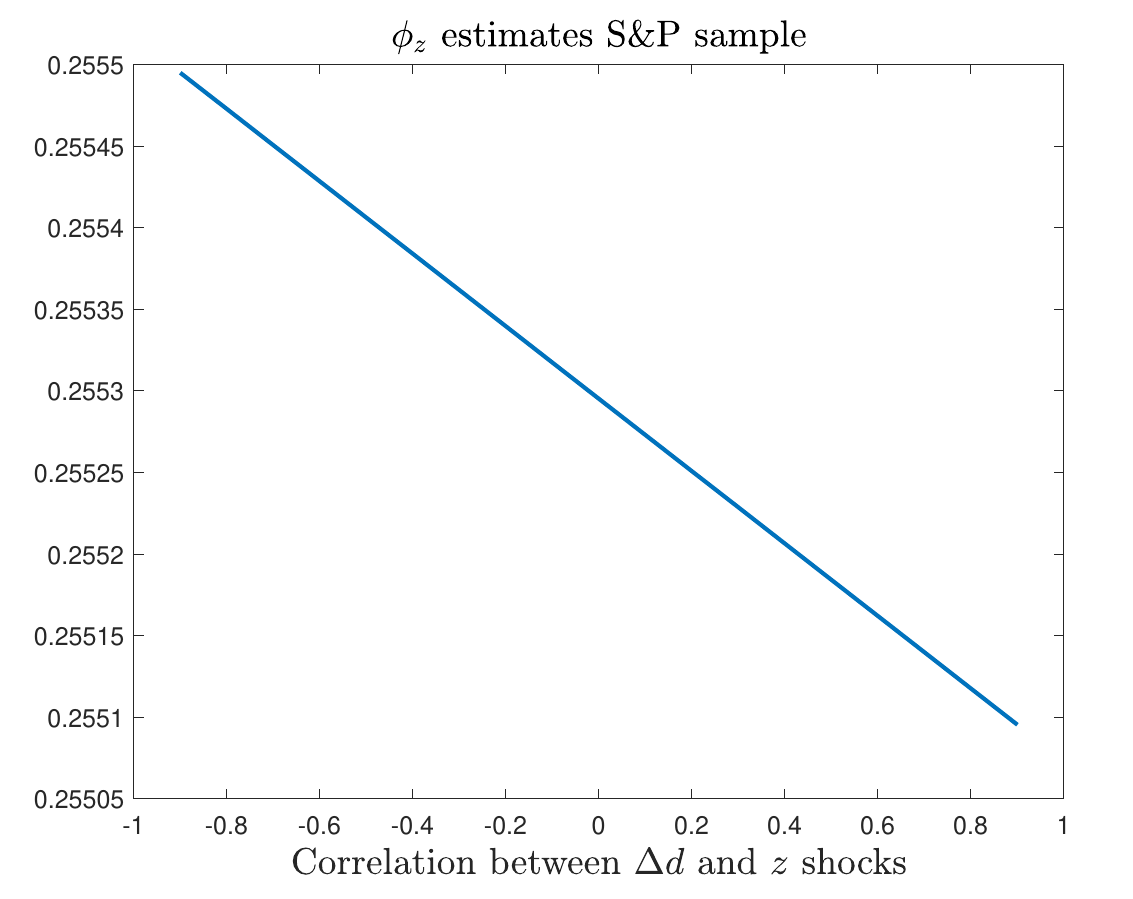}
		\caption{S\&P 500 dividend}
	\end{subfigure}
	\quad
	\begin{subfigure}[b]{.6\textwidth}
		\includegraphics[width=\textwidth]{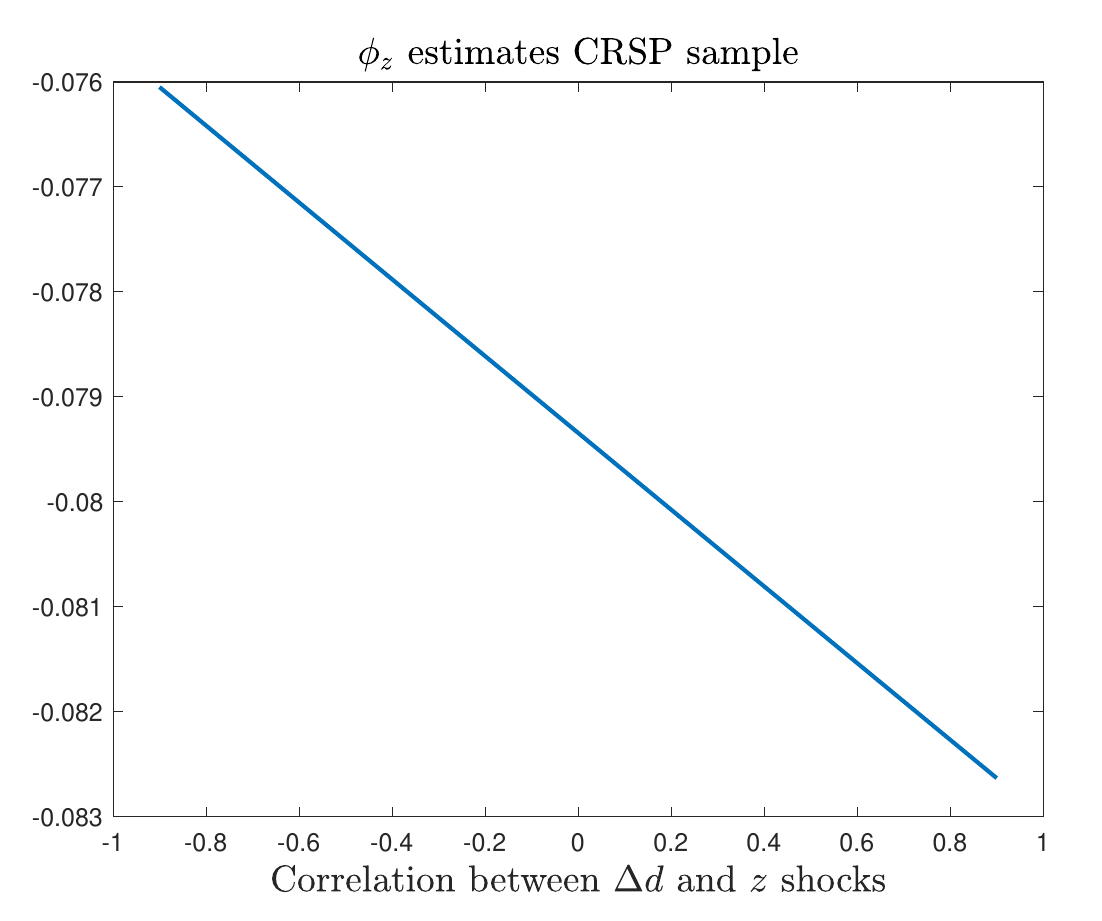}
		\caption{MKT dividend}
	\end{subfigure}
	\caption[fig: rhoz sensitivity to shock correlations]{{$ \rho_z $ Estimates from the State-Space Model with Correlated Shocks.} This figure reports the expected dividend growth autoregressive coefficient $ \rho_z $ point estimates in unrestricted state-space models as in Section \ref{sec:model} with different correlations of $ \Delta d $ and $ z $ shocks. The correlations of $ \Delta d $ and $ z $ shocks range from -0.9 to 0.9 and the volatility of $ \Delta d $ shock is calibrated to the estimated $ \hat{\sigma}_D $ from the state-space model with uncorrelated shocks. Panel A uses the annual dividend growth (non-overlapping) of the S\&P 500 index, and Panel B uses the annual dividend growth (non-overlapping) of the Fama-French market portfolio.}
	\label{fig:ssm_est}
\end{figure}

\clearpage
\begin{figure}[!t]
	\centering
	\includegraphics[scale=0.7]{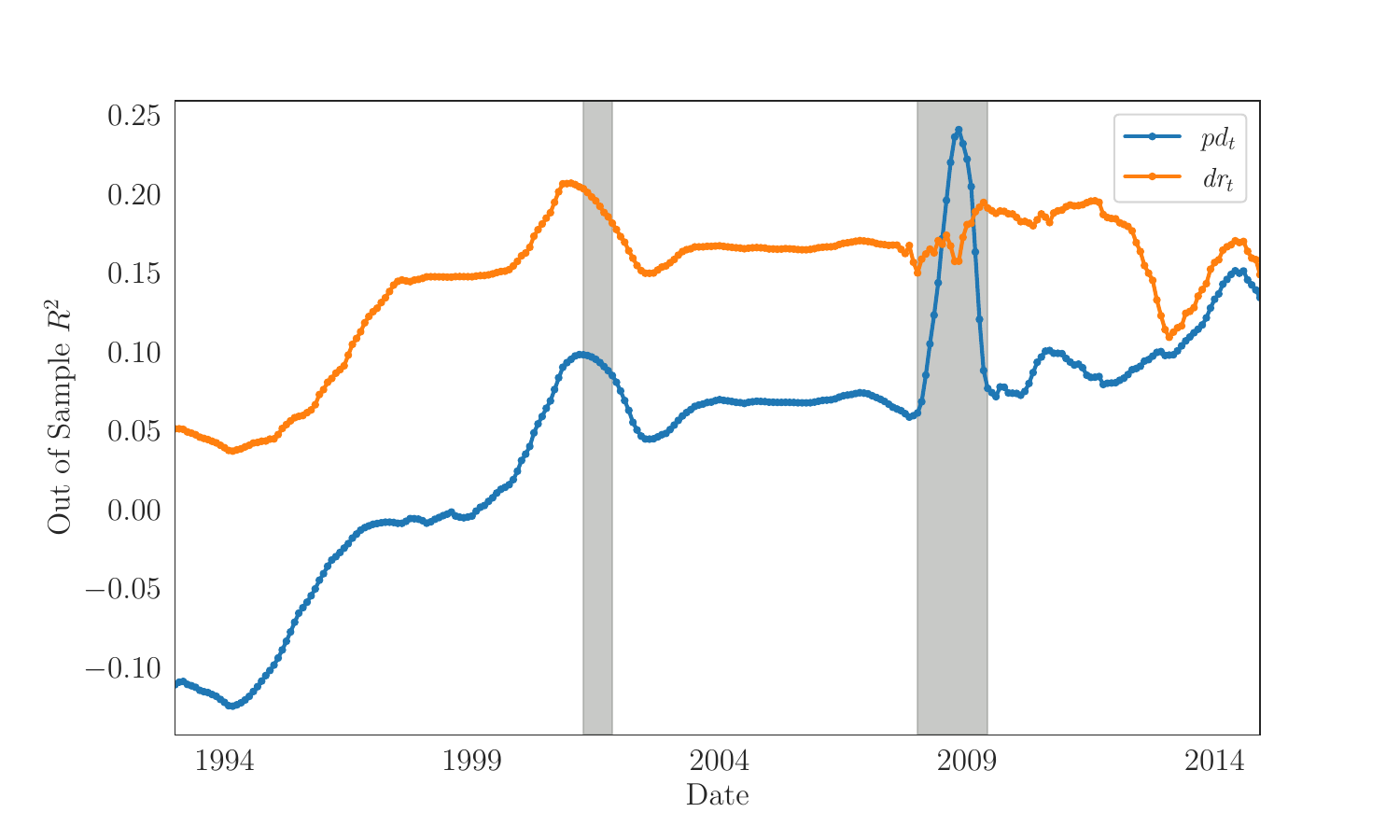}
	\caption[fig: OOS R2 with different split dates - s&p500 return]{{Out-of-sample $R^2$ by Sample Split Date.}

		\footnotesize
		This graph reports the out-of-sample $ R^2 $ of 1-year return prediction with different sample split dates.
		The first and last out-of-sample split dates are Jan 1993 and Jun 2015, respectively.}
	\label{fig:R2split}
\end{figure}

\begin{figure}[ht]
	\centering
	\begin{subfigure}[b]{0.8\textwidth}
		\centering
		\includegraphics[width=\textwidth]{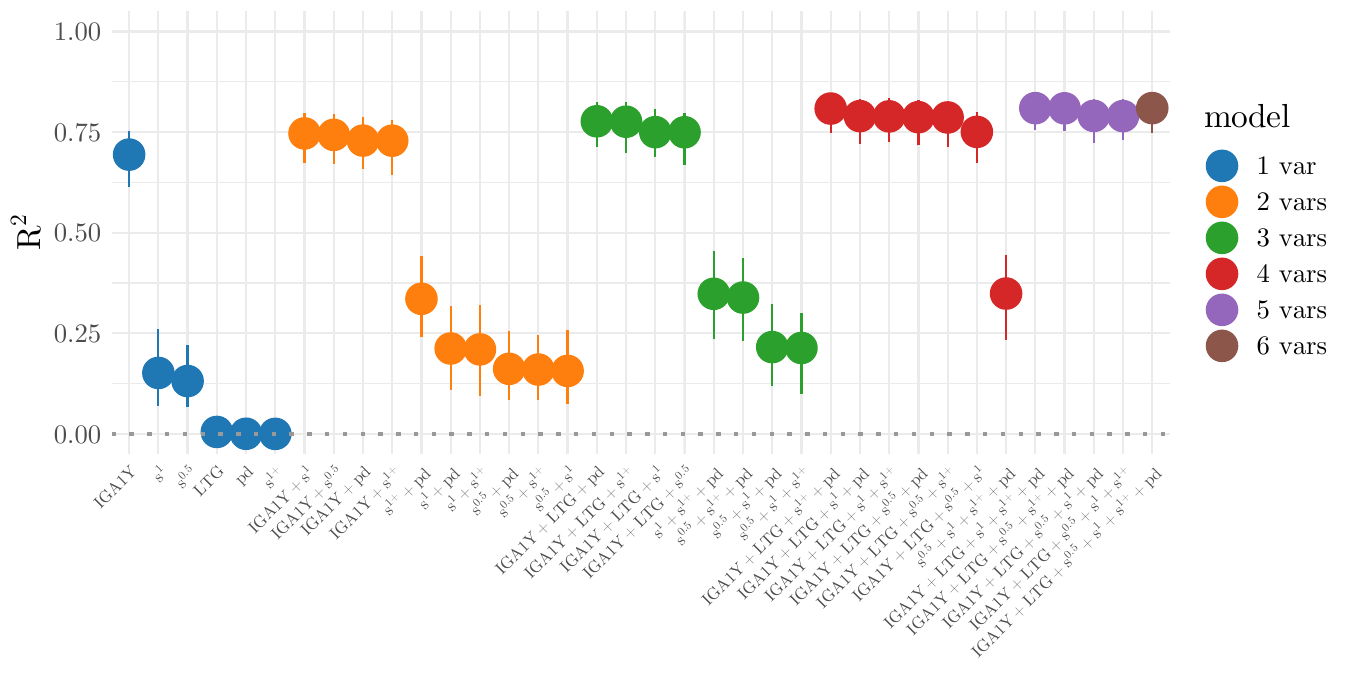}
		\caption{IGA one-year earnings growth $\Delta e_{t+1}$}
	\end{subfigure}
	\\
	\begin{subfigure}[b]{0.8\textwidth}
		\centering
		\includegraphics[width=\textwidth]{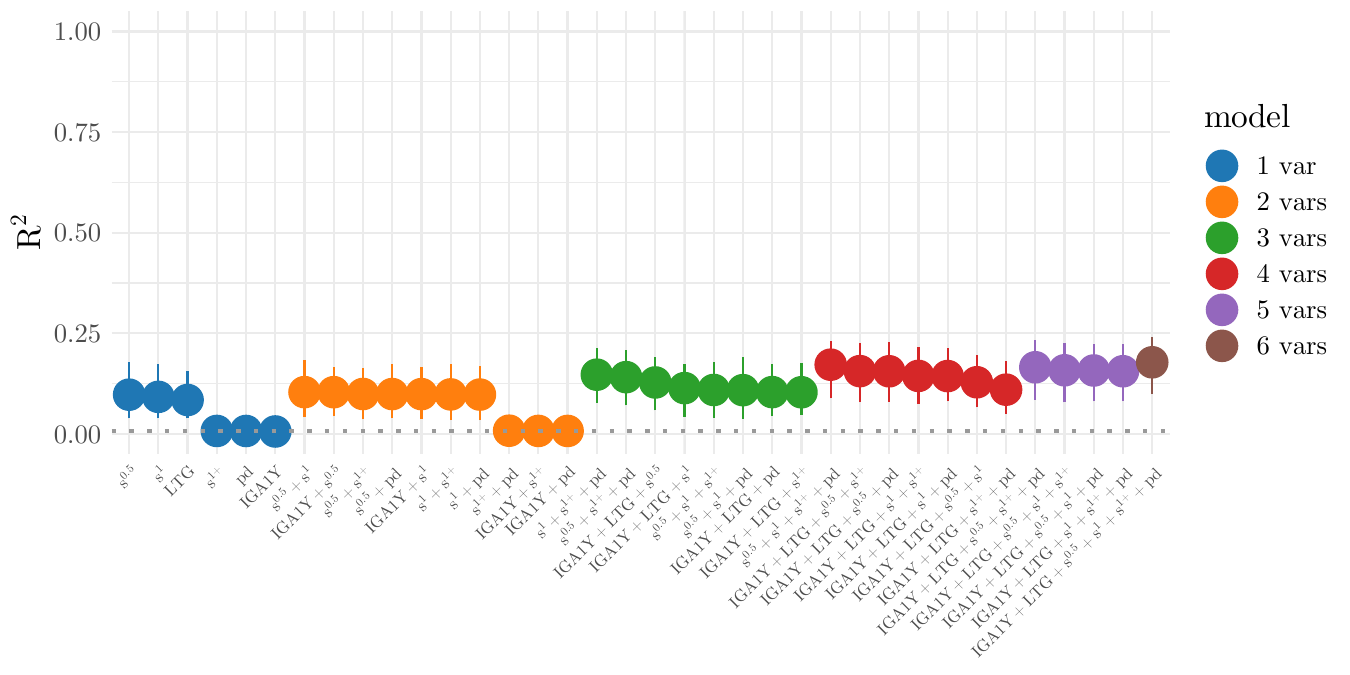}
		\caption{IGA earnings growth between years 1 and 2 $\Delta e_{t+2}$}
	\end{subfigure}
	\\
	\begin{subfigure}[b]{0.8\textwidth}
		\centering
		\includegraphics[width=\textwidth]{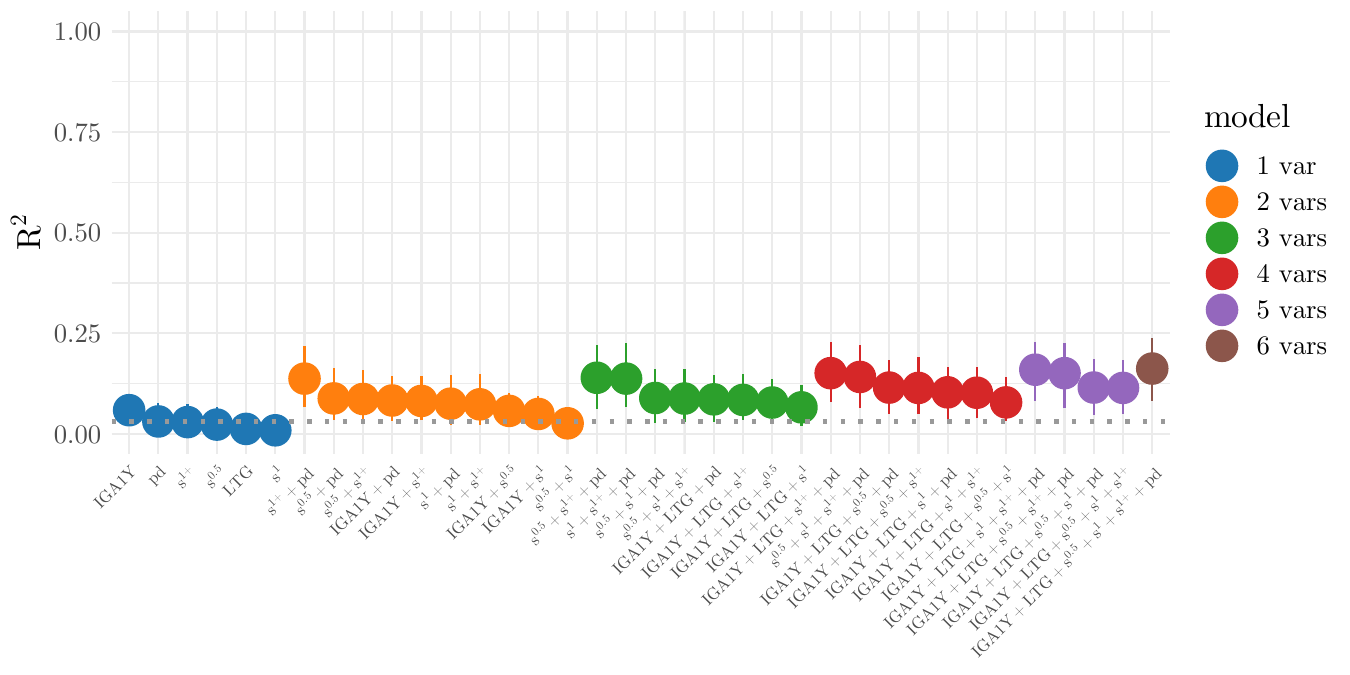}
		\caption{IGA earnings growth between years 2 and 3 $\Delta e_{t+3}$}
	\end{subfigure}
	\caption[fig: bootstrap r2 dividend growth]{$R^2$ from Earnings Growth Predictive Regressions at Various Horizons with Bootstrapped Confidence Interval.

	}
	\label{fig:bootstrap iga earnings}%
\end{figure}

\clearpage

\clearpage

\end{document}

\paragraph{International sample.} We examine the performance of the return predictor outside of the United States. The return and futures price of indices in other countries are from Datastream, and index dividends and zero-coupon bond prices are from Bloomberg. We start with all developed countries with index futures and drop a country from the sample if one of the following criteria is met: (1) futures contract with a maturity longer than or equal to one year does not exist (Germany, Hong Kong, Switzerland) or exist for less than five years (Norway); 
(2) futures price exhibits strong seasonality (Italy, Netherlands, and Switzerland) or structural break (Canada).\footnote{In the appendix, Figure \ref{fig:intlsample} plots the futures-to-spot ratio for these four countries.} For each country, our sample starts from the earliest date when index return, dividend, and futures data are all available. We end up with 1,469 country-month observations: UK (FTSE100, starting in 1993), France (CAC40, starting in 1998), Spain (IBEX35, starting in 1994), Australia (ASX200, starting in 2002), Canada (S\&P/TSX, starting in {\color{red} XXXX}) and Japan (Nikkei225, starting in 1993).

\begin{figure}[H]
	\centering
	\includegraphics[scale=0.65]{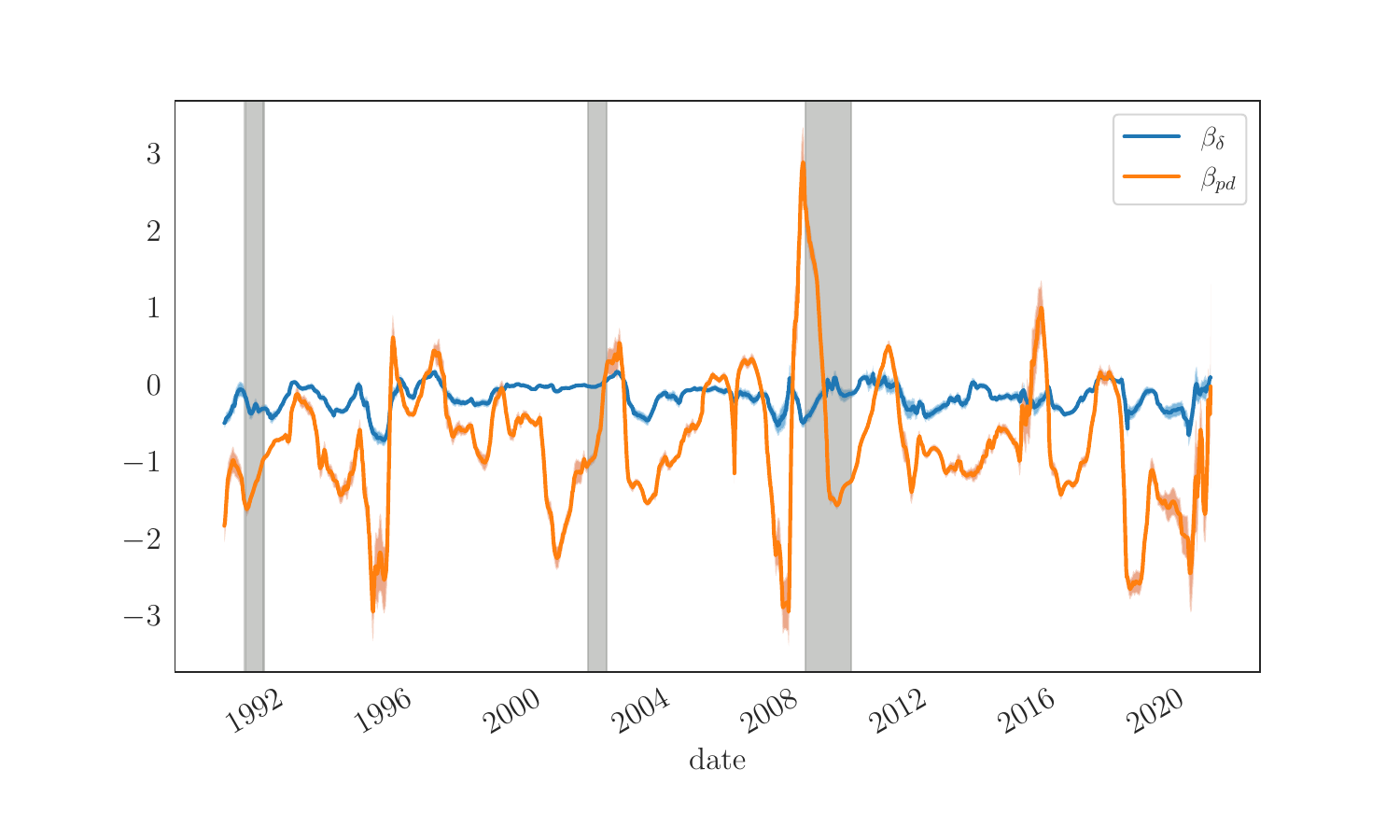}
	\caption[fig: predictive coefficient stability]{{Predictive Coefficient Stability.}

		\footnotesize
		This graph plots the coefficients from one-year return predictive regression of the S\&P 500 Index and their $95\%$ confidence interval (in shade.)
		The coefficients are estimated using a one-year rolling window of daily observations.
		The first rolling window ends in December 1988.}
	\label{fig:stability}
\end{figure}